\documentclass[useAMS,usenatbib]{mn2e}    	  %Use for submitting
\usepackage{times}
\usepackage{amssymb}
\usepackage{mathptmx}  		
\usepackage{amsmath}			
\usepackage{changepage}
\usepackage{graphicx}
\usepackage[section]{placeins}
				
\usepackage[normalsize,compatibility=false]{caption}
\usepackage{subcaption}			

\usepackage{pgfplots}			
\usepackage{psfrag}

\usepackage{enumitem}
\usepackage{balance}			
\setlist[enumerate]{leftmargin=5.5mm} 	%Keeps enumeration numerals w/in left margin

\begin{document}

\title[Star Formation and Gas Phase History of the Cosmic Web]{Star Formation and Gas Phase History of the Cosmic Web}

\author[A. Snedden et al.]{Ali Snedden$^{1}$\thanks{E-mail: asnedden@nd.edu}, Jared Coughlin$^{1}$, Lara Arielle Phillips$^{1}$, Grant Mathews$^{1}$, \newauthor In-Saeng Suh$^{1}$\\
$^{1}$University of Notre Dame, Notre Dame, IN 46556, USA}

%\date{Submitted}
%\pagerange{\pageref{firstpage}--\pageref{lastpage}} \pubyear{2014}
\maketitle
%\label{firstpage}

%================================== ABSTRACT =============================%
\begin{abstract}
We present a new method of tracking and characterizing the environment in which galaxies and their associated circumgalactic medium evolve.
We use a structure finding algorithm we developed to self-consistently parse and follow the evolution of poor clusters, filaments and voids in large scale simulations.  We trace the complete evolution of the baryons in the gas phase and the star formation history within each structure in our simulated volume.  We vary the structure measure threshold to probe the complex inner structure of star forming regions in poor clusters, filaments and voids.  We find the majority of star formation occurs in cold, condensed gas in filaments at intermediate redshifts (z $\sim$ 3).  We also show that much of the star formation above a redshift z = 3 occurs in low contrast regions of filaments, but as the density contrast increases at lower redshift star formation switches to the high contrast regions, or inner parts, of filaments.  Since filaments bridge the void and cluster regions, it suggests that the majority of star formation occurs in galaxies in intermediate density regions prior to the accretion onto poor clusters.  We find that at the present epoch, the gas phase distribution is 43.1\%, 30.0\%, 24.7\% and 2.2\% in the diffuse, WHIM, hot halo and condensed phases, respectively. The majority of the WHIM is associated with filaments. However, their multiphase nature and the fact that the star formation occurs predominantly in the condensed gas both point to the importance of not conflating the filamentary environment with the WHIM.  Moreover, in our simulation volume 8.77\%, 79.1\%, 2.11\% of the gas at z = 0 is located in poor clusters, filaments, and voids, respectively.  We find that both filaments and poor clusters are multiphase environments distinguishing themselves by different distribution of gas phases.   

\end{abstract}

%===========================================================================
%=========================            BACKGROUND              =========================
%===========================================================================
\section{Introduction}

The large scale structure (LSS) is composed of dark matter, gas and stars strung together in a system of clusters, filaments, sheets with vast voids spanning the regions between them.  From numerical simulations, we know that the filaments and sheets act as conduits vacating matter from low density regions (e.g. voids) and accreting it onto high density regions (e.g. clusters) \citep{klypin83, davis85, bertschinger91}.  These intricately woven sheets and filaments connecting high density regions together has been dubbed the ``cosmic web" \citep{bond96}.  

The structural features of the cosmic web are observed in large galaxy redshift surveys \citep{joeveer78, delapparent86, geller89, colless01, gott05}.  These surveys provide information on the distribution and morphology of galaxies within each structure.  
Initial detections of the underlying dark matter structure (e.g.~filaments) \citep{massey07, heymans08, jauzac12} were challenging due to the requirement for precise ($\sim$ 1\%) measurements of weak lensing distortions of background galaxies by foreground large scale structure.  In recent years, however, individual dark matter filaments have been confirmed and studied through weak lensing experiments \citep{dietrich05, jauzac12}.  
Motivated by measuring the matter power spectrum in a regime not affected by baryonic physics, this weak lensing technique has also been used to measure the cosmic shear due to the underlying dark matter structure \citep{bacon00, kaiser00, vanwaerbeke00, wittman00}.  More expansive surveys measuring cosmic shear are proposed (LSST, JDEM).  These surveys will, in the near future, provide a more complete map of the underlying dark matter structure (e.g.~3-D mass tomography) and constrain the time evolution of dark energy \citep{ivezic08, albrecht09}.  

The characterization of the baryonic matter in the IGM near density peaks is more complete.
 Hot gas in clusters and in some higher density filaments has been directly detected through X-ray emission \citep{gursky71, sarazin86, tittley01, nicastro03, rosati02, akamatsu11}.  Because gas emission is proportional to the density squared, this method of detecting hot gas in the LSS works well in and around galaxy clusters.  
In the low density regime, the IGM is observed via quasar absorption spectra.  
Observers rely on the absorption of high ionization species (e.g.~OVI, OVII and NeVIII) in the Ly-$\alpha$ spectrum as the primary method of detection 
\citep[and references therein]{savage98, tripp00, richter04, sembach04, danforth08, fang10, narayanan11, gupta12}. 
Since the sampling of quasars is sparse, it is challenging to produce a complete image of how the IGM is distributed within the LSS and its phase structure. 

The large scale structure environment has important consequences for the evolution and formation of galaxies.  In rich galaxy clusters, elliptical galaxies dominate \citep{abell65, oemler74, dressler80}.  They make up about 80\% of the cluster galaxies, while in the field they are only 30\% of galaxies.  These cluster galaxies tend to be redder, have less star formation \citep{dressler85, balogh97, balogh00a}, and have less gas than their field galaxy counterparts.  This suggests that field galaxies undergo a morphological transition when falling into a galaxy cluster \citep{mo10}.  There are several likely contributing factors to this morphological change, including galaxy harassment, galactic cannibalism, ram-pressure stripping and strangulation.  These factors disrupt the galactic disc, cause satellite mergers with the central dominant galaxy, shock-heat the interstellar medium (ISM) and strip accreting halo gas from in-falling galaxies \citep{gunn72, farouki81, moore98, aragon98, balogh00a, kauffmann93, vangorkom04, mo10}.  These morphological effects have been observed to exhibit a redshift dependence \citep{butcher78}. 
The evolution history of galaxies and their circumgalactic medium (CGM) will therefore be linked to the properties and star formation rates of the different environments the galaxies inhabit throughout their evolution.

Filamentary and sheet-like environments are also important in galactic evolution and may affect the accretion of intergalactic gas onto galaxies.  Numerical simulations indicate an environmental and temporal dependance on the orientation of the dark matter halo spin within filaments and sheet-like structures \citep{aragon07b, hahn07, paz08, zhang09}.  It has also been noted that dark matter halos with masses less than 10$^{12}$ $h^{-1}$M$_{\odot}$ tend to have their spin vector oriented parallel to the filament, while masses larger than 10$^{12}$ $h^{-1}$M$_{\odot}$ tend are oriented perpendicularly \citep{aragon07b, hahn07}.  

There is also observational evidence for an environmental effect on galaxy spin vector orientation \citep{navarro04, trujillo06, jones10}.  It has been suggested \citep{jones10} that tidal forces on galaxies within filaments can lead to a tendency for the angular momentum vectors of galaxies to be preferentially torqued perpendicular to the axis of the filament.  This preferred orientation affects how gas is accreted onto galaxies along with the subsequent star formation and/or AGN activity.  

Filamentary and sheet-like environments also have a morphological effect on galaxies as observed in the Sloan Great Wall \citep{gott05} and the CFA Great Wall \citep{geller89}.  These features are composed of intricate systems of interconnected super-clusters that are hundreds of $h^{-1}$Mpc in length and tens of $h^{-1}$Mpc in width.  The morphological differences between the galaxies in these large structures and the field is well documented \citep{einasto11}. 

Filamentary structure is not only important for the evolution, morphology and properties of galaxies, but it may also play a role in the ``Missing Baryon Problem" \citep{fukugita98, cen99, fukugita04a, fukugita04b, cen06a, cen06b, mathews14}. 
\cite{fukugita04b} revised their original baryon census \citep{fukugita98} and determined that the missing fraction was 35\%.  Numerical simulations have suggested that a corresponding amount of gas may in fact be located in shock heated, moderately overdense ($\delta$ $\sim$ 10 - 30) filamentary structures with temperatures in the range of $10^{5}$ - $10^{7}$K  \citep{cen99, dave01, cen06a, cen06b, shull12}.

Galaxies within large, underdense regions (e.g. voids) evolve quiescently, experiencing very few large merging events.  Because of this, void regions provide a pristine laboratory for the study of galaxy evolution \citep{peebles01b, peebles01a}.  Since the advent of large galaxy redshift surveys, voids have been studied in detail \citep{kirshner81, delapparent86, vogeley94, hoyle04, ceccarelli06}.  They have been found to occupy a large spatial component of the universe and galaxies located in these underdense regions are morphologically different when compared to their counterparts in clusters and filaments.  Void galaxies tend to be bluer, gas rich, have a higher star formation rate and be of a later galaxy type \citep{grogin99, grogin00, rojas04, kreckel11}. 

% ALI: check the font size in CIV
There is abundant evidence that the neighborhood history of the galaxies and their circumgalactic medium affects their evolution. This coupled with recent evidence that the intergalactic medium may be providing enriched (e.g. C{\small IV}) material to the circumgalactic region \citep{rubin14b}, highlights the importance of beginning to understand how the CGM interacts with the larger scale structures. We must first address the nature of these structures: the redshift history of the temperature and matter distributions, whether these structure are single-phase vs. multi-phase environments, the star formation rate history and efficiency vs. location within these structures. The phase and spatial distribution intergalactic gas will affect the gas accretion and subsequent star formation of embedded galaxies. The properties of the IGM in different structures will have direct ramifications for the galaxies and their circumgalactic medium.

To study the IGM and associated star formation, we have run cosmological simulations with radiative cooling, star formation, stellar winds, chemical enrichment and supernova feedback.  We analyze the evolution of the stars and gas in poor clusters, filaments and voids using a new structure finding algorithm \citep{snedden12,snedden14} we developed that allows us to study the properties of the IGM and dark matter structure underpinning the galaxy distribution. We create a catalogue of the particles belonging to each of these structures and obtain the structure properties.  From this catalogue, we can characterize the density-temperature evolution, star formation rate, and gas phase evolution within the different structures as a function of redshift.

We gain insight into how these cosmic structures affect the properties of the intergalactic medium. We also study the location of star formation and the environments that are conducive to it and thus begin to probe the the role that structures play in the formation and evolution of galaxies and their associated circumgalactic medium

%===========================================================================
%======================                METHODS                         =========================
%===========================================================================
\section{Large Scale Simulations}
\label{sec:simulations}
%\FloatBarrier
We have used a modified version of GADGET-2 \citep{springel01, springel05c} to generate the simulations.  It is a smoothed particle hydrodynamics (SPH) code that utilizes the TreePM method \citep{xu95} to solve for the gravity.  The computation of the gravitational force is broken up into two components, one short-range and the other long-range.  The short-range component is computed exactly by walking an oct-tree and finding the force from nearby particles.  The long-range component is approximated by computing the gravitational potential on a grid and then calculating the contribution of the force \citep{springel05c}.  

The dynamics of the system is determined by evolving the momentum and entropy equations, which expressly conserves both energy and entropy \citep{springel02}. The entropy evolution equation for the $i$th particle is then written as 
\begin{multline}
\label{eq:m1}
\frac{\text{d}A_{i}}{\text{d}t} = -\frac{\gamma - 1}{\rho_{i}^{\gamma}}\Lambda(\rho_{i}, u_{i}) + \\ \frac{1}{4}\frac{\gamma - 1}{\rho_{i}^{\gamma - 1}}\sum_{\substack{j=1}}^{N_{\text{gb}}}m_{j}\Pi_{ij}{\mathbfit{v}}_{ij}\cdot {{\bf \nabla}_{i}}\left(W(r_{ij}, h_{i} + W(r_{ij}, h_{j}\right),
\end{multline}
where $A$ is the entropic function, $\rho$ is the density,  $\gamma$ is the adiabatic index, $\Lambda(\rho_{i}, u_{i})$ is the radiative heating and cooling per unit volume, $u$ is the internal energy,  $\Pi_{ij}$ is the artificial viscosity, ${\textbf{\textit{v}}}_{ij}$ is the relative velocity between the $i$th and $j$th particle, $W(r_{ij}, h)$ is the SPH smoothing kernel, $r_{ij}$ is the relative distances between the $i$th and $j$th particles and $h_{i}$ is the particle smoothing length.  The change in entropy is interpolated from a number of nearest neighbor particles (i.e.~$N_{\text{gb}}$).  The standard version of GADGET-2 solves the hydrodynamic and gravitational equations without radiative cooling, leaving it up to the user to include the $\Lambda(\rho_{i}, u_{i})$ term while solving equation~\ref{eq:m1}.

%==========================  COOLING  SECTION =============================
\subsection{Cooling}
\label{sec:cooling}
%\FloatBarrier
The cooling timescales are comparatively smaller than the dynamic integration timescales.  So correctly resolving the radiative cooling term in equation~\ref{eq:m1} using an explicit integration technique (such as the leap-frog method that GADGET-2 already uses) would require very fine time stepping.  This is computationally prohibitive, so a semi-implicit integration scheme is required.  We use the GADGET-2 standard solver to find the adiabatic change in entropy, $\dot A^{\text{ad}}$ (i.e. the second term in equation~\ref{eq:m1}).  Then, using the isochoric approximation \citep{springel01}, we can write the new updated entropy as
\begin{equation}
\label{eq:m2}
A^{n+1} = A^{n} + \dot A^{\text{ad}} \Delta t - \frac{(\gamma - 1) \Lambda(\rho^{n}, A^{n+1})}{\rho^{\gamma}} \Delta t.
\end{equation}
The isochoric approximation allows one to ignore the change in density over a time step.  This simplifies solving for the net gas cooling rate, $\Lambda(\rho^{n}, A^{n+1})$.  We then iteratively solve equation~\ref{eq:m2} by using Brent's method from the GNU Scientific Library \citep{galassi11}. 

The net gas cooling rate, $\Lambda(\rho^{n}, A^{n+1})$, in equation~\ref{eq:m2} is linearly interpolated from a table of cooling ($\Lambda_{\text{cool}}(\rho^{n}, A^{n+1})$) and heating ($\Lambda_{\text{heat}}(\rho^{n}, A^{n+1})$) values, generated by CLOUDY version 10.0, last described by \cite{ferland98}.  At redshifts z $>$ 8.5, we assume collisionally ionized equilibrium and interpolate the cooling rate from a two dimensional cooling table that is a function of metallicity and temperature.  

\begin{figure}
	%\begin{center}
	\psfrag{Y}[cc][cc][1.2]{$| \Lambda_{\text{cool}}$ - $\Lambda_{\text{heat}} |$ / n$_{\text{H}}^{2}$ (ergs cm$^3$ / s)}
	\psfrag{X}[cc][cc][1.2]{T (K)}
	\includegraphics[width=3.25in]{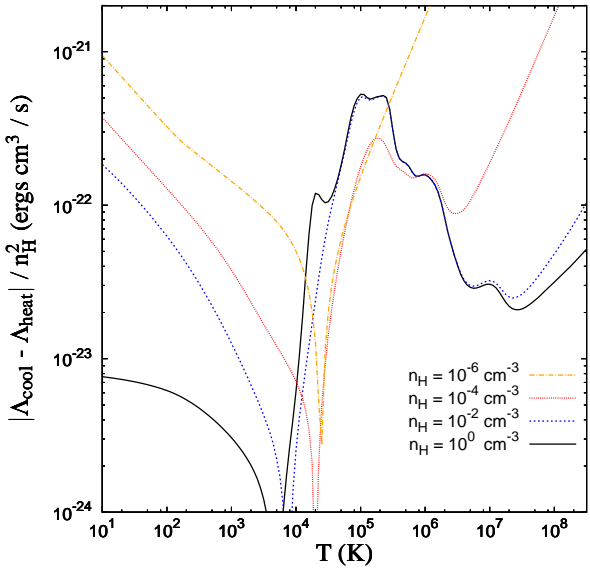}
	%\end{center}
	\vspace{-0.25in}
	\caption{Cooling curves for various hydrogen densities generated by CLOUDY version 10.0 for solar metallicity at a redshift z = 3.0 with the UV-background.  This shows the net heating and cooling of the gas.  For this plot we used the cosmological helium abundance, $n_{\text{He}}$ = 0.0789 $n_{\text{H}}$}.  
	\label{fig:uv}
\end{figure}

At redshifts z $<$ 8.5, we include the UV-background and interpolate the gas cooling rate over a four dimensional table.  The table contains over 1.6 million elements, spanning redshift, density, temperature and metallicity.  The density dependence when the UV-background is enabled is evident in Fig.~\ref{fig:uv}.  As the density gets lower, the UV heating increases, while above about T $\sim$ $3\times10^{4}$ K the cooling $also$ increases while the density decreases.  For simplicity (and so that CLOUDY could converge to a solution), we do not include molecular cooling from either $\text{H}_{2}$ or CO in the high density ($n_{\text{H}}$ $>$100 cm$^{-3}$) and low temperature regime (T $<$10$^{4}$K).   The input radiation spectrum is spatially uniform and composed of the CMB and the Haardt \& Madau (2005) UV-background as provided by CLOUDY \citep{haardt01}.  

When calculating the temperature of a gas particle, it is necessary to compute the mean molecular weight, $\mu$.  To do this, we assume that the gas is photoionized.  This is a reasonable assumption given the transparent nature of the low redshift universe \citep{schaye08}.  We neglect the small metal contribution to the mean molecular weight and associated free electron density.  This assumption is reasonable given the relatively low abundance of metals and obviates the need to solve  computationally challenging ionization species equations. 

Since we track a finite number of metal species (e.g. C, O, Ca, Cr, Mn and Fe), we must have a way to compute the metallicity of the gas particles of interest.  Because we do not track all the metals, we compare the metal mass fraction for each particle with the corresponding metal mass fraction (using the same metals) at solar metallicity.  The ratio of these two mass fractions gives the metallicity of a particle.  This is the ``particle method" for determining the metallicity of the gas.  It solely uses the mass of the metals deposited into the gas particle by stars to determine its metallicity.  This is different from the ``smoothed" metallicity method of \cite{wiersma09b}.  In the smoothed method, the SPH kernel is used to interpolate the metallicity. We use the particle method because it is less computationally expensive.

%\FloatBarrier

%==========================  STAR FORMATION  =============================
\subsection{Star Formation}
\label{sec:sf}
Star formation is an intricate process dependent on many underlying physical processes spanning a wide range of scales, from galaxy-galaxy interactions to the radiative feedback from proto-stellar cores.  The primary location of star formation is in dense giant molecular clouds (GMC).  In fact, in the solar neighborhood $all$ GMCs have some amount of star formation \citep{blitz93, williams00}.  The rate of star formation within these GMCs is inefficient and a complicated process in its own right \citep{mo10}.

The processes that governs the formation of stars span over 20 orders of magnitude in density, from galaxy halo densities ($\sim$ 10$^{-24}$ g cm$^{-3}$) to the gas densities inside stars ($\sim$ 1 g cm$^{-3}$).  Cosmological simulations simply cannot resolve such a large dynamic range and are too coarse to even resolve individual star forming regions.  Hence, we adopt a sub-grid model to follow star formation.  

We follow the familiar stochastic star formation recipe presented in \cite{katz92} and \cite{kobayashi04}.  We have three primary star formation criteria: the gas particle must be rapidly cooling relative to the dynamic timescale (equation~\ref{eq:m3}), it must be in a converging flow (equation~\ref{eq:m4}) and it must be Jeans unstable (equation~\ref{eq:m5}).  This implies:
\begin{equation}
\label{eq:m3}
t_{\text{cool}} < t_{\text{dyn}},
\end{equation}
\begin{equation}
\label{eq:m4}
\left(\nabla \cdot {\mathbfit{v}}\right) < 0,
\end{equation}
\begin{equation}
\label{eq:m5}
t_{\text{dyn}} < t_{\text{sound}},
\end{equation}
where $t_{\text{cool}}$, $t_{\text{dyn}}$ and $t_{\text{sound}}$ are defined as the cooling, dynamical (i.e. free fall) and the sound crossing times respectively.  These are given by: 
\begin{equation}
\label{eq:m6}
t_{\text{cool}} = \frac{\rho u}{\Lambda_{\text{cool}}},
\end{equation}
\begin{equation}
\label{eq:m7}
t_{\text{dyn}} = \frac{1}{\sqrt{4\pi G \rho}},
\end{equation}
\begin{equation}
\label{eq:m8}
t_{\text{sound}} = \frac{h_{i}}{c_{\text{s}}},
\end{equation}
where $h_{i}$ is the particle SPH smoothing length, G is the gravitational constant, $u$ is the thermal energy per unit mass and $c_{\text{s}}$ is the speed of sound.  The thermal energy and sound speed are found from
\begin{equation}
\label{eq:m9}
u = \frac{A(s)}{\gamma - 1} \rho^{\gamma - 1}
\end{equation}
\begin{equation}
\label{eq:m10}
c_{\text{s}} = \sqrt{\frac{\gamma P}{\rho}}
\end{equation}
where $\gamma$ is the adiabatic index, $P$ is the pressure and $A(s)$ is the entropic function (c.f. equation~\ref{eq:m1}).  

The empirical \cite{schmidt59} law is
\begin{equation}
\label{eq:m11}
\dot \Sigma_{\star} \propto \Sigma_{\text{gas}}^{N},
\end{equation}
where the power $N$ is observationally determined to be 1.4 $\pm$ 0.15 \citep{kennicutt98a}.
This suggests that the formation rate of stars is determined by the gravitational collapse of the gas.  This leads to a star formation rate that is proportional to some power of the surface gas density.  The rate at which a cloud collapses is determined by its free-fall time (i.e. dynamical time), so we assume that the star formation rate (SFR) is of the form
\begin{equation}
\frac{\text{d}\rho_{\star}}{\text{d}t} = -\frac{\text{d}\rho_{\text{gas}}}{\text{d}t} = -\frac{c_{\star} \rho_{\text{gas}}}{t_{\text{dyn}}} = -c_{\star} \sqrt{4 \pi G} \rho^{3/2}
\label{eq:m12}
\end{equation}
 where the star formation parameter, $c_{\star}$, is tuned to match observations.   
 Separating and integrating equation~\ref{eq:m12} to get the change in gas density in a time step $\Delta t$ implies that the probability ($P_{\star}$) of  forming a star in an interval $\Delta t$ is 
\begin{equation}
\label{eq:m13}
P_{\star} = 1 - \text{exp}\left( - \frac{c_{\star}}{t_{\text{dyn}}} \Delta t \right) = 1 - \text{exp}\left( -c_{\star} \sqrt{4 \pi G \rho} \Delta t \right),
\end{equation}
\citep{katz92, katz96}.  To determine if a star forms we check that our three star formation criteria (see equation~\ref{eq:m20} -~\ref{eq:m22}) are met.  If they are, we then draw a random number from a uniform deviate and compare it to $P_{\star}$.  If $P_{\star}$ is greater than the random number, then the gas particle forms a star particle.  We also set a minimum limit on time between star formation events of 2 Myr.  Many groups \citep{katz92, thacker01,  mosconi01, springel03b, okamoto05, stinson06} do not conserve particle number, they permit a single gas particle to spawn multiple star particles.  This increases the resolution of the star formation events, but can also significantly increase the run time and  memory requirements.  To save on run time and memory usage, we instead conserve particle number \citep{schaye10} by converting the entire gas particle into a star particle.  The star particle then represents a population of stars and we follow the evolution of the stellar population using our feedback routines.

%=============================  FEEDBACK  ===================================
\subsection{Feedback}
\label{sec:feedback}
The relative inefficiency of star formation is partly due to stellar feedback through AGB winds, supernovae and ionizing radiation from O and B stars.  These forms of energetic feedback can quench star formation by heating the condensed molecular gas.  This is illustrated by the fact that GMCs are often associated with star clusters that are 10 million years old or younger \citep{leisawitz89, fukui99}.  This suggests that young stellar populations, once formed, heat the ISM and kinetically drive off the cooling gas through photoionization, stellar winds and Type II supernovae.  

These feedback processes, like the star formation routines, are particularly challenging to model in cosmological simulations.  Feedback is necessary to include in cosmological simulations because, without it, excessive gas cooling will lead to exaggerated star formation \citep{larson74, white78}.  The simplest method of feedback, where the supernova and stellar wind energy is thermally deposited onto nearby gas particles, is ineffective at suppressing star formation.  This is because the gas particles near the newly formed star are still very dense and can efficiently radiate the newly deposited energy \citep{katz96, balogh01}.

There are several common ways to solve this ``overcooling problem".  One method is to distribute the feedback energy in both thermal and kinetic energy forms.  The kinetic energy takes the form of a kick in velocity of nearby gas particles \citep{navarro93, springel03a, oppenheimer06, dalla_vecchia08, booth09, dave10, oppenheimer10}.  To encourage galactic winds, the hydrodynamical force is disabled long enough for the wind particles to escape the galaxy \citep{springel03a, dalla_vecchia08, oppenheimer06, oppenheimer10, dave10}.  Another method is to inject the feedback thermally to nearby gas particles and turn off the cooling for one or more of the gas particles \citep{gerritsen97, thacker00, sommer03, stinson06, christensen10, shen10, piontek11}.  By suppressing cooling, there is time for the gas particles to adiabatically expand and for the supernova bubble to become resolved during the Sedov phase.  This attempts to account for the fact that the simulation cannot resolve the multiphase medium.  

In this work, we adopt the latter method and follow the algorithm outlined by \cite{stinson06}.  We take advantage of the SPH kernel and distribute the energy, mass and metal feedback in a weighted manner using 
\begin{equation}
\label{eq:m14}
\Delta Q_{\text{SN}} = \frac{m_{i} W(| {{\bf r}_{s}} - {{\bf r}_{j}}|, h_{\text{s}}) \Delta Q_{\text{SN}}}{\sum_{j=1}^{N_{\text{gb}}} m_{j} W(| {{\bf r}_{s}} - {{\bf r}_{j}}|, h_{\text{s}})},
\end{equation}
where $ \Delta Q_\text{SN}$ represents the quantity (e.g.~metals, mass or energy) being distributed by a star particle to its nearest $N_\text{gb}$  gas particles, $h_\text{s}$ is the star particle smoothing length, ${\bf r}_{i} - {\bf r}_{s}$ is the displacement between the star and gas particle and the denominator is the normalization.  Since the feedback only occurs when star particles (i.e. collisionless particles) deposit their feedback onto gas particles, the star smoothing length is determined by the number of nearby gas particles ($N_\text{gb}$).  We determine the star smoothing length separately from the gas smoothing length.   

Since a simulation star particle represents a population distribution of stars, we approximate all the feedback energy from Type II SNe as creating an effective Sedov blast wave.  This is a reasonable approximation because the short lifetimes of massive stars guarantee that most of the SNe will occur relatively close together (both spatially and temporally). We approximate the effective Sedov blast radius \citep{mckee77, chevalier74}  as:
\begin{equation}
\label{eq:m15}
R_{\text{E}} = 10^{1.74} E_{\text{51}}^{0.32}n_{\text{0}}^{-0.16}P_{\text{04}}^{-0.20} \text{pc}, 
\end{equation}
where $E_{\text{51}}$ is the total kinetic Type II energy in units of $10^{51}$ ergs, $n_{\text{0}}$ is the ambient hydrogen number density in cm$^{-3}$ and $P_{\text{04}} = 10^{-4} P_{\text{0}} k^{-1}$ in K/cm$^3$.  $P_{\text{0}}$ is the pressure at the location of the star particle and $k$ is the Boltzman constant (in cgs units).  If a gas particle is within the adiabatic blast wave, we temporarily (e.g.~30 Myr) disable cooling to permit the gas time to adiabatically expand.  This was the optimal time determined by \cite{stinson06} and is approximately the age of a star forming region.  We do not suppress the cooling for gas particles within the blast wave of Type Ia supernovae because they are generally not associated with active star forming regions and occur over a much larger timespan.

%=============================  CHEMODYNAMICS ==============================
\subsection{Chemodynamics}
\label{sec:chemodynamics}
In our simulations, we follow \cite{kobayashi04}'s method for chemodynamics with a few modifications.  We can trace the metal enrichment of up to 23 different elements (see \cite{kobayashi04}) along with the energy feedback from stellar winds, Type Ia and core-collapse (i.e. called ``Type II" in this paper but including Type II, Ib and Ic) supernovae.  We also follow the mass and metals returned (but not the newly synthesized metals) from AGB and super AGB (sAGB) stars.  Following the mass returned is necessary because over the lifetime of a stellar population a substantial amount of the mass ($\geq$ 33\% depending on initial mass function) is returned to the gas phase by AGB and sAGB stars with initial masses between 1 - 8 M$_{\odot}$.  Not following the newly synthesized metals contributed by AGB and sAGB stars is justified because few metals besides carbon and the s-process elements are produced.

In our simulations, each star particle represents a population of stars following the initial mass function (IMF).  Each star particle has some initial metallicity and obeys a stellar IMF.  We keep track of the maximum stellar mass by using its stellar age to determine the turnoff mass threshold.  This is done by inverting and solving for the maximum mass possible based upon the mass-age relation.  Specifically, we use
\begin{equation}
\label{eq:m16}
\text{log}_{\text{10}}\tau_{m} = 10.0 + (-3.42 + 0.88 \:\text{log}_{\text{10}}m_{t})\: \text{log}_{\text{10}}m_{t}
\end{equation}
from \citep{david90} where $\tau_{m}$ is the main sequence lifetime in years and $m_{t}$ is the turnoff mass in solar masses.  To maintain real roots when solving equation~\ref{eq:m16}, a minimum time of 5 Myr is required before evolving the turnoff mass. The energy ejected by a star particle with initial mass $m_{\star}$ is 
\begin{equation}
\label{eq:m17}
E_{\text{e}}(t) = m_{\star} \left[e_{e, \text{SW}} R_{\text{SW}}(t) + (e_{e,II} + e_{\text{SW}}) R_{\text{II}}(t) + e_{e, \text{Ia}} R_{\text{Ia}}\right]
\end{equation}
where $R_{\text{SW}}$, $R_{\text{II}}$ and $R_{\text{Ia}}$ are the rates of stellar winds, core collapse supernovae and Type Ia supernovae, respectively, in units of number per $M_{\odot}$ and $e_{e, \text{SW}}$, $e_{e, \text{II}}$ and $e_{e, \text{Ia}}$ are their respective energies.  The energy ejected by AGB and sAGB stars is relatively small and not included.  The energies per event (e.g.~stellar wind, supernovae etc.) are defined  \citep{kobayashi04}
\begin{equation}
\label{eq:m18}
e_{e, \text{SW}} =  \begin{cases} 
										0.2 \times 10^{51} \left(\dfrac{Z}{Z_{\odot}}\right)^{0.8} & (m_{2, u} < m \leq m_{u})\\
										 0.2 \times 10^{51} & (m_{2, l} < m \leq m_{2, u})
					\end{cases},
\end{equation}
\begin{equation}
\label{eq:m19}
e_{e, \text{II}} = 1.4 \times 10^{51}(\text{erg}) \quad \quad \\(m_{2, l} < m \leq m_{2, u}) 
\end{equation}
and
\begin{equation}
\label{eq:m20}
e_{e, \text{Ia}} = 1.3 \times 10^{51} (\text{erg}) \quad \quad \\(m_{1,l} < m < m_{1, u}).  
\end{equation}
The metallicity dependence in equation~\ref{eq:m18} is due to the metallicity dependence of the strength of the wind \citep{leitherer92}. The respective mass limits for equation~\ref{eq:m18} are listed in Table~\ref{tab:m2}.  The rates (in units of number per $M_{\odot}$) of stellar winds (SW), Type II SN and AGB stars are given by
\begin{equation}
\label{eq:m21}
R_{\text{SW}} = \int^{m_{t}(t)}_{m_{t}(t + \Delta t)} \phi(m) dm \quad \quad  (m_{2, u} < m_{t} \leq m_{u}),
\end{equation}
\begin{equation}
\label{eq:m22}
R_{\text{II}} = \int^{m_{t}(t)}_{m_{t}(t + \Delta t)} \phi(m)dm \quad \quad  (m_{2, l} < m_{t} \leq m_{2, u})
\end{equation}
and
\begin{equation}
\label{eq:m23}
R_{\text{AGB}} = \int^{m_{t}(t)}_{m_{t}(t + \Delta t)} \phi(m) dm \quad \quad  (m_{l} \leq m_{t} \leq m_{2, l}).
\end{equation}
respectively.  The Type Ia supernovae rate is more complicated than the AGB and Type II supernova rates which solely depend upon the progenitor's initial mass and composition.   Computing the Type Ia supernovae rate is a difficult task because the progenitors are uncertain, however, there are two favored models.  One is the doubly degenerate model whereby two white dwarfs, in a binary system, merge after losing orbital energy through gravitational radiation.  The other model is singly degenerate whereby a white dwarf accretes mass from a main sequence or red giant companion star \citep{podsiadlowski08}.  A Type Ia explosion occurs as the white dwarf approaches the Chandrasekhar limit and nuclear burning is ignited.  Due to the degenerate nature of the white dwarf, pressure is independent of temperature and a runaway thermonuclear explosion ensues.    

Either (or even both) the single or the doubly degenerate channels may be the physical mechanism that actually generates Type Ia supernovae.  The most common method used in previous simulations follows the singly degenerate model of \cite{greggio83} \citep{portinari98, kobayashi98, kobayashi04, stinson06, christensen10, shen10}.  This requires detailed knowledge of the binary mass fraction, initial mass fraction and makes assumptions on the actual progenitor systems, all of which are fraught with uncertainty \citep{wiersma09b}.  Given these difficulties, we adopt a method based on the observations that does not contain any assumptions about the progenitors \citep{wiersma09b, vogelsberger13}.  It is based upon the observation that Type Ia supernovae simply follow after a delay from a star formation event.

This is parameterized by the delay-time distribution (DTD) \citep{dahlen04, greggio05, mannucci06, matteucci06, maoz12a}.   Following \cite{vogelsberger13}, one can write the Type Ia supernova rate as 
\begin{equation}
\label{eq:m24}
R_{\text{Ia}} = \int^{t + \Delta t}_{t}g(t'-t_{\text{0}})dt' 
\end{equation}
where $t_{0}$ is the birth time of a stellar population's and $g(t'-t_{0})$ is the normalized \citep{maoz12a} power law DTD with
\begin{equation}
\label{eq:m25}
g(t) = \begin{cases} 0 & \text{if } t < \tau_{\text{8} M_{\odot}}\\
		 N_{0} \left(\frac{t}{\tau_{\text{8} M_{\odot}}}\right)^{-s} \frac{s-1}{\tau_{\text{8} M_{\odot}}}  & \text{if } t \geq \tau_{\text{8} M_{\odot}}
		 \end{cases},
\end{equation}
where the normalization is $N_{\text{0}}$ = 1.3 $\times$ 10$^{-3}$ [SN M$_{\odot}$], $\tau_{8 M_{\odot}}$ is the lifetime of an 8 M$_{\odot}$ star and the power law index is $s$ = 1.12 as determined by \cite{maoz12a}.  

\begin{figure}
	\begin{center}
	\includegraphics[width=3.25in]{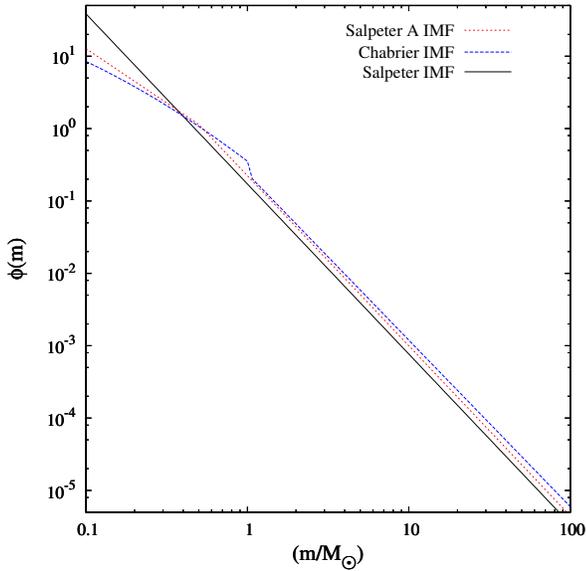}
	\end{center}
	\caption{The normalized Salpeter, Salpeter A and Chabrier IMF's.}
	\label{fig:imf}
\end{figure}

To determine $R_{\text{SW}}$, $R_{\text{II}}$ and $R_{\text{AGB}}$, we specify a stellar initial mass function (IMF).  We have tested several different initial mass functions (IMFs) including the Salpeter \citep{salpeter55}, Salpeter A \citep{baldry03} and the Chabrier IMF \citep{chabrier03a}.  The Salpeter IMF is
\begin{equation}
\label{eq:m26}
\phi (m) \propto m^{-2.35}, 
\end{equation}
the Salpeter A IMF is
\begin{equation}
\label{eq:m27}
\phi (m) = \begin{cases} A \: m^{-1.5} & (m \leq 0.5 M_{\odot})\\
											  B \: m^{-2.35} & (m > 0.5 M_{\odot})
						\end{cases},
\end{equation}
and the Chabrier IMF is
\begin{equation}
\label{eq:m28}
\phi (m) =  \begin{cases} A \: \text{exp}\left(-[ \text{log}_{\text{10}}(m  /  0.079 M_{\odot})]^{2}/0.6\right)m^{-1} & (m \leq 1 M_{\odot})\\
                                                        B \:  m^{-2.3}  & (m > 1 M_{\odot})
                                \end{cases},
\end{equation}
where the constants $A$ and $B$ are chosen such that the IMFs are continuous across the piecewise steps.  The Chabrier and Salpeter A IMFs are top-heavy relative to the Salpeter IMF  (see Fig.~\ref{fig:imf}) and contain approximately twice as many Type II progenitor stars.  The IMF is normalized such that 
\begin{equation}
\label{eq:m29}
\int_{m_{l}}^{m_{u}} m \phi (m) dm = 1.
\end{equation}

In addition to energy feedback, we also calculate the stellar mass and the mass of metals returned to the gas phase via stellar winds and supernovae.  The total amount of mass ejected by a star particle is computed from  
\begin{equation}
\label{eq:m30}
E_{m}(t) = m_{\star}\left[e_{m, \text{SW}}(t) + e_{m, \text{II}}(t) + e_{m, \text{Ia}}(t) + e_{m, AGB}\right],
\end{equation}
where the mass ejection rates for all four processes, stellar winds, Type II supernovae, Type Ia supernovae and AGB  are given by 
\begin{equation}
\label{eq:m31}
e_{m, \text{SW}} = \left(\frac{Z}{Z_{\odot}}\right)^{0.8} \int^{m_{t}(t)}_{m_{t}(t + \Delta t)}(1-w_{m})m\phi (m) dm   \quad   (m_{2, u} < m_{t} \leq m_{u}),
\end{equation}
\begin{equation}
\label{eq:m32}
e_{m, \text{II}} = \int^{m_{t}(t)}_{m_{t}(t + \Delta t)}(1 - w_{m})m\phi(m)dm  \quad \quad  (m_{2, l} < m_{t} \leq m_{2, u}),
\end{equation}
\begin{equation}
\label{eq:m33}
e_{m, \text{Ia}} = m_{\text{CO}}R_{\text{Ia}}(t)
\end{equation}
and
\begin{equation}
\label{eq:m34}
e_{m, \text{AGB}} = \int^{m_{t}(t)}_{m_{t}(t + \Delta t)}(1 - w_{m})m\phi(m)dm \quad \quad  (m_{l} \leq m_{t} \leq m_{2, l})
\end{equation}
respectively, where $w_{m}$ is the remnant mass fraction (see Table~\ref{tab:m3}).  Since we do not follow the metallicity dependance of $w_{m}$ explicitly, we use the remnant mass fractions for stars at solar metallicity.  Thus, the ($Z / Z_{\odot}$) dependence in equation \ref{eq:m18} and~\ref{eq:m31} act as a proxy for the remnant mass dependance on metallicity.  It should be noted that the remnant mass fraction is poorly understood \citep{eldridge04}.  The total metal mass, for the $i$th metal, ejected is
\begin{equation}
\label{eq:m35}
E_{Z_{i}}(t) = m_{\star}\left[e_{Z_{i}, \text{SW}}(t) + e_{Z_{i}, \text{II}}(t) + e_{Z_{i}, \text{Ia}}(t)\right].
\end{equation}
The metals are ejected into the ISM at the using equation~\ref{eq:m36} -~\ref{eq:m39}, which include supernova nucleosynthesis but neglect the nucleosynthesis from AGB stars.  The stellar winds, Type II SNe and Type Ia SNe metal feedback rates are  
\begin{multline}
\label{eq:m36}
e_{Z_{i}, \text{SW}} = \left(\dfrac{Z}{Z_{\odot}}\right)^{0.8} \int^{m_{t}(t)}_{m_{t}(t + \Delta t)}(1 - w_{m})Z_{i}m\phi(m)dm \\ \quad \quad  (m_{2, u} < m_{t} \leq m_{u}),
\end{multline}
\begin{multline}
\label{eq:m37}
e_{Z_{i}, \text{II}} =   \int^{m_{t}(t)}_{m_{t}(t + \Delta t)}(1 - w_{m} - p_{Z_{i}m, \text{II}})Z_{i}m\phi (m) dm\\ 
\phantom{{}=2} +   \int^{m_{t}(t)}_{m_{t}(t + \Delta t)} p_{Z_{i}m, \text{II}} m\phi(m)dm \quad \quad  (m_{2, l} < m_{t} \leq m_{2, u}),
\end{multline}
\begin{equation}
\label{eq:m38}
e_{Z_{i}, \text{Ia}} = m_{\text{CO}}p_{Z_{i}m, \text{Ia}} R_{\text{Ia}}(t)
\end{equation}
and
\begin{equation}
\label{eq:m39}
e_{Z_{i}, \text{AGB}} = \int^{m_{t}(t)}_{m_{t}(t + \Delta t)}  (1 - w_{m})Z_{i}m\phi(m)dm   \quad  (m_{l} \leq m_{t} \leq m_{2, l}),
\end{equation}
respectively, where $p_{Z_{i}m, \text{II}}$ and $p_{Z_{i}m, \text{Ia}}$ are the metal mass fraction yields for Type II and Type Ia SNe respectively.  The Type II SN yield ($p_{Z_{i}m, \text{II}}$) is linearly interpolated from core collapse supernova nucleosynthesis model presented by \cite{kobayashi06} and the Type Ia yield ($p_{Z_{i}m, \text{Ia}}$) is directly taken from \cite{nomoto97a}.  Since we do not consider metallicity dependent metal yields or remnant mass fractions, we have adopted the Type II yields from \cite{kobayashi06} at solar metallicity to stay consistent with the remnant mass fractions used.  The metallicity tables from \cite{kobayashi06} only go down to 13 M$_{\odot}$, while the minimum mass for a core collapse supernova is 8 M$_{\odot}$.  We assume that stars with masses between 8 - 10 M$_{\odot}$ undergo an electron capture supernova and do not produce a significant amounts of metals.  For stars with masses between 10 - 13 M$_{\odot}$ we interpolate from 0 yield to the metal yield at 13 M$_{\odot}$ \citep{nomoto97b}.

%__________________________________________________________________________
\begin{table}
	\centering
	\begin{tabular}{| l | c | l|}
	\hline
		Variable & Mass ($M_{\odot}$) & Description\\
		\hline 
		   m$_{u}$            & 100     & IMF upper limit\\
		   m$_{2, u}$        & 40       & Type II SN progenitor upper limit\\
		   m$_{l, \text{BH}}$     & 25       & Black hole progenitor lower limit\\
             m$_{2, l}$         & 8         & Type II SN progenitor lower limit\\
             m$_{1, u}$        & 8         & Type Ia progenitor upper limit\\
             m$_{l}$             & 0.1   & Lowest stellar mass possible\\ 
		\hline
		\end{tabular}
	\caption[Chemodynamic Mass Limits]{Table of mass limits used in the supernovae feedback scheme.}
\label{tab:m2}
\end{table}
%__________________________________________________________________________
\begin{table*}
	\centering
	\begin{tabular}{| l | l | l|}
	\hline
		Value for w$_m$ & Regime & Reason or Reference\\
		\hline 
		   m            				  & 0.1 $M_{\odot}$ $<$ m $\leq$ 0.8 $M_{\odot}$ & Star on M.S. for $>$ 13.6 Gy\\
		   -0.58m + 1.13      & 0.8 $M_{\odot}$ $<$ m $\leq$ 1.0 $M_{\odot}$   & \cite{marigo01}, Fig. 8 \\
		   -0.058m + 0.662   & 1.0 $M_{\odot}$ $<$ m $\leq$ 8.0 $M_{\odot}$    & \cite{marigo01}, Fig. 8\\
             -4.49E$^{-3}$m + 0.18 & 8.0 $M_{\odot}$ $<$ m $\leq$ 25 $M_{\odot}$ & \cite{kobayashi06}, Table 1\\
             -8.2E$^{-4} + 0.0881$& 25 $M_{\odot}$ $<$ m $\leq$ 40 $M_{\odot}$   & \cite{kobayashi06}, Table 1\\
             -5.68E$^{-4}$ + 0.078 & 40 $M_{\odot}$ $<$ m $\leq$ 100 $M_{\odot}$ &\cite{portinari98}, Fig. 8\\
 		 \hline
		\end{tabular}
	\caption[Remnant Mass Function]{List of remnant mass fractions for progenitor stars of solar metallicity.  The remnant masses were taken from the listed sources, converted to remnant mass fractions and interpolated.}
\label{tab:m3}
\end{table*}

%===========================================================================
%=========================== CURRENT PROGRESS =========================%
%===========================================================================
\section{Verification}
\label{sec:verification}
To determine the best values for the free parameters in our simulation,
we ran a suite of GADGET-2 simulations at smaller volumes (25$h^{-1}$Mpc)$^{3}$ and with 2 $\times$ 128$^{3}$ paticles.  The initial conditions were generated using second-order Lagrangian perturbation theory described by \cite{scoccimarro12}.  The simulations began at a redshift z = 49 with  an initial gas temperature of 34K (see Chapter 9 of \cite{ellis12}).  For the cosmology, we use the 7-year WMAP \citep{komatsu11} results.  We chose a $\Lambda$CDM cosmology with $\Omega_{\text{m}}$ = 0.274, $\Omega_{\Lambda}$ = 0.726, $\Omega_{\text{b}}$ = 0.0456 and $h$ = $H_0$/(100 km s$^{-1}$ Mpc$^{-1}$) = 0.702.

We tested the Salpeter, modified Sapleter A (SalA) and Chabrier IMFs (see Fig.~\ref{fig:imf}) at several different star formation parameter values (see equation~\ref{eq:m12}) with 0.01 $\leq$ $c_{\star}$ $\leq$ 0.1.  The star formation history for each simulation was then compared to the observed star formation history \citep{hopkins06}.  To appropriately compare the results from our simulations, we scaled our star formation history by factors of 0.77 and 0.606 for the Salpeter A and Chabrier IMFs, respectively \citep{hopkins06, wiersma09b}.  

\begin{figure}
	\begin{center}
	\psfrag{SFR}[cc][cc][1.0]{SFR (M$_{\odot}$ yr$^{-1}$ Mpc$^{-3}$)}
	\psfrag{z}[cc][cc][1.0]{Redshift}
	\psfrag{100Mpc   }[cc][cc][0.8]{(100$h^{-1}$Mpc)$^{3}$}
	\psfrag{50Mpc   }[cc][cc][0.8]{(50$h^{-1}$Mpc)$^{3}$}
	\psfrag{25Mpc   }[cc][cc][0.8]{(25$h^{-1}$Mpc)$^{3}$}
	\includegraphics[width=3.25in]{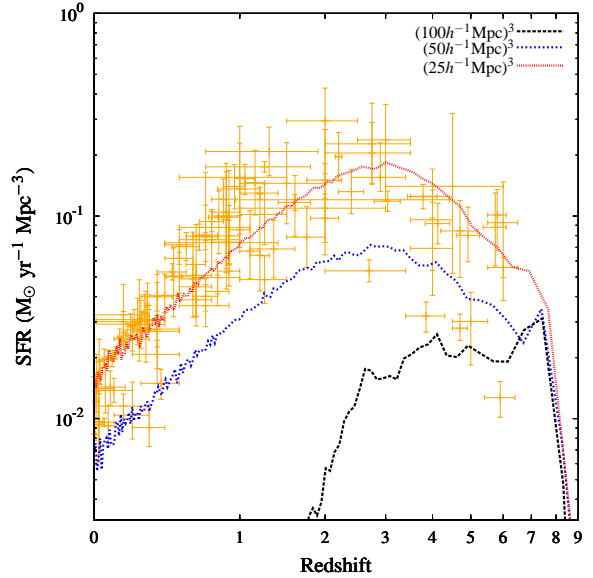}
	\end{center}
	\caption{The cosmic star formation history for three simulations with the star formation parameter c$_{\star}$ = 0.015 and a SalA IMF.  Each simulations has a different volume (e.g. (25$h^{-1}$Mpc)$^{3}$, (50$h^{-1}$Mpc)$^{3}$ and (100$h^{-1}$Mpc)$^{3}$) but each has 2 $\times$ 128$^{3}$ particles.  The observational data is from \protect\cite{hopkins06}.}
	\label{fig:sfd_resol}
\end{figure}

\begin{figure}
	\begin{center}
	\psfrag{100Mpc}[rc][cc][0.8]{2$\times$512, (100$h^{-1}$Mpc)$^{3}$}
	\psfrag{50Mpc}[rc][cc][0.8]{2$\times$256, (50$h^{-1}$Mpc)$^{3}$}
	\psfrag{25Mpc}[rc][cc][0.8]{2$\times$128, (25$h^{-1}$Mpc)$^{3}$}
	\psfrag{SFR}[cc][cc][1.0]{SFR (M$_{\odot}$ yr$^{-1}$ Mpc$^{-3}$)}
	\psfrag{z}[cc][cc][1.0]{Redshift}
	\includegraphics[width=3.25in]{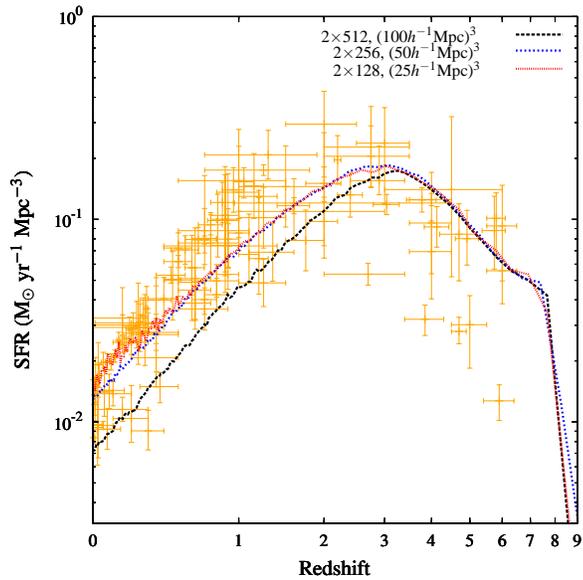}
	\end{center}
	\caption{The cosmic star formation history in three simulations with identical resolution, star formation parameter c$_{\star}$ = 0.015 and a SalA IMF.  The three runs are : 2 $\times$ 128$^{3}$ particles in a (25$h^{-1}$Mpc)$^{3}$ volume, 2 $\times$ 256$^{3}$ particles in a (50$h^{-1}$Mpc)$^{3}$ volume and 2 $\times$ 512$^{3}$ particles in a (100$h^{-1}$Mpc)$^{3}$ volume.  The observational data is from \protect\cite{hopkins06}.}
	\label{fig:sfd_resol_256}
\end{figure}

\begin{figure}
	\begin{center}
	\includegraphics[width=3.25in]{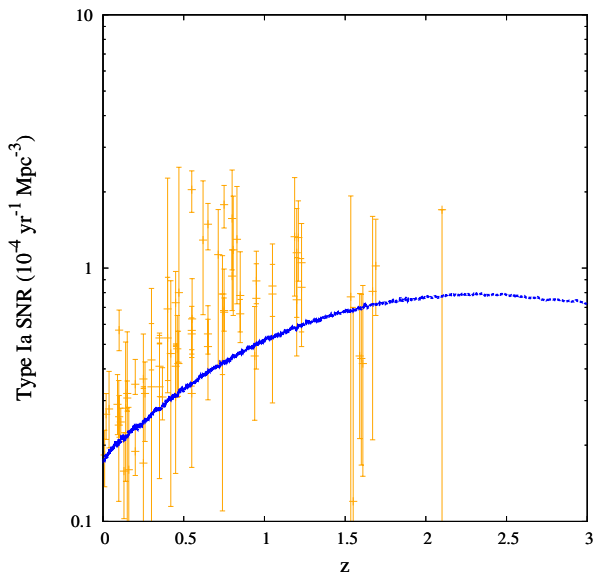}
	\end{center}
	\caption{The volumetric Type Ia SN rate for a simulation with 2$\times$128$^{3}$ particles using the SalA IMF and c$_{\star}$ = 0.015.  The observational data is taken from \protect\cite{graur14}}.
	\label{fig:typeIa}
\end{figure}

Using a $\chi^{2}$ minimization fit with the observational data of the star formation history, we found that the SalA IMF with a star formation parameter value equal to 0.015 is the optimal choice.  This ensures (see Figs.~\ref{fig:sfd_resol} and~\ref{fig:sfd_resol_256}) that our star formation and feedback routines can reproduce a reasonable star formation history of the universe.
We also kept track of the number of Type Ia supernovae in our simulation and compared it the observed data (see Fig.~\ref{fig:typeIa}).  In spite of not explicitly tuning our simulations to fit the number of Type Ia SN, we still get a rate that is within the observational range.  This independently confirms that our adopted star formation parameters are reasonable.

In Fig.~\ref{fig:sfd_resol}, we plot the three simulations, each with 2 $\times$ 128$^{3}$ particles, with different volume sizes (25$h^{-1}$Mpc)$^{3}$, (50$h^{-1}$Mpc)$^{3}$ and (100$h^{-1}$Mpc)$^{3}$.  The mass resolution has a strong effect on the star formation rate.  So to scale up the spatial size of the simulations, we decided to keep the same mass resolution as the (25$h^{-1}$Mpc)$^{3}$ run.  However, keeping the mass resolution the same and increasing the volume does not guarantee that the larger simulation will have a star formation rate that fits the observed data as well as the (25$h^{-1}$Mpc)$^{3}$ run.  This is seen in Fig.~\ref{fig:sfd_resol_256} where the (50$h^{-1}$Mpc)$^{3}$ run with 2 $\times$ 256$^{3}$ particles does not fit the observed data as well as the (25$h^{-1}$Mpc)$^{3}$ run with 2 $\times$ 128$^{3}$ particles, in spite of having the same mass resolution.  Following the resolution dependence of the feedback and star formation algorithm are beyond the scope of this paper.  In spite of the challenges of scaling up simulations while keeping the mass resolution constant, it is often the best that can be done given the finite computational resources available. 

We use this method of holding the mass resolution constant to scale up our simulations,   
to (50$h^{-1}$Mpc)$^{3}$ with 2 $\times$ 256$^{3}$ using the optimal IMF and $c_{\star}$ values.  This scaled up run is presented in Section~\ref{sec:defining_structure} and~\ref{sec:temp_and_rho_evol}.

%__________________________________________________________________________
%__________________________________________________________________________

%===========================================================================
%=============================== RESULTS ==================================%
%==========================================================================
\section{Defining Structure}
\label{sec:defining_structure}

To study the redshift history of the temperature, density and phase distributions of the baryonic gas, as well as the star formation rate and efficiency, in groups/poor clusters, filaments and voids, we ran a GADGET-2 simulation with 256$^{3}$ gas and 256$^{3}$ dark matter particles in a (50$h^{-1}$Mpc)$^{3}$ volume using the optimal parameters and initial conditions described in Section~\ref{sec:verification} and the physics described in Sections~\ref{sec:feedback} and~\ref{sec:chemodynamics}.     

We output our data in 31 logarithmically spaced snapshots spanning redshifts 0 $\geq$ z $\geq$ 10.  To analyze our simulations we used the segmentation algorithm described in \cite{snedden14}. Following the logic from \cite{frangi98} and \cite{descoteaux04} and discussed in detail in \cite{snedden14}, the structure measures for the clusters, filaments and voids are defined by 3 quantities:
\begin{equation}
V_{\text{c}} = (1- e^{-2 |\lambda_1/\lambda_3|^2})(1 - e^{-2 |F_{\text{norm}}|^2})~~,
\label{eq:r1}
\end{equation}
\begin{equation}
V_{\text{f}} = (1- e^{-2 |\lambda_1/\lambda_2|^2})(1 - e^{-2 |F_{\text{norm}}|^2})~~,
\label{eq:r2}
\end{equation}
\begin{equation}
V_{\text{v}} = (1- e^{-2 |\lambda_1/\lambda_3|^2})(e^{-2 |F_{\text{norm}}|^2}),
\label{eq:r3}
\end{equation}
respectively.  The Frobenius norm, $F_{\text{norm}}$, is
\begin{equation}
F_{\text{norm}} = ({\lambda_1}^2 + {\lambda_2}^2 + {\lambda_3}^2)^{\frac{1}{2}}
\label{eq:r4}
\end{equation}
and $\lambda_1$, $\lambda_2$ and $\lambda_3$ are the eigenvalues resulting from solving the Hessian matrix constructed from the second order 
spatial derivatives of the gas density.  In conjunction with these definitions for the structure measure, \cite{snedden14} used a ``density criterion" to enforce the fact that filaments and clusters are in overdense regions, while voids are in underdense regions.  At each snapshot, the structure measures are normalized on a scale of 0 to 1 to facilitate comparison.  It is important to note that given the size of our simulations, the clusters probed are likely poor clusters / Local group galaxy systems.  The filaments probed were on the order of a few megaparsecs in size.

\begin{figure*}
	%\begin{center}
	$\begin{array}{cc}
	\begin{subfigure}{.5 \textwidth}
		\begin{center}
		\psfrag{GasMass}[cc][cc][0.8]{Gas Mass (M$_{\odot}$)}
		\psfrag{z}[cc][cc][0.8]{Redshift}
		\includegraphics[width=3.25in]{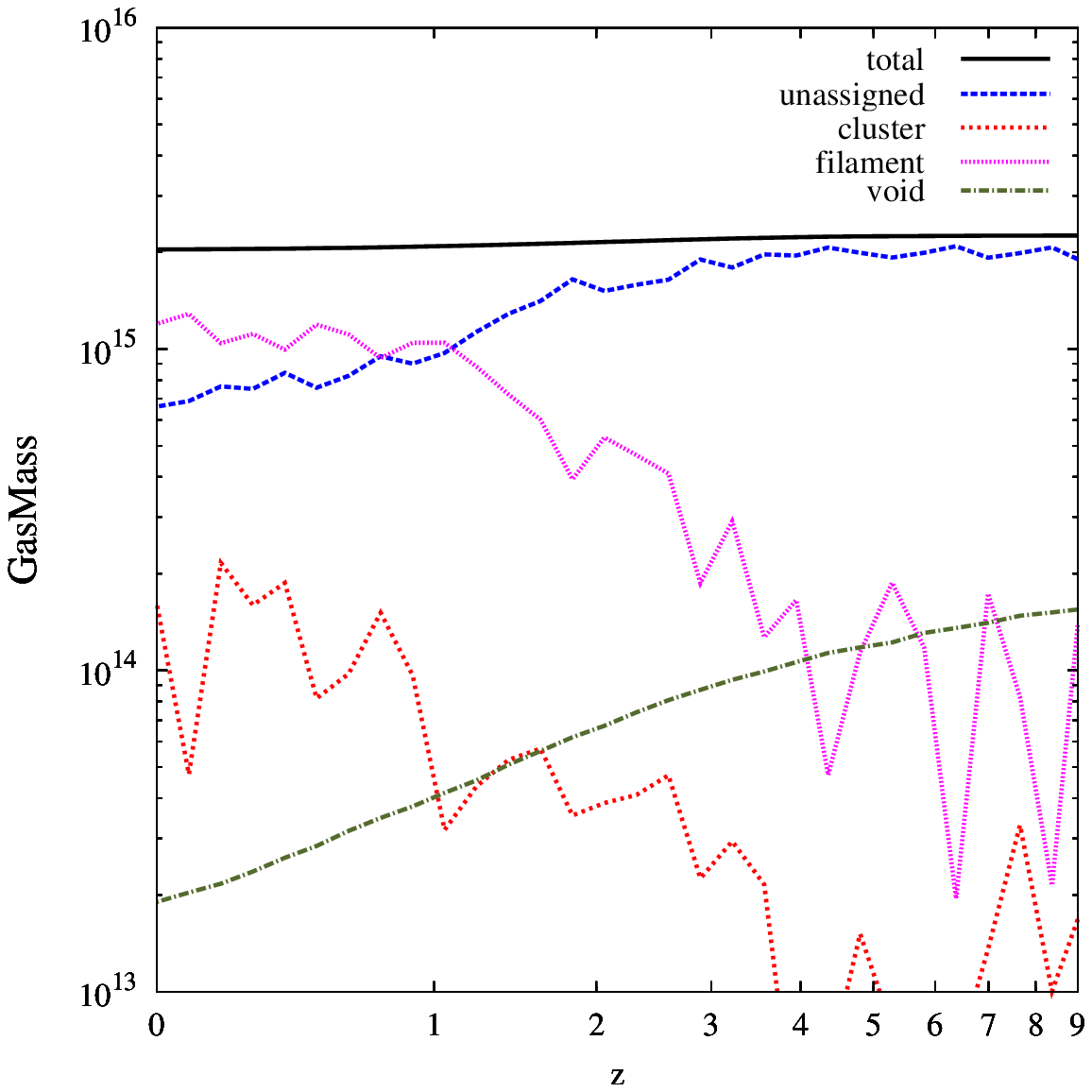}
		\caption{Gas Mass with criterion 1.}
		\label{fig:gas_mass_1}
		\end{center}
	\end{subfigure}%
	\begin{subfigure}{.5 \textwidth}
		\begin{center}
		\psfrag{GasMass}[cc][cc][0.8]{}		
		\psfrag{z}[cc][cc][0.8]{Redshift}
		\includegraphics[width=3.25in]{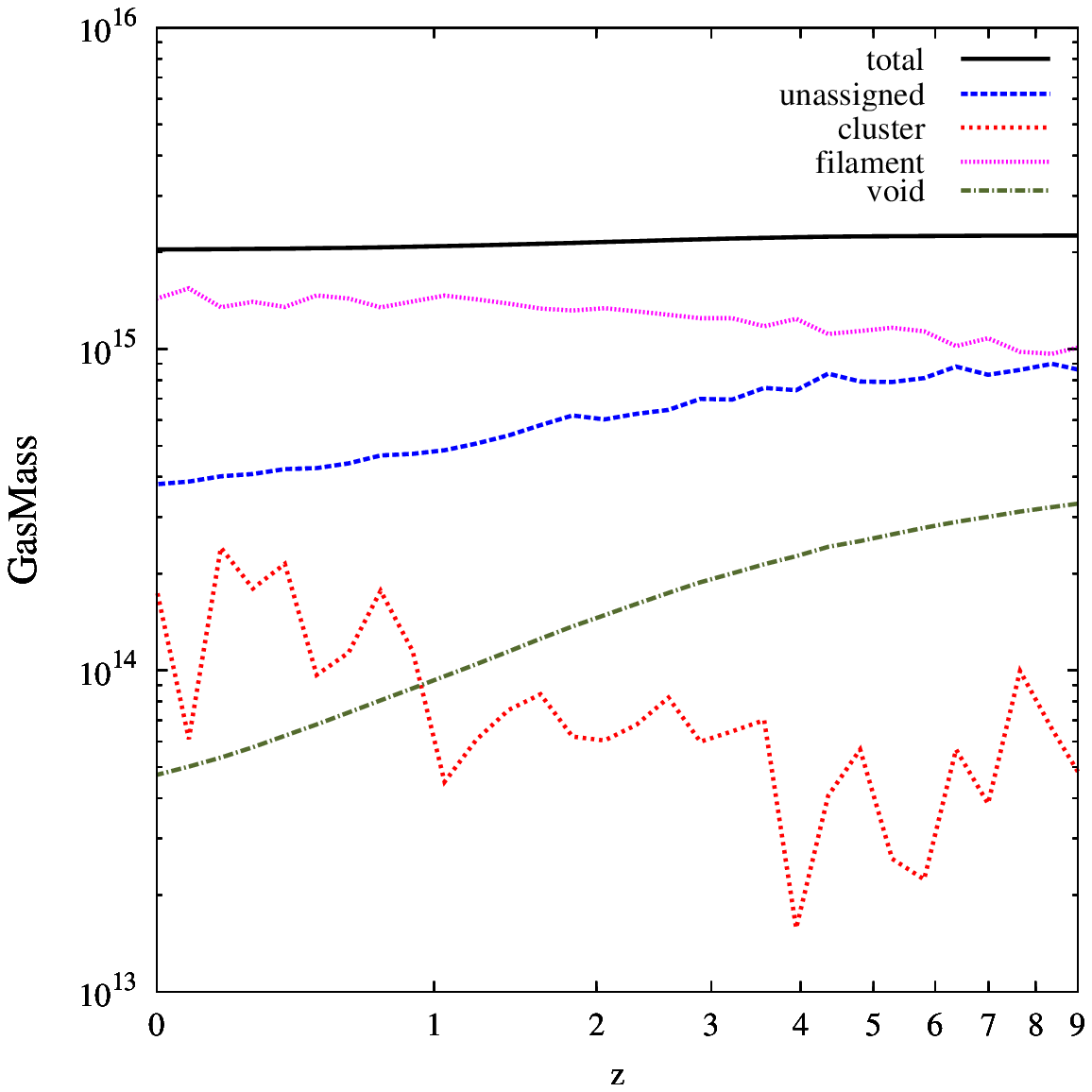}
		\caption{Gas Mass with criterion 2.}
		\label{fig:gas_mass_2}
		\end{center}
	\end{subfigure}\\
	\begin{subfigure}{.5 \textwidth}
		\begin{center}
		\psfrag{GasMass}[cc][cc][0.8]{}		
		\psfrag{z}[cc][cc][0.8]{Redshift}
		\includegraphics[width=3.25in]{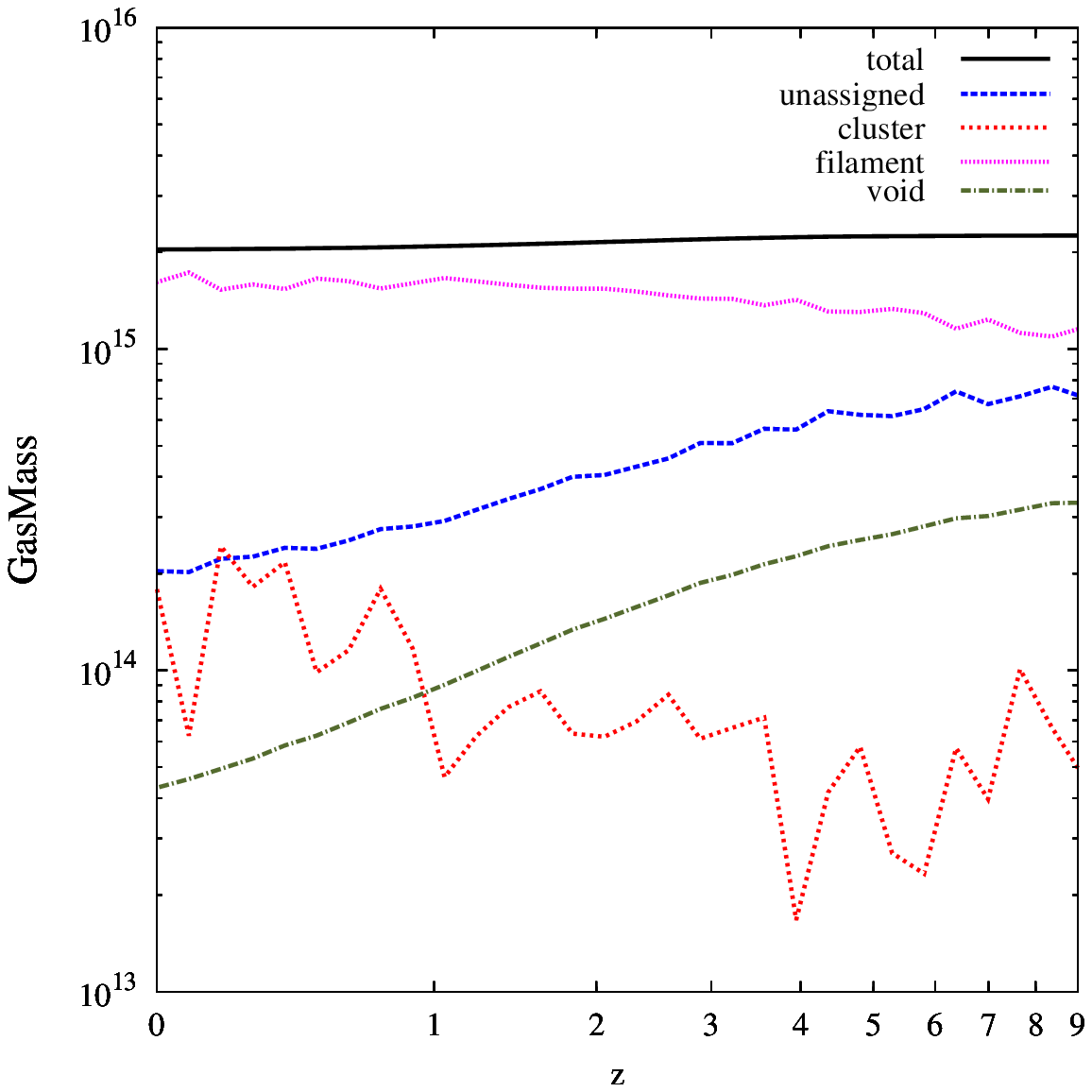}
		\caption{Gas Mass with criterion 3.}
		\label{fig:gas_mass_3}
		\end{center}
	\end{subfigure}%\\
	\end{array}$
\caption{The gas mass distribution as a function of structure type, redshift and structure criteria.}
\label{fig:gas_mass}
\end{figure*}

\begin{figure*}
	%\begin{center}
	$\begin{array}{cc}
	\begin{subfigure}{.5 \textwidth}
		\begin{center}
		\psfrag{StellarMass}[cc][cc][0.8]{Star Mass (M$_{\odot}$)}
		\psfrag{z}[cc][cc][0.8]{Redshift}
		\includegraphics[width=3.5in]{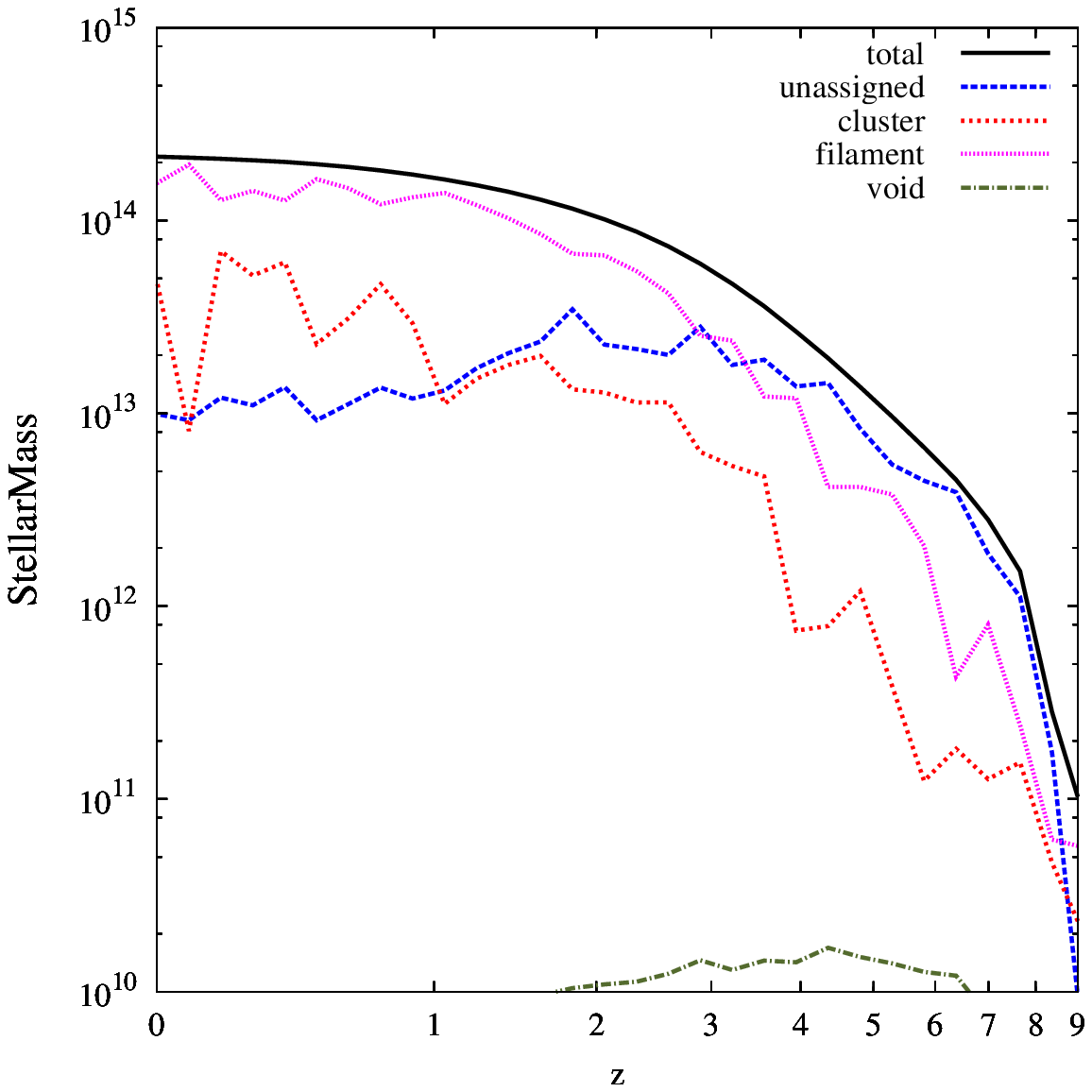}
		\caption{Star Mass with criterion 1.}
		\label{fig:star_mass_1}
		\end{center}
	\end{subfigure}%
	\begin{subfigure}{.5 \textwidth}
		\begin{center}
		\psfrag{StellarMass}[cc][cc][0.8]{}
		\psfrag{z}[cc][cc][0.8]{Redshift}
		\includegraphics[width=3.5in]{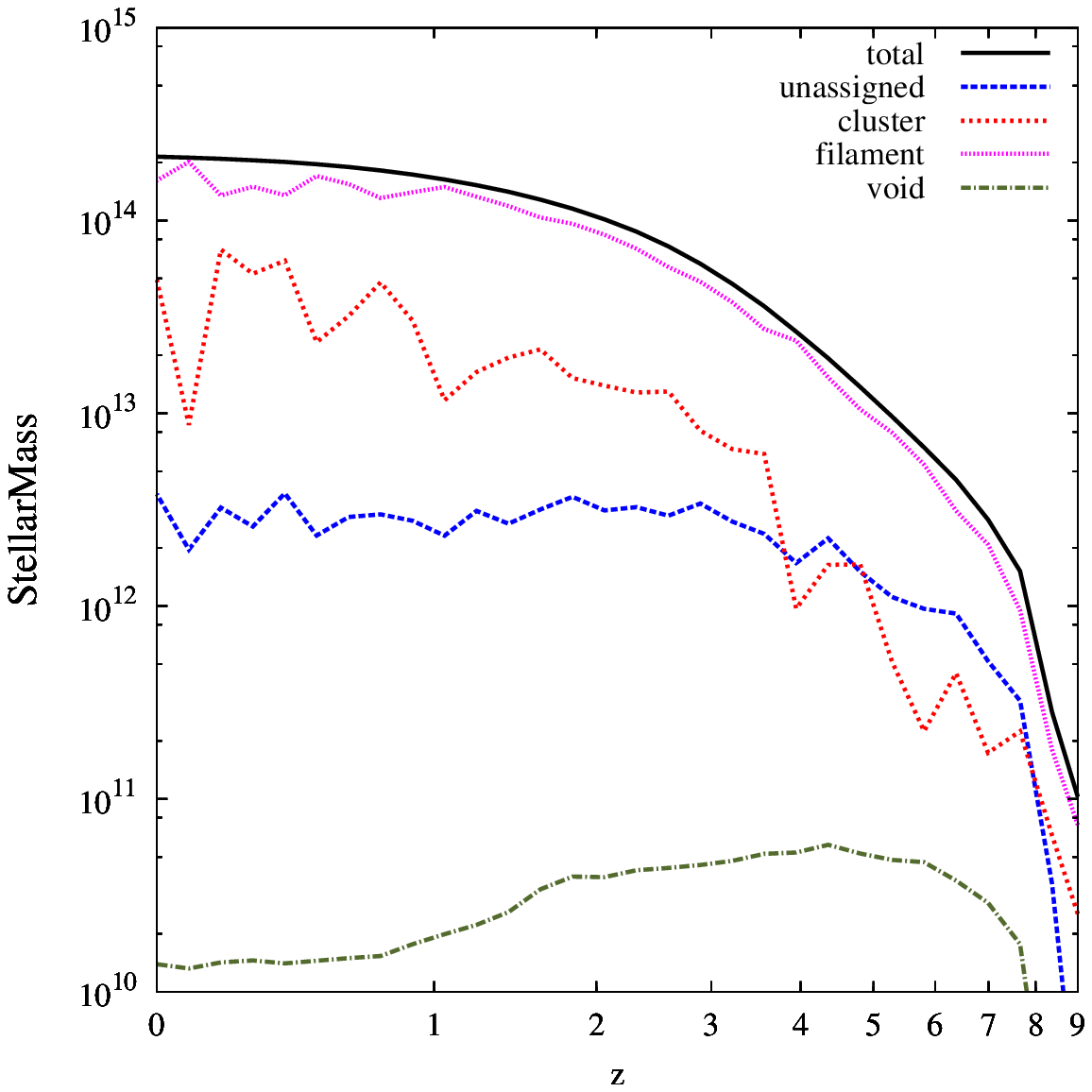}
		\caption{Star Mass with criterion 2.}
		\label{fig:star_mass_2}
		\end{center}
	\end{subfigure}\\
	\begin{subfigure}{.5 \textwidth}
		\begin{center}
		\psfrag{StellarMass}[cc][cc][0.8]{}
		\psfrag{z}[cc][cc][0.8]{Redshift}
		\includegraphics[width=3.5in]{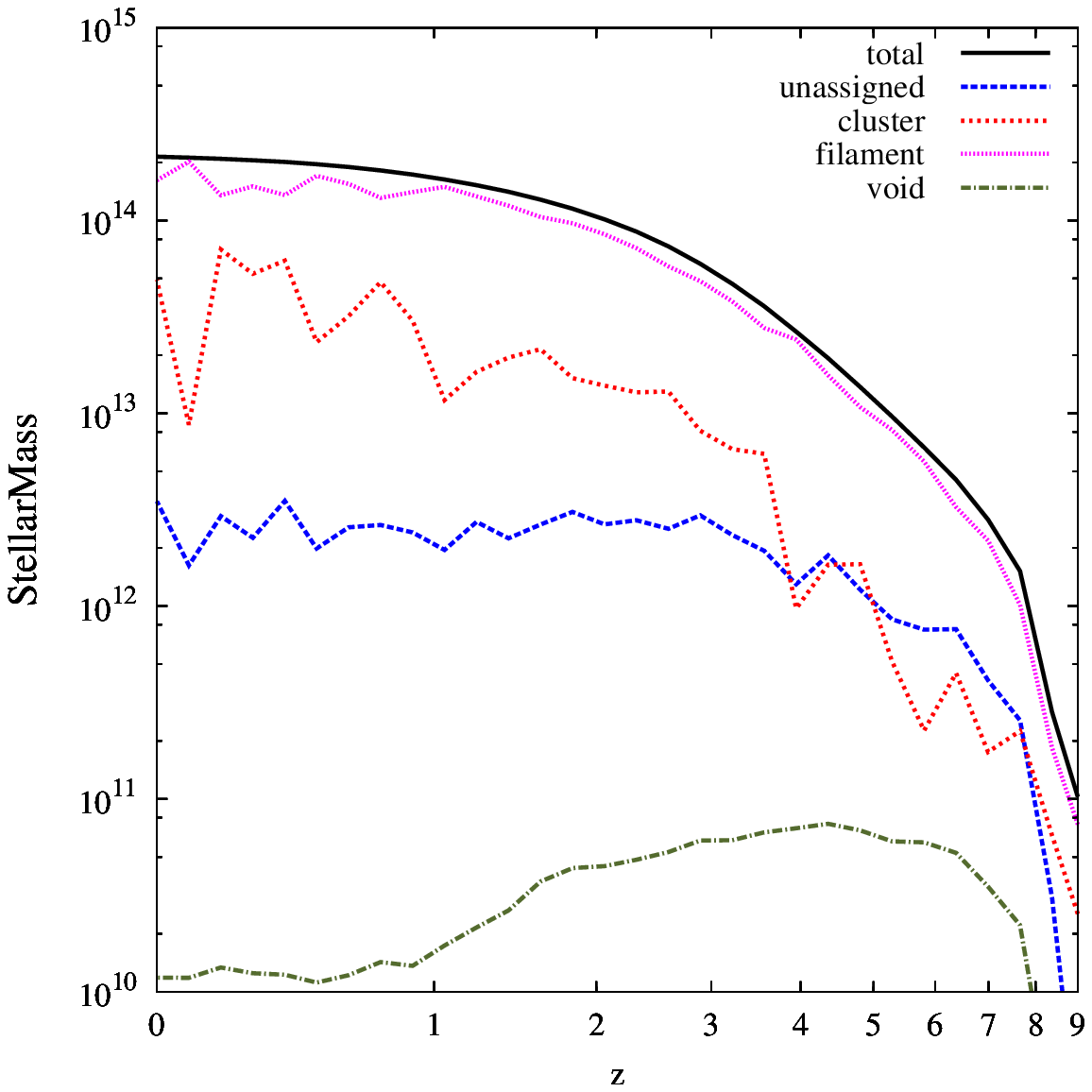}
		\caption{Star Mass with criterion 3.}
		\label{fig:star_mass_3}
		\end{center}
	\end{subfigure}%\\
	\end{array}$
	%\end{center}
\caption{The star mass distribution as a function of structure type, redshift and structure criteria.}
\label{fig:star_mass}
\end{figure*}

\begin{figure*}
	%\begin{center}
	$\begin{array}{cc}
	\begin{subfigure}{.5 \textwidth}
		\begin{center}
		\psfrag{SFR}[cc][cc][0.8]{SFR (M$_{\odot}$ yr$^{-1}$ Mpc$^{-3}$)}
		\psfrag{z}[cc][cc][0.8]{Redshift}
		\includegraphics[width=3.5in]{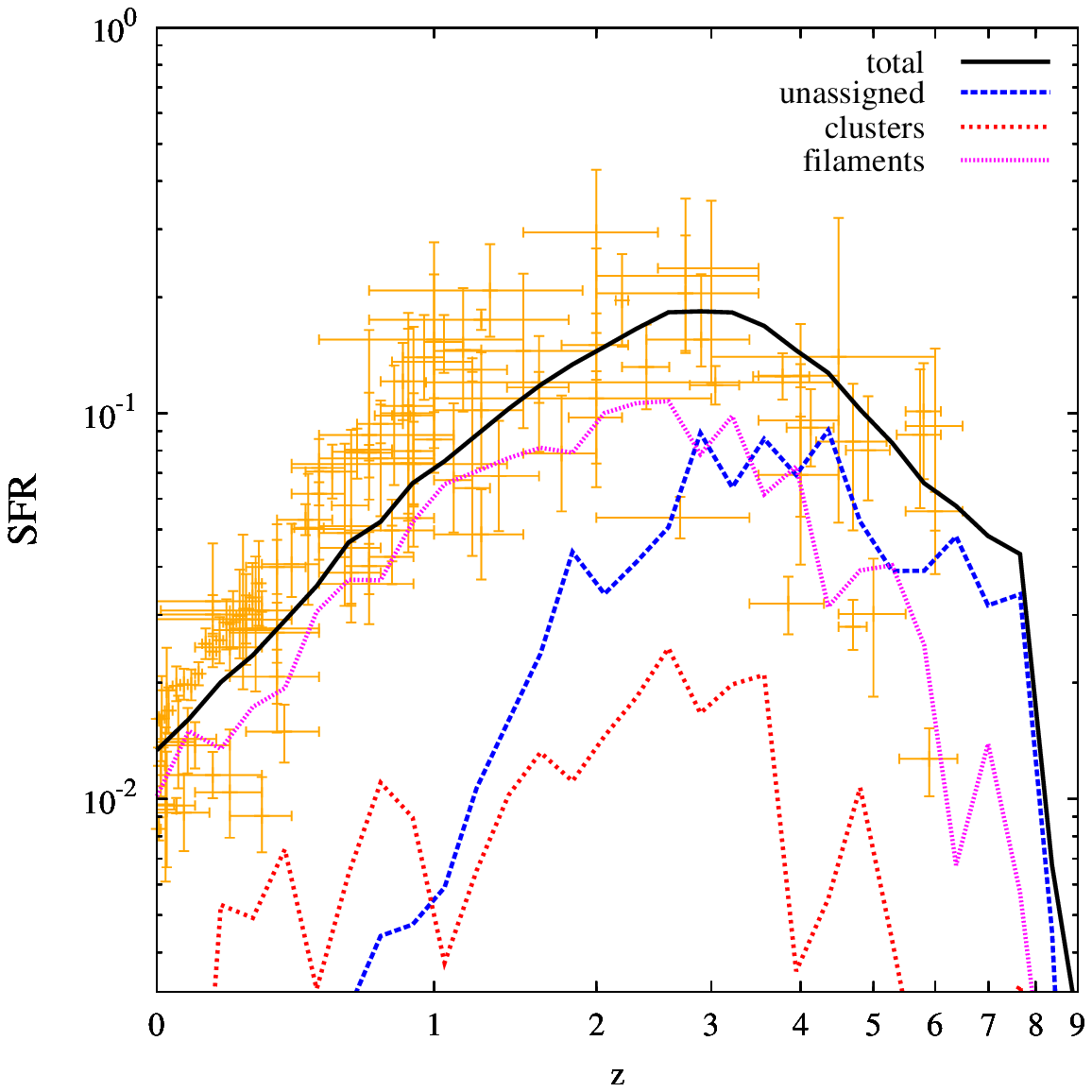}
		\caption{SFR with criterion 1.}
		\label{fig:sfh_1}
		\end{center}
	\end{subfigure}%
	\begin{subfigure}{.5 \textwidth}
		\begin{center}
		\psfrag{SFR}[cc][cc][0.8]{}
		\psfrag{z}[cc][cc][0.8]{Redshift}
		\includegraphics[width=3.5in]{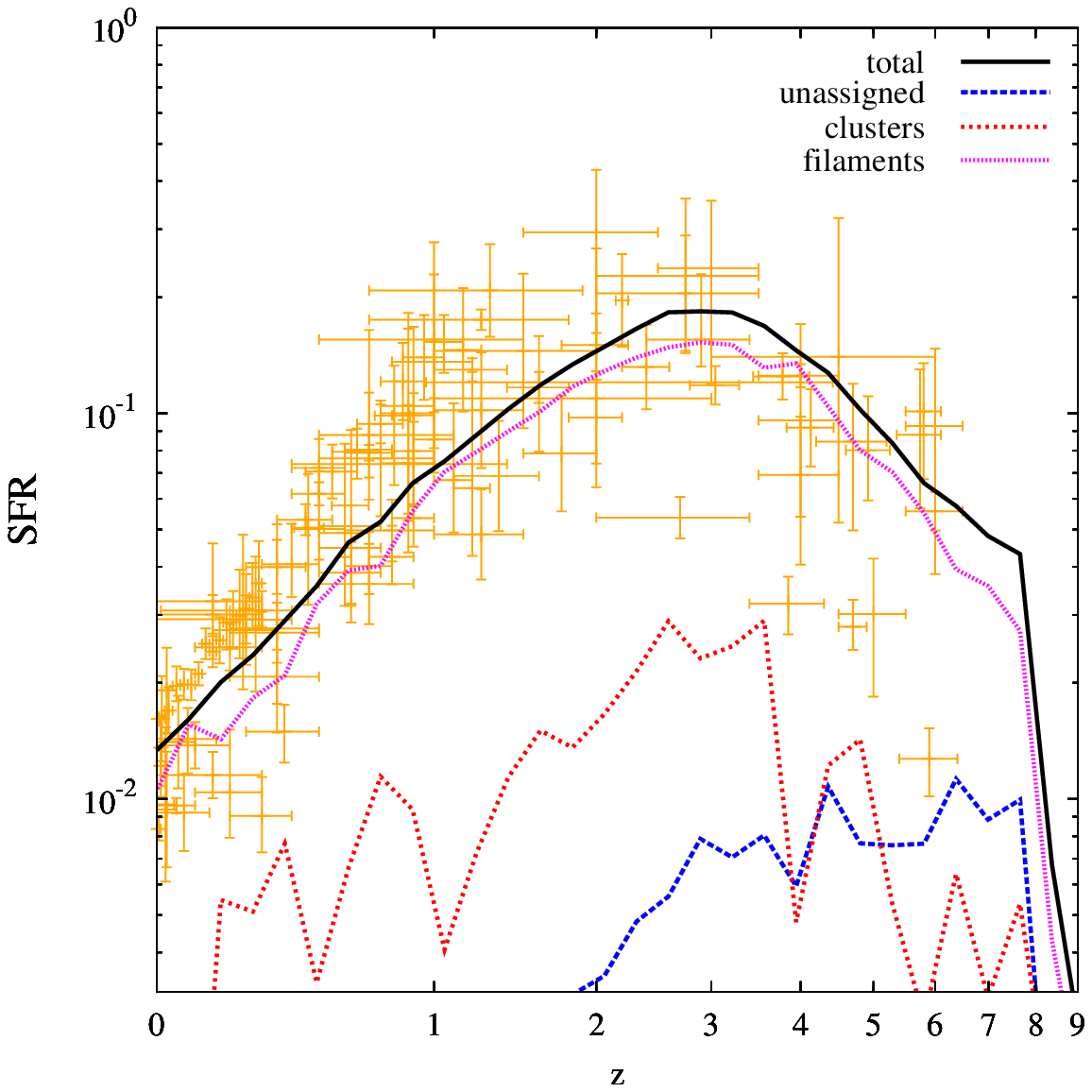}
		\caption{SFR with criterion 2.}
		\label{fig:sfh_2}
		\end{center}
	\end{subfigure}\\
	\begin{subfigure}{.5 \textwidth}
		\begin{center}
		\psfrag{SFR}[cc][cc][0.8]{}
		\psfrag{z}[cc][cc][0.8]{Redshift}
		\includegraphics[width=3.5in]{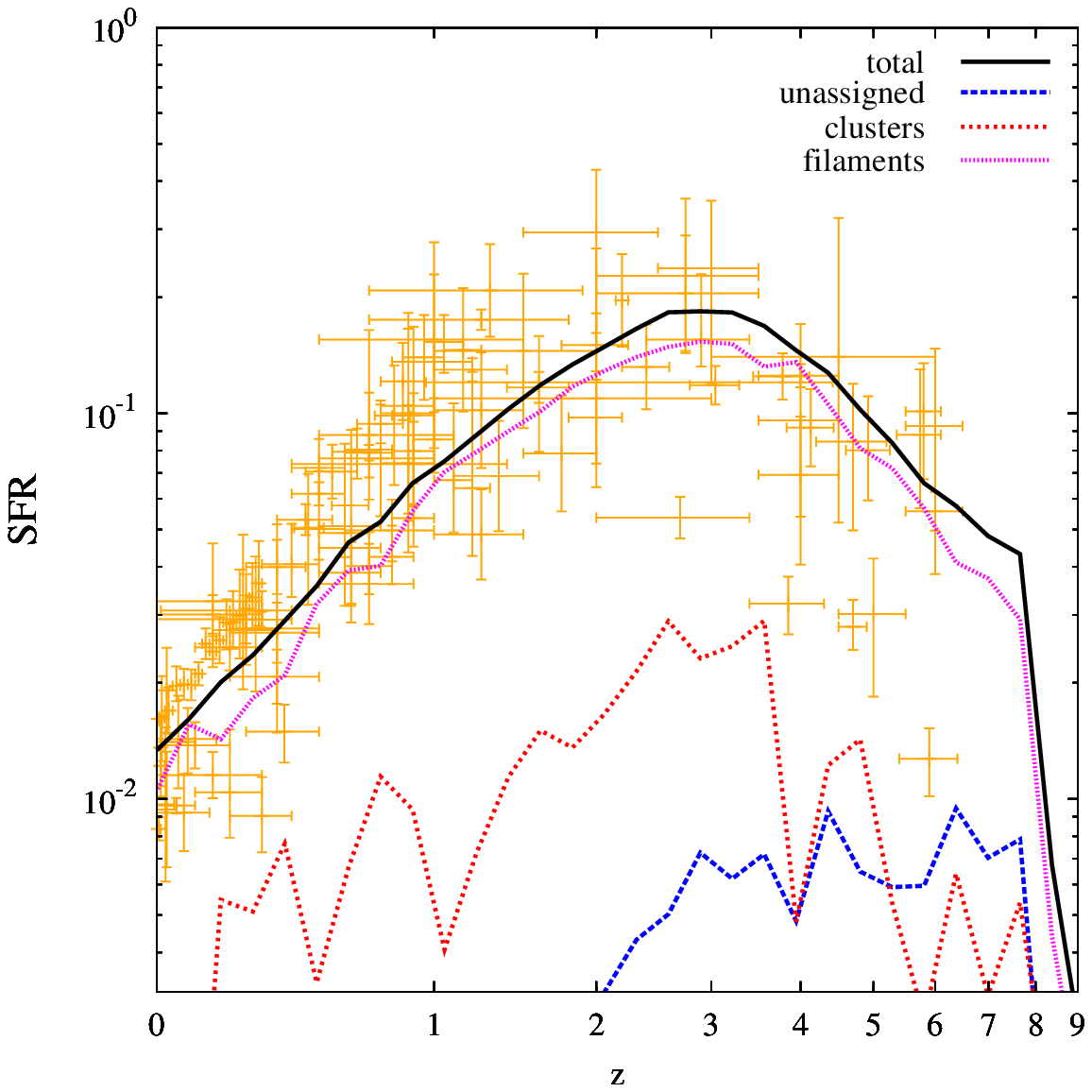}
		\caption{SFR with criterion 3.}
		\label{fig:sfh_3}
		\end{center}
	\end{subfigure}%\\
	\end{array}$
	%\end{center}
\caption{The star formation history as a function of structure type, redshift and structure criteria.}
\label{fig:sfh}
\end{figure*}

We expand our work from \cite{snedden12,snedden14} by following the evolution of gas mass, star mass and star formation rate as a function of structure type in Figs.~\ref{fig:gas_mass} -~\ref{fig:sfh} respectively.  We also consider the evolution using different structure measure criteria and we parse our simulations using the scales listed in Table~\ref{table:6}.  This gives us insight into the nature of structures at different redshifts.  In each of the figures (e.g. Figs.~\ref{fig:gas_mass} -~\ref{fig:sfh}) we consider the three different structure finding measure criteria:
\begin{enumerate}
	\item {\bf Criterion 1} : use the density criterion and the structure measure must be greater than 0.1, same as \cite{snedden14}
	\item {\bf Criterion 2} : use the density criterion and the structure measure must be greater than 0.0038
	\item {\bf Criterion 3} : does \emph{not} use the density criterion and the structure measure must be greater than 0.0038.
\end{enumerate}
The gas and stars are classified as being in clusters, filaments or voids using the maximum structure measure from equations~\ref{eq:r1} -~\ref{eq:r3} and the above criteria.  Gas or stars that do not meet the density criterion (if used) and the minimum structure measure threshold for any structure type, are classified as unassigned.

The evolution of gas mass in different structures as a function of redshift for different structure finder criteria is shown in Fig.~\ref{fig:gas_mass}.  Comparing Fig.~\ref{fig:gas_mass_1} to Fig.~\ref{fig:gas_mass_2}, we see that by keeping the density criterion and  relaxing the structure measure threshold, we significantly decrease the amount of unassigned material.  Most of this previously unassigned gas becomes classified as filaments.  This suggests that much of the unassigned material in Fig.~\ref{fig:gas_mass_1} is really in the \emph{low contrast} regions, or the outskirts, of filaments, i.e. filamentary regions with a relatively low structure measure.  In Fig.~\ref{fig:gas_mass_1} there is a crossover between the unassigned and filament material.  This is not seen in Fig.~\ref{fig:gas_mass_2}.  This suggests that the higher contrast filamentary regions begin to dominate in mass by a redshift of z $\sim$1.  By loosening the criteria further and eliminating the density criterion (see Fig.~\ref{fig:gas_mass_3}), we see that amount of unassigned material decreases further, while increasing the material in filaments.  The effect of the elimination of the density criterion is much smaller than that of relaxing the structure threshold and confirms that the dominating effect is the change in structure measure threshold.  These really are low contrast regions of the filaments, located mostly on the outskirts of higher contrast centers.

The evolution of star mass as a function of redshift is shown in Fig.~\ref{fig:star_mass}.  Comparing Fig.~\ref{fig:star_mass_1} to Fig.~\ref{fig:star_mass_2}, we see that by keeping the density criterion and relaxing the structure measure threshold, we significantly decrease the amount of unassigned material.  In Fig.~\ref{fig:star_mass_1} there is a switch between unassigned and filament star mass at a redshift z = 3.  This turnover does not exist in Fig.~\ref{fig:star_mass_2}, where the structure measure threshold has been relaxed.  This also suggests that much of the unclassified material in Fig.~\ref{fig:star_mass_1} exists in low contrast regions of filaments.   Comparing Fig.~\ref{fig:star_mass_2} and Fig.~\ref{fig:star_mass_3}, we see that once again removing the density criterion provides less of an effect than loosening the structure measure.   It is interesting to note that in Figs.~\ref{fig:star_mass_1} -~\ref{fig:star_mass_3}, as structure grows, the star mass accumulates in filaments, clusters and unassigned.  While these structures are increasing in stellar mass, the voids are vacated and the material is transferred to other structure types.  The increase in star mass in voids at very high redshifts (e.g. z = 5 - 7) suggests that some early star formation occurs in voids, but these stars are then vacated as the voids grow.  Voids are the only structure type that have a net decrease in material over time.  

Not only is most of the gas and star mass located within filaments, but most of the star formation also occurs within filaments as illustrated in Fig.~\ref{fig:sfh}.  Fig.~\ref{fig:sfh_1} illustrates that with criterion 1, the star formation begins primarily in the unassigned regions and then switches to filaments at a redshift z = 3.  When taken with Fig.~\ref{fig:sfh_2} or Fig.~\ref{fig:sfh_3} which are both dominated by filaments at all epochs, it becomes clear that at high redshift, the star formation is located in the low contrast regions of filaments.  It appears that the turning point is at redshift z $\sim$3 where the low contrast regions of filaments no longer dominate the SFR and the higher contrast regions prevail as structure forms.  This is the same turning point seen in Figs.~\ref{fig:star_mass_2} and~\ref{fig:star_mass_3}.  This suggests that most of the star mass that formed within filaments stays within the filaments.

Figs.~\ref{fig:sfh_2} and~\ref{fig:sfh_3} show that the star formation history is dominated by these filamentary regions and that the star formation rate in the filaments and clusters peaks at the same time as the overall star formation rate.  At lower redshifts, the overall fraction of the star formation occurring in clusters is increasing as the overall cosmic star formation rate decreases.

\begin{table}
	\centering
	%\begin{adjustwidth}{-1.5cm}{}
	\begin{tabular}{| c c | c c | c c |}
	\hline
		clusters 	& 	 			& filaments 			& 		& voids		\\
		\hline
		voxel 	& $h^{-1}$Mpc 		& voxel 	& $h^{-1}$Mpc 	& voxel 	& $h^{-1}$Mpc	\\
		\hline
		2		&	0.78			& 1		&	0.39		& 5		&	1.9		\\
		4		&	1.6			& 2		&	0.78		& 7.5		&	2.9		\\
		6		&	2.3			& 3		&	1.2		& 10		&	3.9		\\
		8		&	3.1			& 4		&	1.6		& 15		&	5.9		\\
		10		&	3.9			& 5		&	1.9		& 20		&	7.8		\\
		12		&	4.7			& 6		&	2.3		& 25		&	9.8		\\
		14		&	5.5			& 8		&	3.1		& 30		&	11.7		\\
		16		&	6.3			& 10		&	3.9		& 35		&	13.7		\\
		18		&	7.0			& 12		&	4.7		& 40		&	15.6		\\
		20		&	8.8			& 	14	&	5.5		& 50		&	19.5		\\
	\hline
		\end{tabular}
		%\end{adjustwidth}
	\caption{Scales probed by the structure finding algorithm for the 256$^3$ grid in both voxels and comoving size.}  
\label{table:6}
\end{table}

%==================== Temperature and Density Evolution===========================%
\section{The Temperature and Density Evolution}
\label{sec:temp_and_rho_evol}
We analyze the density vs.~temperature distribution for the different structures in Figs.~\ref{fig:gas_phase_z3} -~\ref{fig:gas_phase} using criterion 3.  We approximate the relative density as $\rho_{\text{gas}}$ / $\rho_{\text{baryon}}$.  This is a valid approximation because most of the baryons are in the gas phase and the deviation caused by the creation of stars is on the order of 10\%.  Since we are looking at several times the overdensity, this small deviation does not significantly affect these plots. 

We use the gas phase definitions found in \citep{dave10, oppenheimer12} to study the multiphase nature of the structures:
\begin{enumerate}
\item Diffuse ($T < T_{\text{th}}$, $\delta < \delta_{\text{th}}$)
\item WHIM ($T > T_{\text{th}}$, $\delta < \delta_{\text{th}}$)
\item Hot halo ($T > T_{\text{th}}$, $\delta > \delta_{\text{th}}$)
\item Condensed ($T < T_{\text{th}}$, $\delta > \delta_{\text{th}}$)
\end{enumerate} 
 where $T_{\text{th}}$ = 10$^{5}$ K and $\delta_{\text{th}}$ is defined \citep{kitayama96} as 
\begin{equation}
\label{eq:r5}
\delta_{\text{th}} = 6 \pi^{2} [1 + 0.4093(1 / f_{\Omega} - 1)^{0.9052}] - 1
\end{equation}
and 
\begin{equation}
\label{eq:r6}
f_{\Omega} = \frac{\Omega_{\text{m}} (1 + z)^{3}}{\Omega_{\text{m}}(1+z)^{3} + (1 - \Omega_{\text{m}} - \Omega_{\Lambda})(1 + z)^{2} + \Omega_{\Lambda}}
\end{equation}
$\delta_{\text{th}}$ is the overdensity at the boundary (r$_{\text{200}}$) of a collapsing dark matter halo.  It demarcates the threshold between gas bound to dark matter halos and the unbound intergalactic medium.  Note that the WHIM definition is significantly different than that found in the original papers describing this component \citep{cen99,dave01} where there was no upper density threshold. Readers wishing to compare our WHIM results with other simulations using the original definition should look at both the WHIM and hot halo components.  The gas phase distribution of clusters, filaments, and voids identified by our structure finder, as well as in unassigned regions, is considered.

\begin{description}[leftmargin=*]
\item {\bf Clusters :}  The total gas fraction in clusters peaks at 3-4\% between redshifts z = 2.5 - 3.6.   At a redshift z = 3.2 (where the cosmic star formation rate is at a maximum) the phase distribution (as a fraction of the total mass) in the cluster environment is 1.53\%, 0.26\%, 0.40\% and 0.81\% in the diffuse, WHIM, hot halo and condensed phases respectively (see Fig.~\ref{fig:gas_phase_z3_b}).  By a redshift z = 0, the phase distribution becomes 0.55\%, 2.99\%, 5.02\% and 0.21\% in the diffuse, WHIM, hot halo and condensed phases respectively (see Fig.~\ref{fig:gas_phase_z0_b}). In the present era the total mass fraction of cluster gas has increased to 9\% up from 3\% at redshift z = 3. The diffuse material has collapsed unto higher density regions.  The cluster hot halos increased from 0.40\% of the total gas mass (at z = 3) to 5.02\% at z = 0, while the diffuse material has decreased from 1.53\% to 0.55\% over the same time interval.

The choppy nature in Fig.~\ref{fig:gas_phase_b} is due to the relatively small volume of the simulation is classified as cluster environments.  
Small fluctuations in the structure finder's classification between snapshots lead to large fluctuations in the overall cluster gas fraction as gas particles are moved from clusters into the unassigned category. Fig.~\ref{fig:gas_phase_b} illustrates a moderate transition at a redshift z = 2 from the cluster gas dominated by the diffuse and condensed phases, to being dominated by hot halo and WHIM gas.  In clusters the condensed phase stays relatively constant, pointing to a sustained fraction of gas available for star formation within the clusters.  This, combined with the fact that the total mass in clusters is increasing, suggests that as gas is accreting onto clusters, there is some constant fraction that cools and becomes available for star formation. 
\smallskip

\item {\bf Filaments :} At a redshift z = 3.2 the distribution (as a fraction of the total mass) in the filament environment is 52.8\%, 2.90\%, 1.83\% and 7.37\% in the diffuse, WHIM, hot halo and condensed phases respectively (see Fig.~\ref{fig:gas_phase_z3_c}).  By a redshift z = 0, the distribution becomes 33.1\%, 24.7\%, 19.3\% and 1.97\% in the diffuse, WHIM, hot halo and condensed phases respectively (see Fig.~\ref{fig:gas_phase_z0_c}). As was the case with the clusters, the gas is multiphase. The filamentary gas has a more populated WHIM component than hot halo component by a factor of 1.58 (1.28) at a redshift z = 3.2 (z = 0.) By contrast, the cluster gas has 1.54 (1.68) times less WHIM gas than hot halo gas at z = 3 (z = 0.) The peak in the condensed phase occurs at the same time as the overall star formation rate is peaking (compare Fig.~\ref{fig:gas_phase_c} to Fig.~\ref{fig:sfh_3}).  This suggests that the majority of the star formation in the universe is occurring in the condensed, filaments phase.  This is confirmed by Fig.~\ref{fig:sfr_phase_c}, which peaks in the same redshift range.
\smallskip

\item {\bf Voids :} The total gas fraction in voids peaks at 8.98\% at a redshift z = 3.6 and drops to 2\% of the total gas at a redshift z = 0.  The vast majority of the gas begins and stays in the diffuse phase (see Fig.~\ref{fig:gas_phase_d}).  At a redshift z = 3.2 the gas is single phase with 8.98\% (as a fraction of the total mass) in the diffuse phase (see Fig.~\ref{fig:gas_phase_z3_d}).  By a redshift z = 0, this transitions to a two-phase distribution with  2.01\%, 0.11\% in the diffuse and WHIM, respectively (see Fig.~\ref{fig:gas_phase_z0_d}). At redshift z = 0, the temperature dispersion has increased so that about 5\% of the total void gas is in the WHIM.  The increase in temperature dispersion is likely due to the vacating of the voids and possible shock heating occurring at the edges of voids.  It is most likely not due to feedback heating which mostly occurs within the first 40 Myr (i.e. the oldest star that can produce a Type II supernova) of the formation time of a star particle.  If feedback heating was the dominant source of this dispersion, it would likely peak at z = 3, where the star formation is more intense, instead of at z = 0.
\smallskip

\item {\bf Unassigned :}  The total unassigned gas fraction at 34\% (99\% of which is in the diffuse phase) at a redshift z = 8.37 and drops to 10\% by a redshift z = 0.  At a redshift z = 3.2 the phase distribution (as a fraction of the total mass) in the unassigned environment is 22.7\%, 0.10\%, 0.05\% and 0.19\% in the diffuse, WHIM, hot halo and condensed phases respectively (see Fig.~\ref{fig:gas_phase_z3_e}).  By a redshift z = 0, the phase distribution becomes 7.46\%, 2.15\%, 0.34\% and 0.02\% in the diffuse, WHIM, hot halo and condensed phases respectively (see Fig.~\ref{fig:gas_phase_z0_e}).  This substantial depletion indicates that most of the unclassified material at high redshift has been reclassified into poor clusters or filaments as more structures form.

\end{description}

\section{The Star Formation Rate and Efficiency}
\label{sec:sfr}

We investigate (see Figs.~\ref{fig:sfr_z3} -~\ref{fig:sfr_phase}) the density and temperature of star forming regions in the simulation.  To get the temperature and density of a star particle, we use the SPH kernel and interpolate from the nearby gas particle temperatures and densities.  We then find all the star particles within a particular density and temperature regime which formed within the last 100 Myr and take the average star formation over the time interval.  This is a reasonable approximation to the temperature and density of the gas when the star particle formed because it has not yet had a chance to move away from the star forming region.  A video of the star formation evolution as a function of structure and gas phase is available online.

\begin{figure*}
	\psfrag{-7}[cc][cc][0.8]{-7}
	\psfrag{-6}[cc][cc][0.8]{-6}
	\psfrag{-5}[cc][cc][0.8]{-5}
	\psfrag{-4}[cc][cc][0.8]{-4}
	\psfrag{-3}[cc][cc][0.8]{-3}
	\psfrag{-2}[cc][cc][0.8]{-2}
	\psfrag{2}[cc][cc][0.8]{2}
	\psfrag{3}[cc][cc][0.8]{3}
	\psfrag{4}[cc][cc][0.8]{4}
	\psfrag{5}[cc][cc][0.8]{5}
	\psfrag{6}[cc][cc][0.8]{6}
	\psfrag{7}[cc][cc][0.8]{7}
	\psfrag{8}[cc][cc][0.8]{8}
	\psfrag{overdensity}[tc][tc][0.9]{log$_{\text{10}}$($\rho_{\text{gas}} / \bar{\rho}$)}
	\psfrag{Log(T)}[cc][cc][0.90]{log$_{\text{10}}$(T)}

	$\begin{array}{cc}
	\begin{subfigure}{.33 \textwidth}
		\begin{center}
		\includegraphics[width=2.7in]{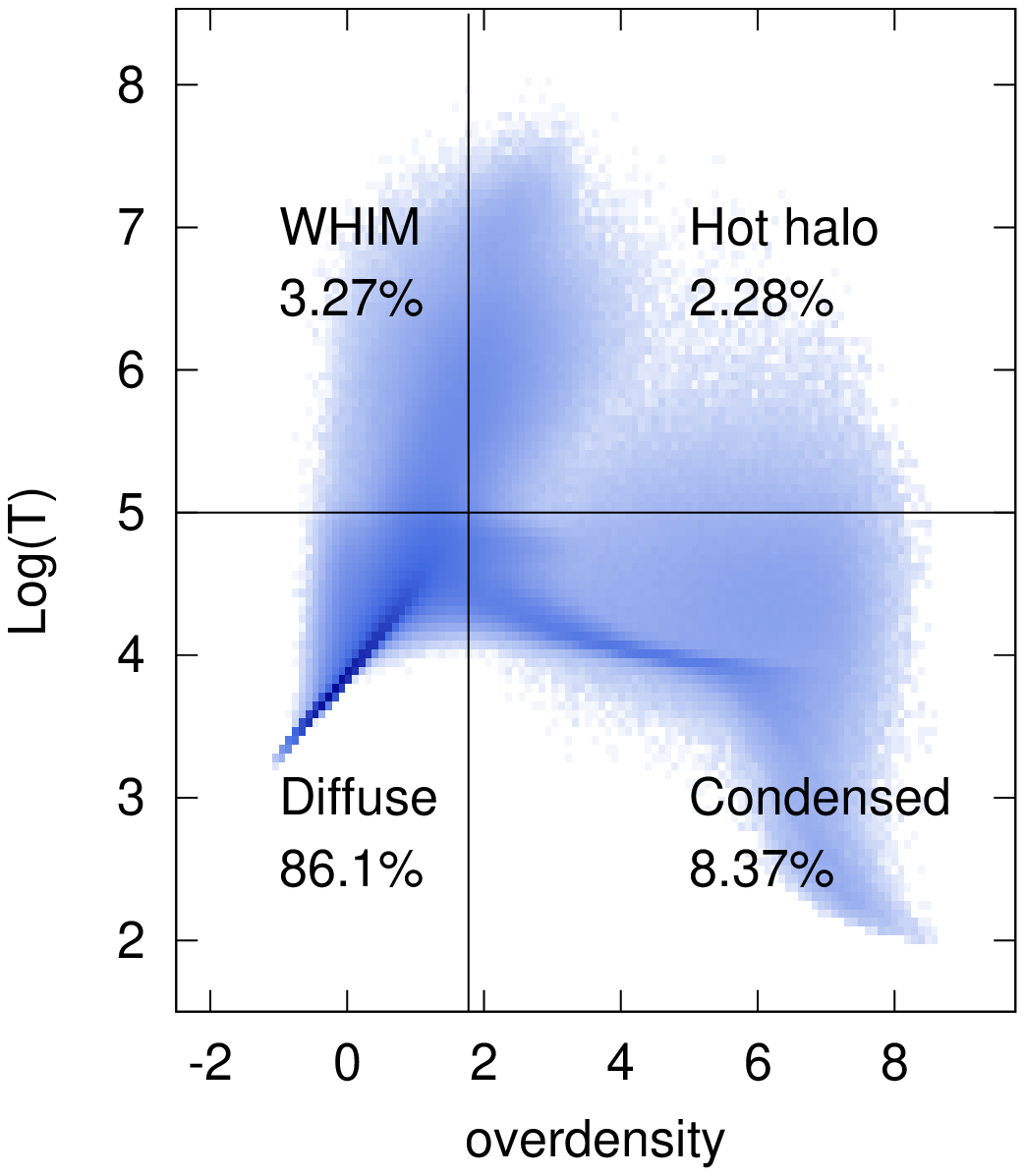}
		\caption{Total}
		\label{fig:gas_phase_z3_a}
		\end{center}
	\end{subfigure}%
	\begin{subfigure}{.33 \textwidth}
		\begin{center}
		\includegraphics[width=2.7in]{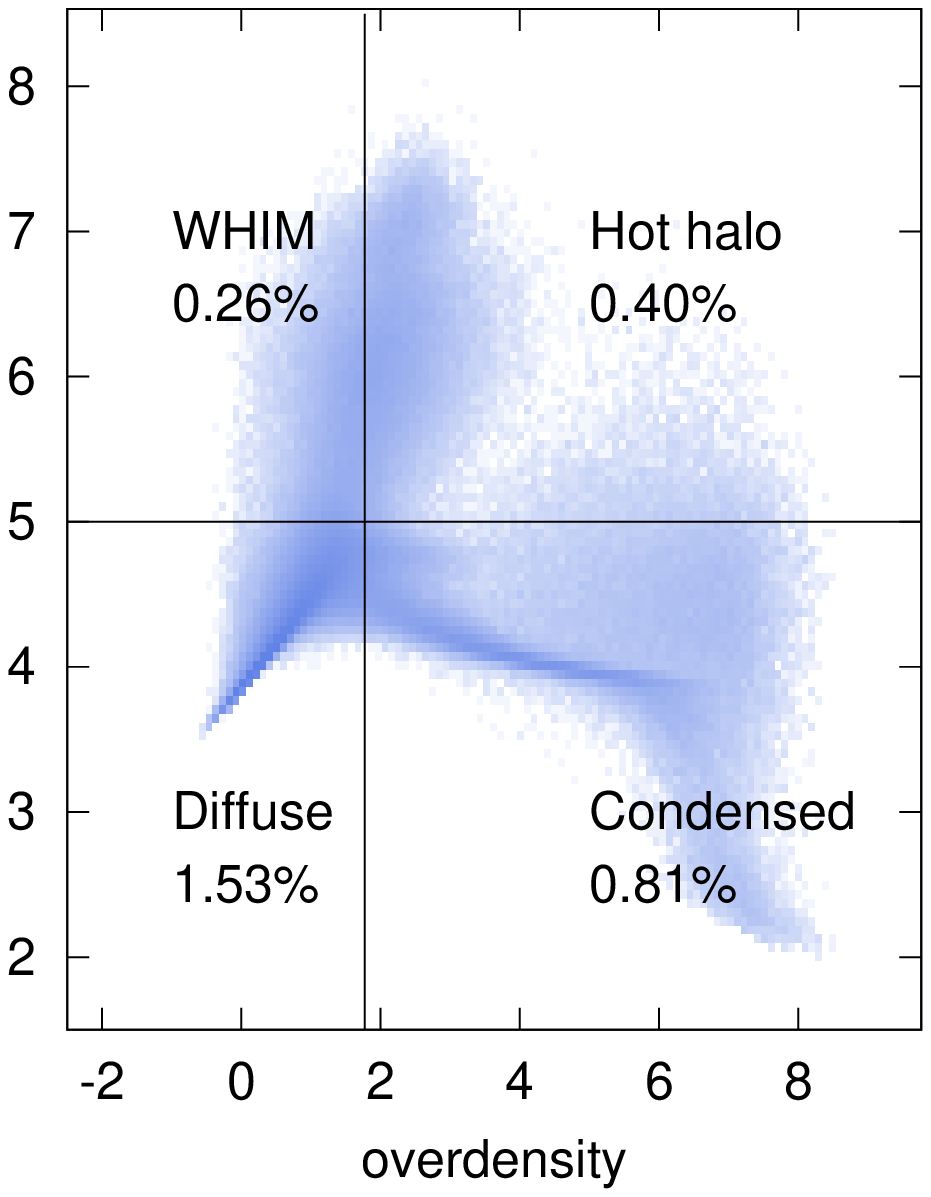}
		\caption{Clusters}
		\label{fig:gas_phase_z3_b}
		\end{center}
	\end{subfigure}
	\quad
	\begin{subfigure}{.33 \textwidth}
		\begin{center}
		\includegraphics[width=2.8in, height=2.7in]{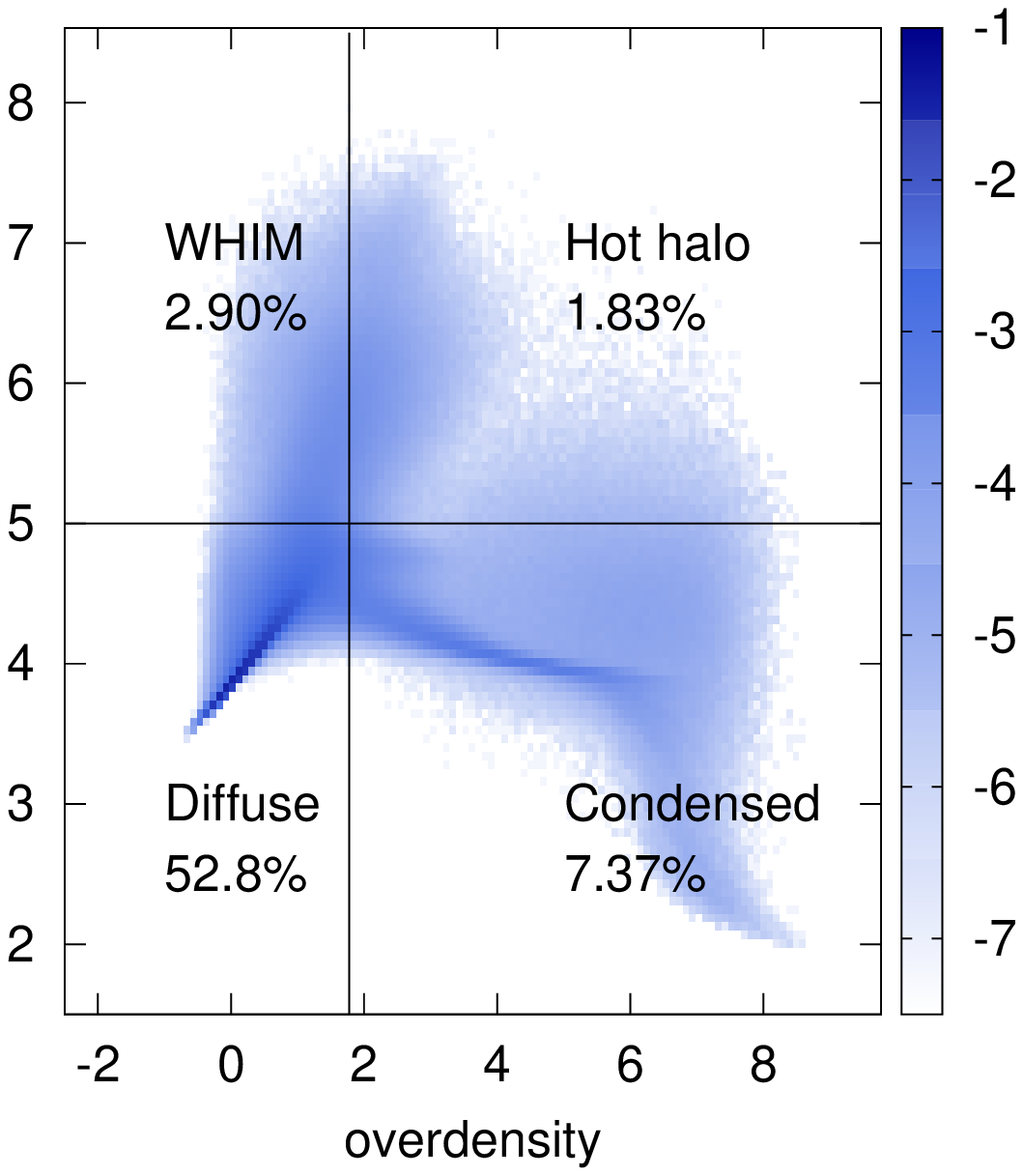}
		\caption{Filaments}
		\label{fig:gas_phase_z3_c}
		\end{center}
	\end{subfigure}
	\\
	\begin{subfigure}{.33 \textwidth}
		\begin{center}
		\includegraphics[width=2.7in]{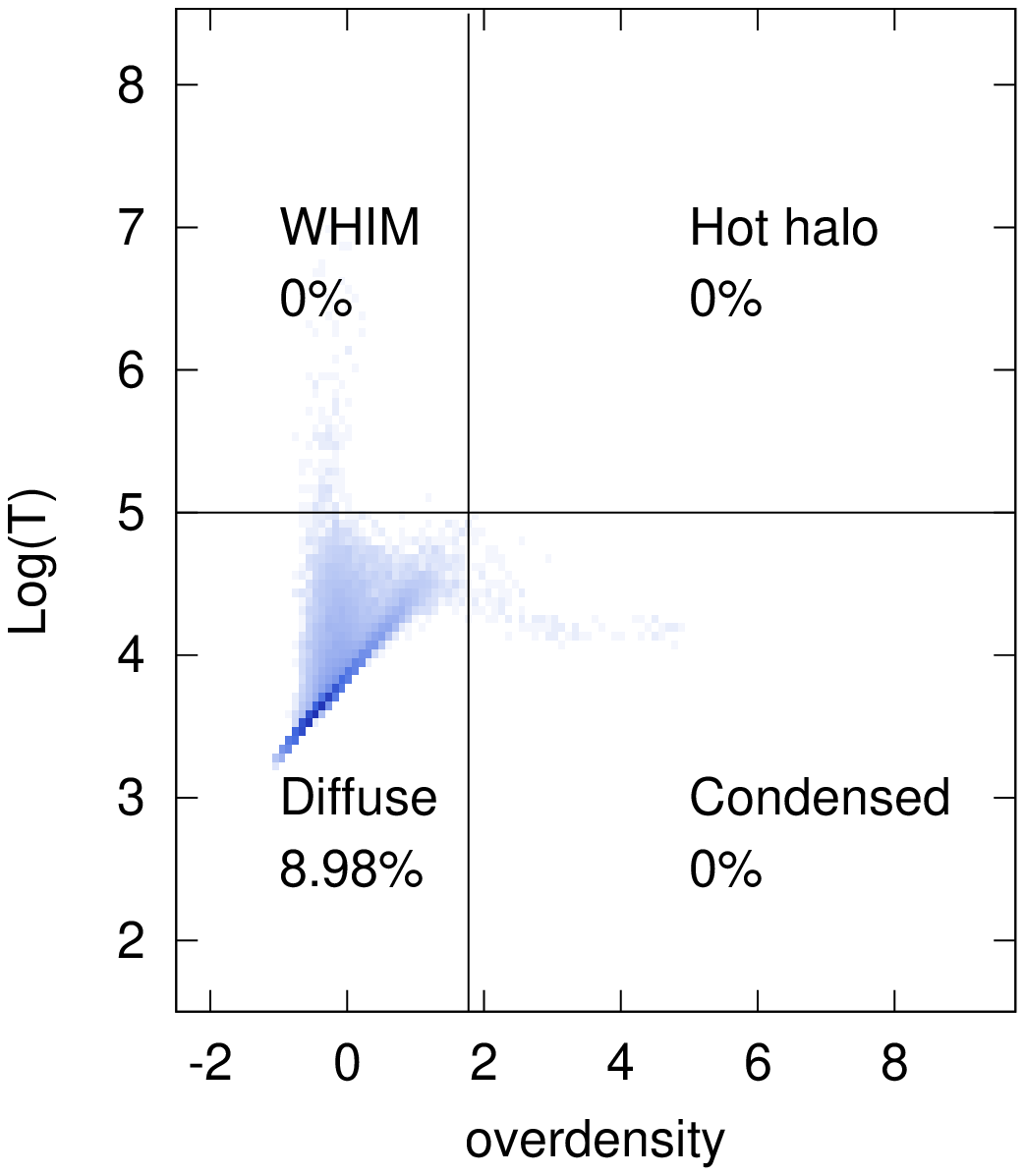}
		\caption{Voids}
		\label{fig:gas_phase_z3_d}
		\end{center}
	\end{subfigure}%
	\quad
	\begin{subfigure}{.33 \textwidth}
		\begin{center}
		\includegraphics[width=2.8in, height=2.7in]{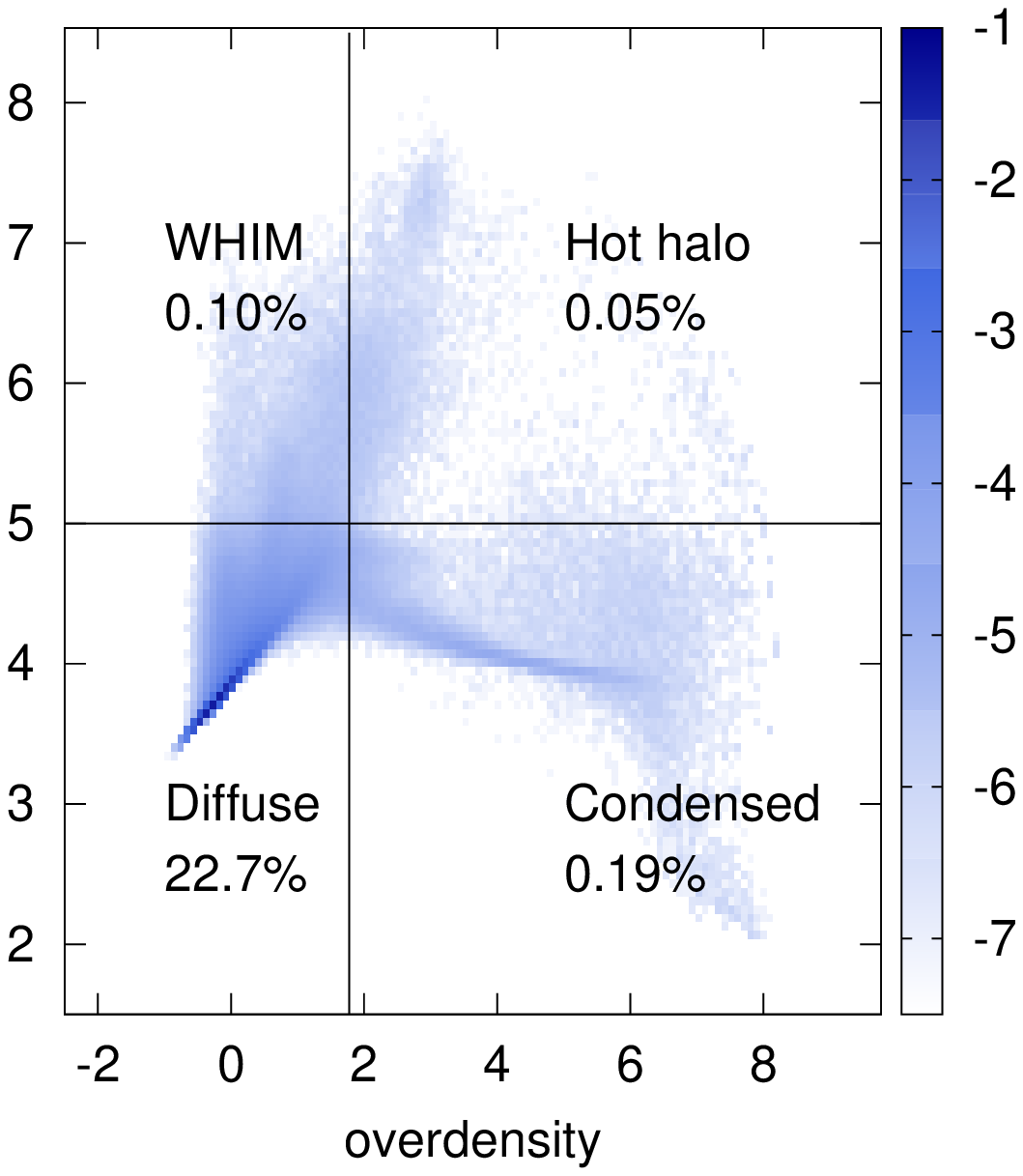}
		\caption{Unassigned}
		\label{fig:gas_phase_z3_e}
		\end{center}
	\end{subfigure}
	\end{array}$
	%\end{center}
\caption{Gas Phase History at Redshift z = 3.2 using criterion 3.  The color bar represents the log$_{\text{10}}$ of the gas mass fraction at a particular density and temperature.  The total gas mass fraction in each phase and structure is listed.  3\% of the gas is in clusters, 64.9\% is in filaments, 8.98\% is in voids and 23.04\% is in unassigned.}

\label{fig:gas_phase_z3}
\end{figure*}

\begin{figure*}
	\psfrag{-7}[cc][cc][0.8]{-7}
	\psfrag{-6}[cc][cc][0.8]{-6}
	\psfrag{-5}[cc][cc][0.8]{-5}
	\psfrag{-4}[cc][cc][0.8]{-4}
	\psfrag{-3}[cc][cc][0.8]{-3}
	\psfrag{-2}[cc][cc][0.8]{-2}
	\psfrag{2}[cc][cc][0.8]{2}
	\psfrag{3}[cc][cc][0.8]{3}
	\psfrag{4}[cc][cc][0.8]{4}
	\psfrag{5}[cc][cc][0.8]{5}
	\psfrag{6}[cc][cc][0.8]{6}
	\psfrag{7}[cc][cc][0.8]{7}
	\psfrag{8}[cc][cc][0.8]{8}
	\psfrag{overdensity}[tc][tc][0.9]{log$_{\text{10}}$($\rho_{\text{gas}} / \bar{\rho}$)}
	\psfrag{Log(T)}[cc][cc][0.90]{log$_{\text{10}}$(T)}

	$\begin{array}{cc}
	\begin{subfigure}{.33 \textwidth}
		\begin{center}
		\includegraphics[width=2.7in]{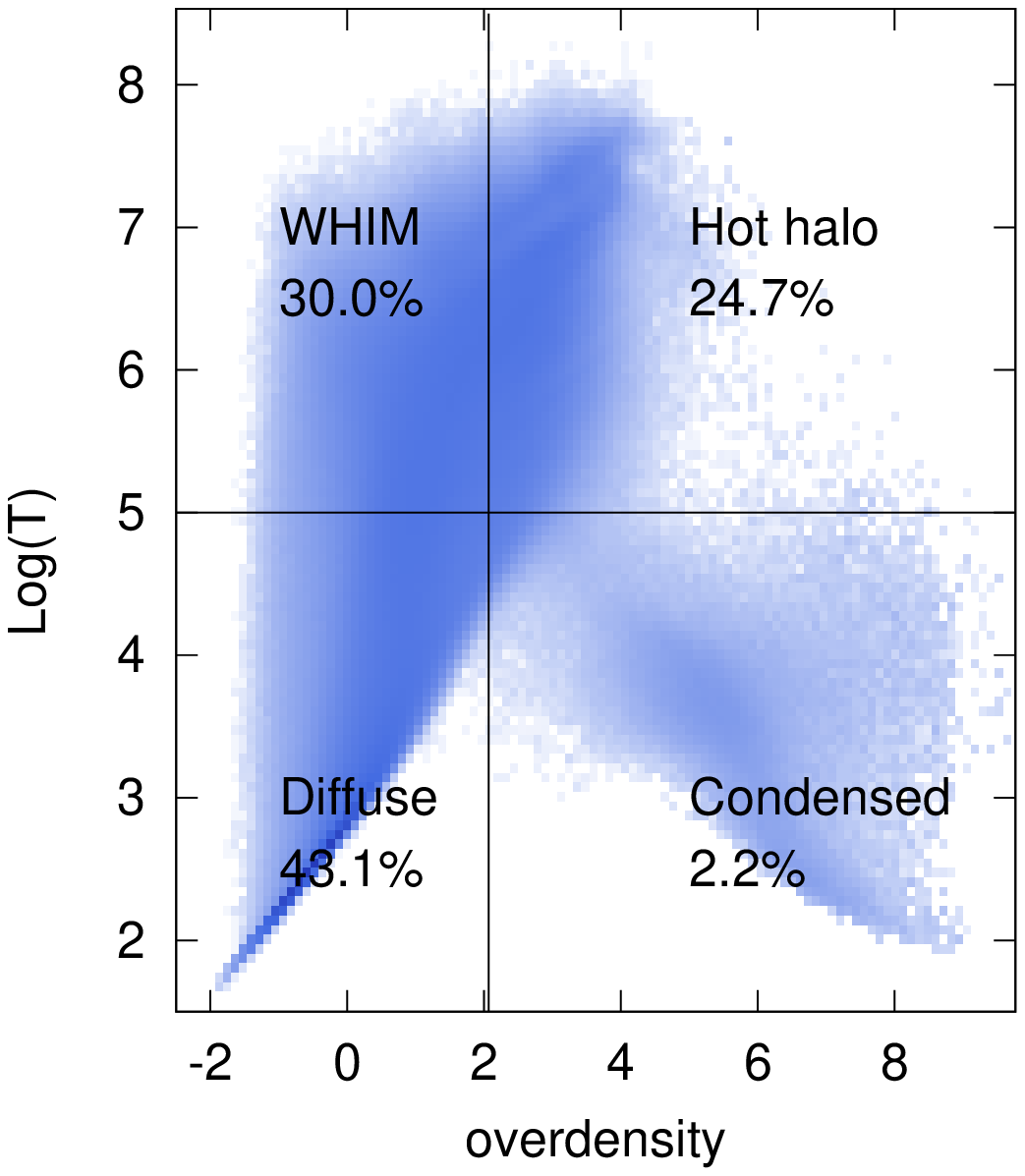}
		\caption{Total}
		\label{fig:gas_phase_z0_a}
		\end{center}
	\end{subfigure}%
	\begin{subfigure}{.33 \textwidth}
		\begin{center}
		\includegraphics[width=2.7in]{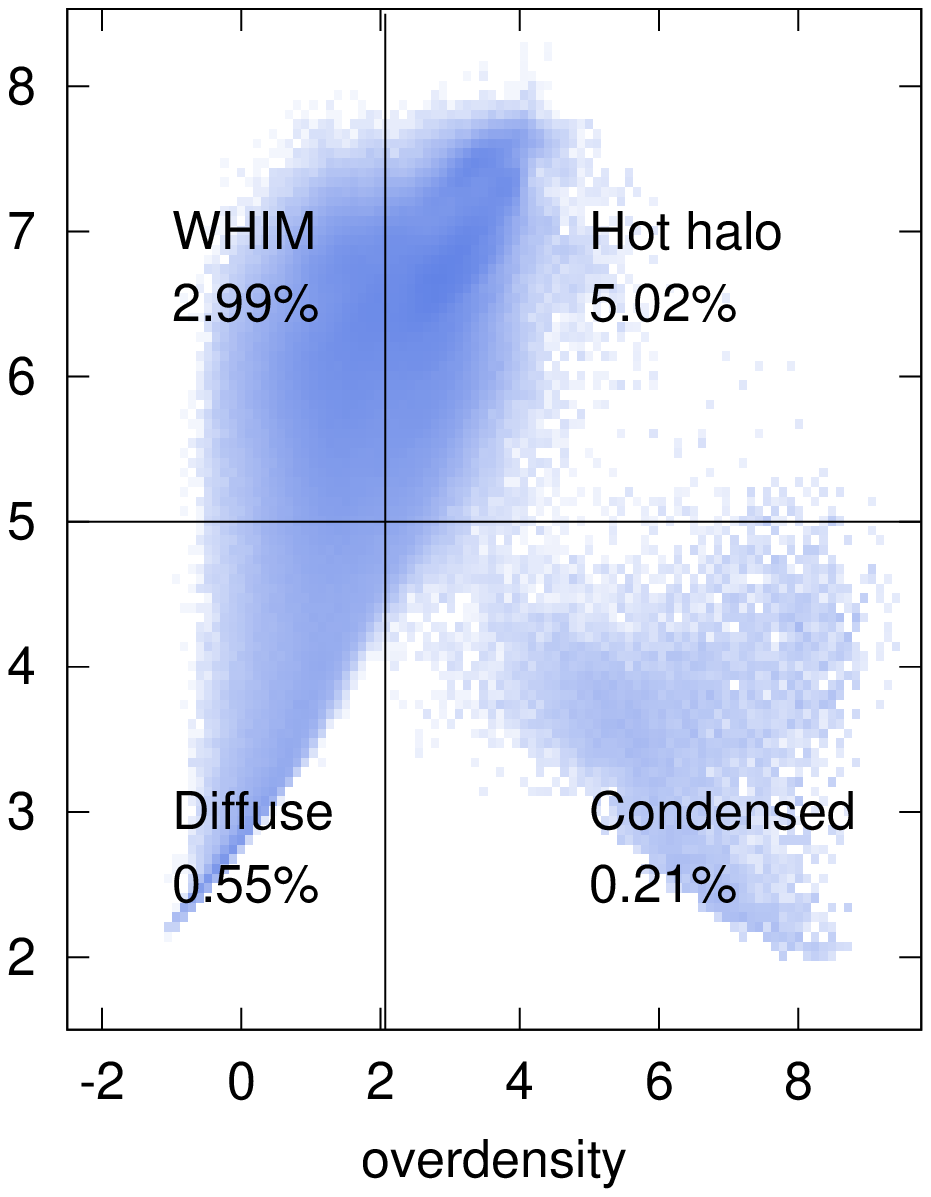}
		\caption{Clusters}
		\label{fig:gas_phase_z0_b}
		\end{center}
	\end{subfigure}
	\quad
	\begin{subfigure}{.33 \textwidth}
		\begin{center}
		\includegraphics[width=2.8in, height=2.7in]{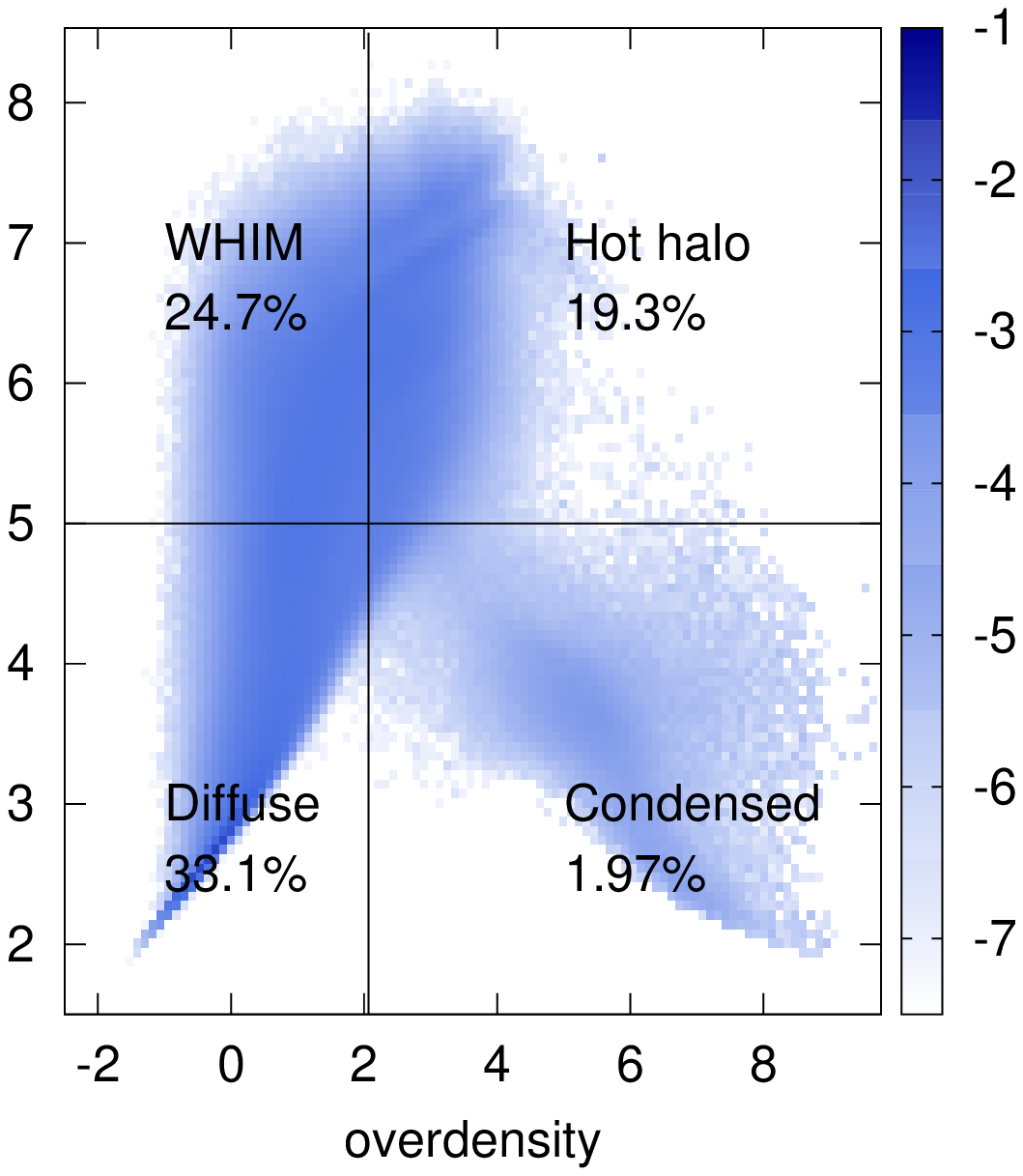}
		\caption{Filaments}
		\label{fig:gas_phase_z0_c}
		\end{center}
	\end{subfigure}
	\\
	\begin{subfigure}{.33 \textwidth}
		\begin{center}
		\includegraphics[width=2.7in]{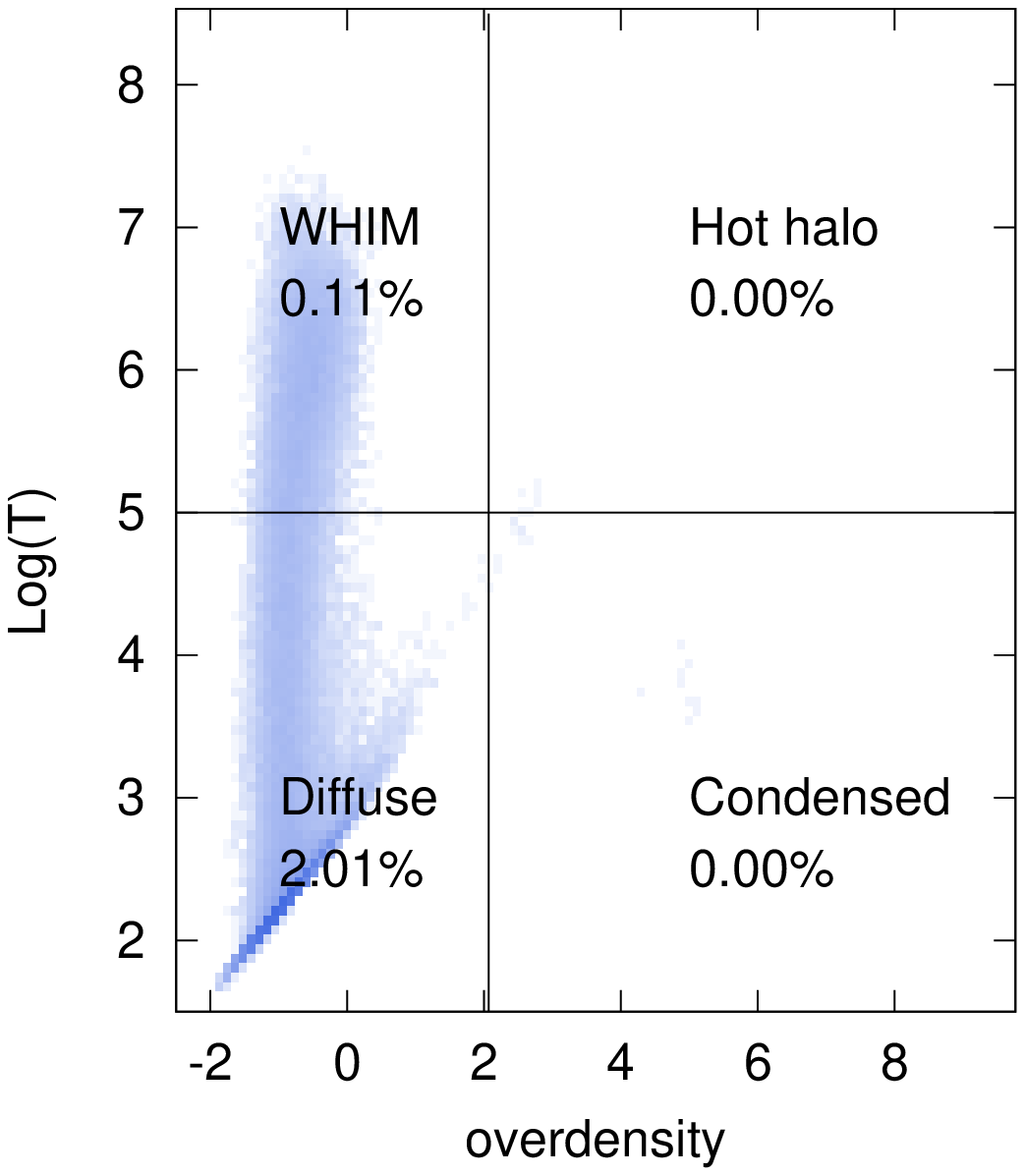}
		\caption{Voids}
		\label{fig:gas_phase_z0_d}
		\end{center}
	\end{subfigure}%
	\quad
	\begin{subfigure}{.33 \textwidth}
		\begin{center}
		\includegraphics[width=2.8in, height=2.7in]{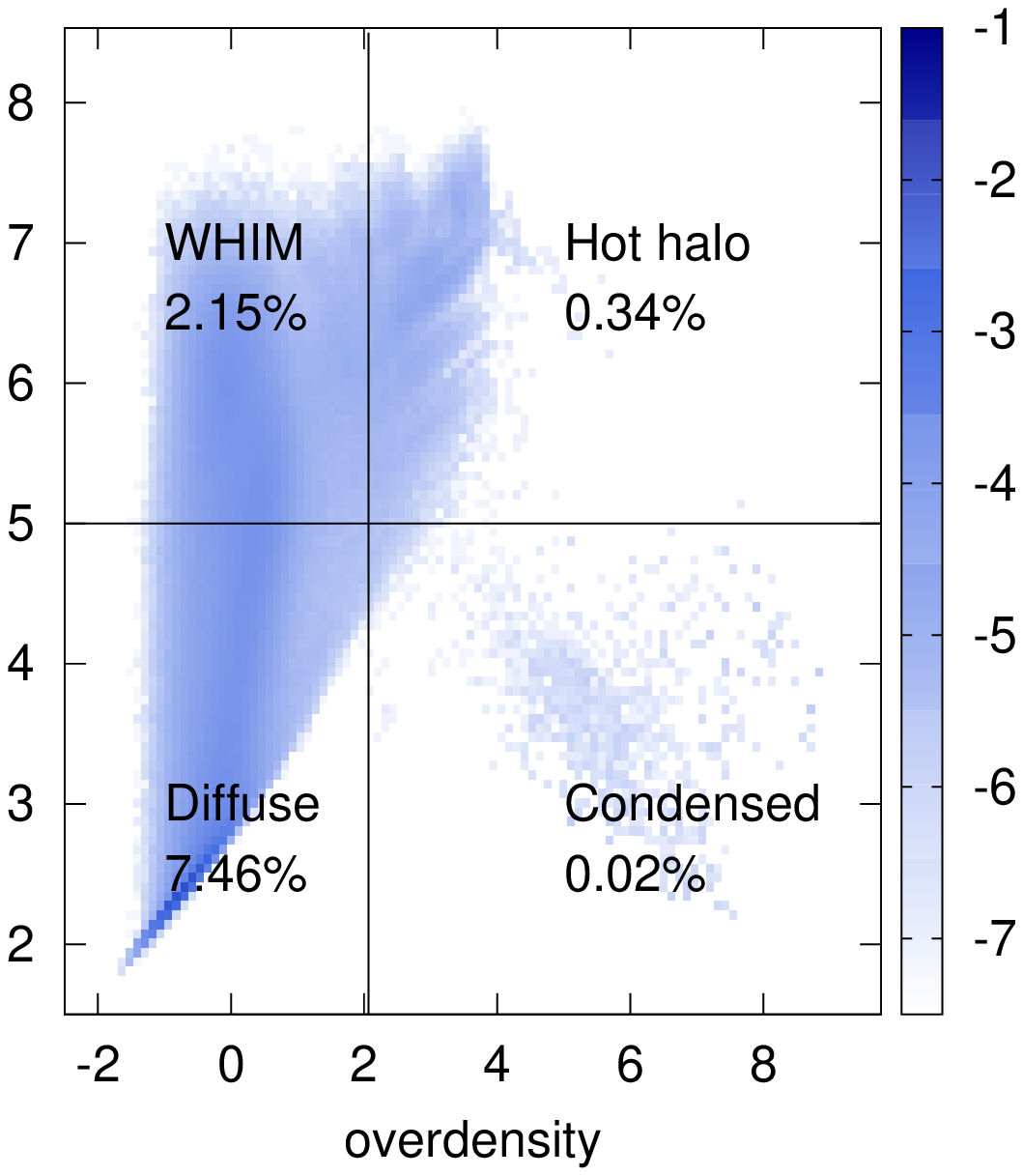}
		\caption{Unassigned}
		\label{fig:gas_phase_z0_e}
		\end{center}
	\end{subfigure}
	\end{array}$
	%\end{center}
\caption{Gas Phase History at Redshift z = 0 using criterion 3.  The color bar represents the log$_{\text{10}}$ of the gas mass fraction at a particular density and temperature. The total gas mass fraction in each phase and structure is listed.  8.77\% of the gas is in clusters, 79.07\% is in filaments, 2.12\% is in voids and 9.97\% is in unassigned.}
\label{fig:gas_phase_z0}
\end{figure*}

\begin{figure*}
%	\psfrag{overdensity}[tc][tc][0.9]{$\rho_{\text{gas}} / \bar{\rho}$ - 1}
%	\psfrag{Log(T)}[tc][tc][0.90]{Log$_{\text{10}}$(T)}	
	$\begin{array}{cc}
	\begin{subfigure}{.33 \textwidth}
		\begin{center}
		\includegraphics[width=2.7in]{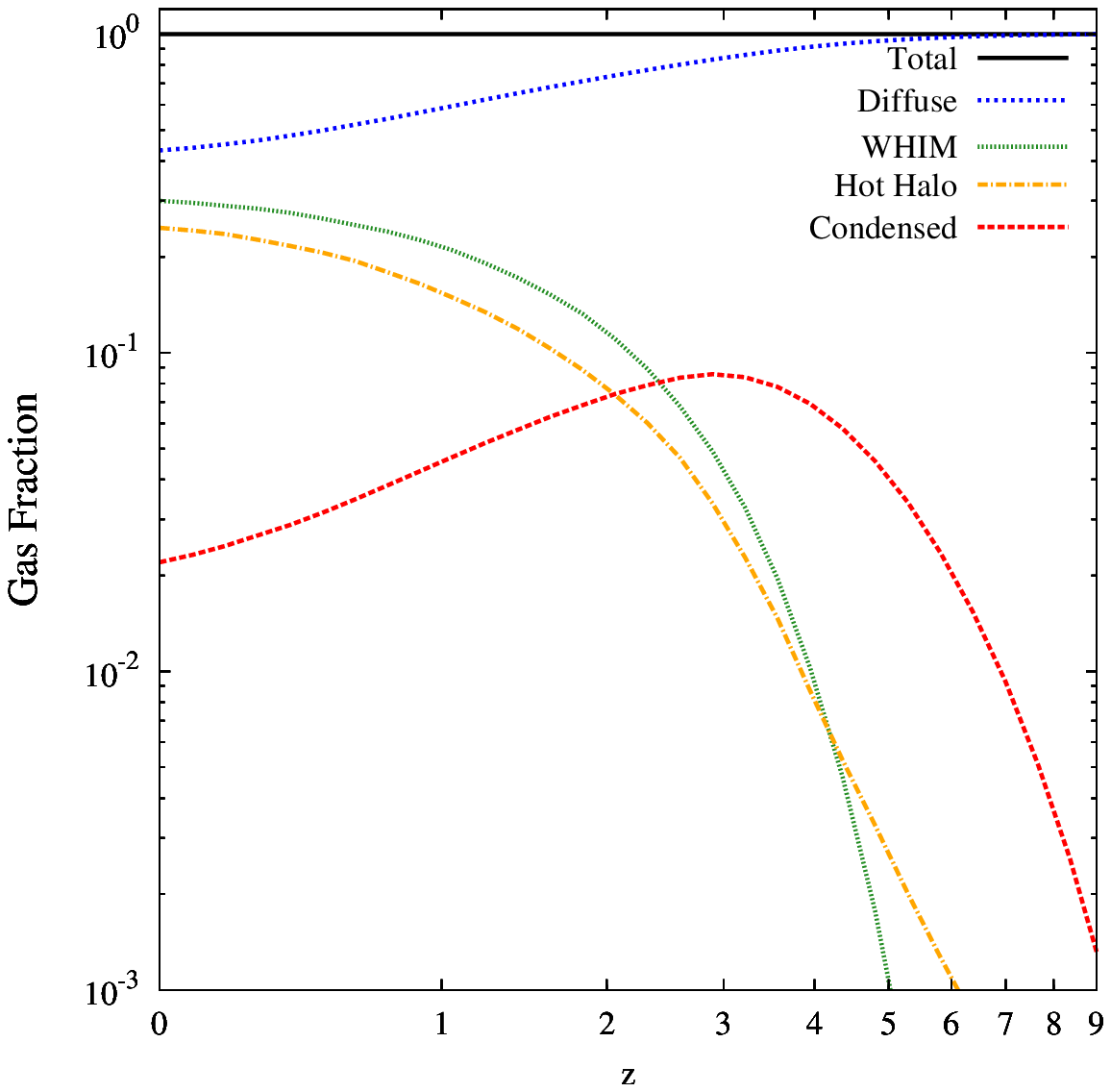}
		\caption{Total}
		\label{fig:gas_phase_a}
		\end{center}
	\end{subfigure}%
	\quad
	\begin{subfigure}{.33 \textwidth}
		\begin{center}
		\includegraphics[width=2.7in]{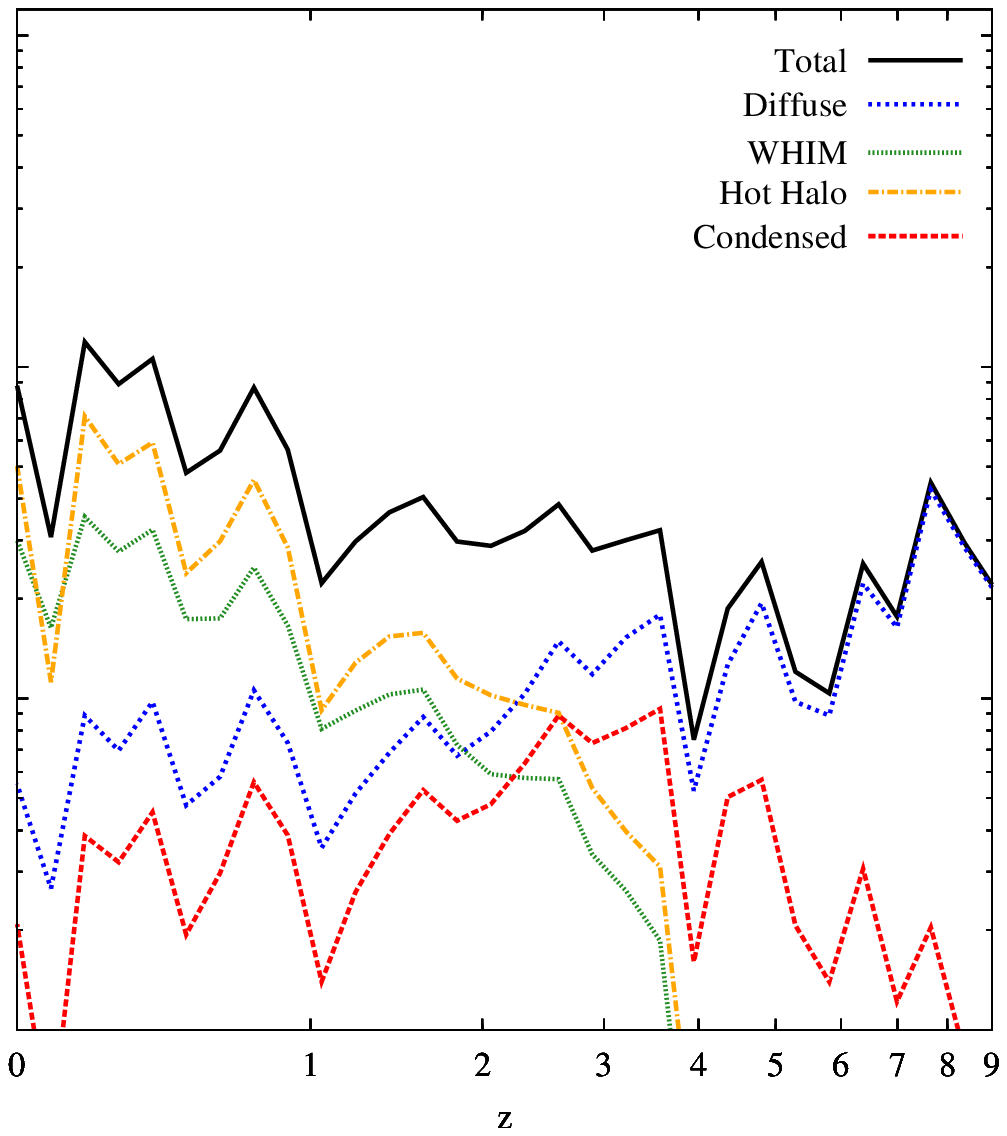}
		\caption{Clusters}
		\label{fig:gas_phase_b}
		\end{center}
	\end{subfigure}
	\quad
	\begin{subfigure}{.33 \textwidth}
		\begin{center}
		\includegraphics[width=2.8in, height=2.7in]{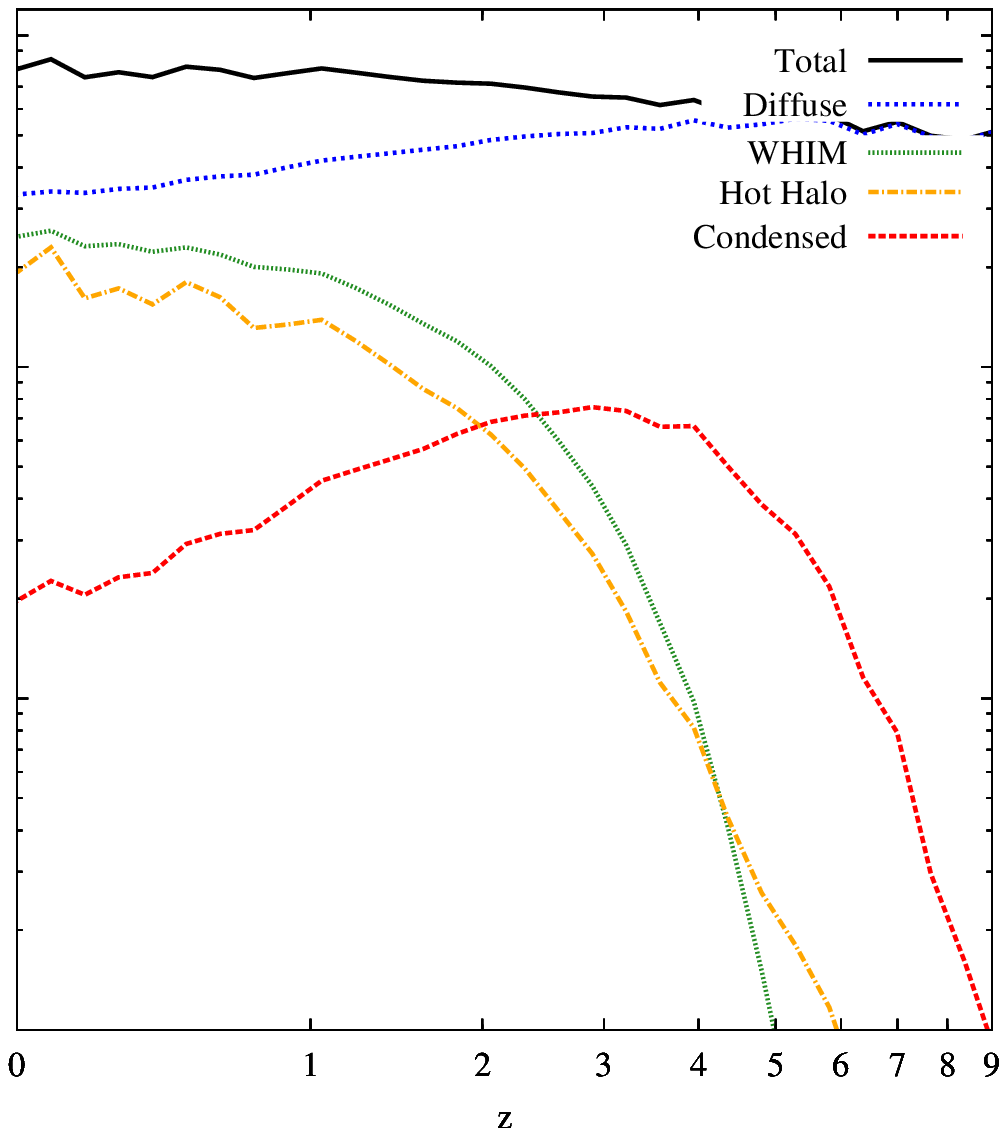}
		\caption{Filaments}
		\label{fig:gas_phase_c}
		\end{center}
	\end{subfigure}
	\\
	\begin{subfigure}{.33 \textwidth}
		\begin{center}
		\includegraphics[width=2.7in]{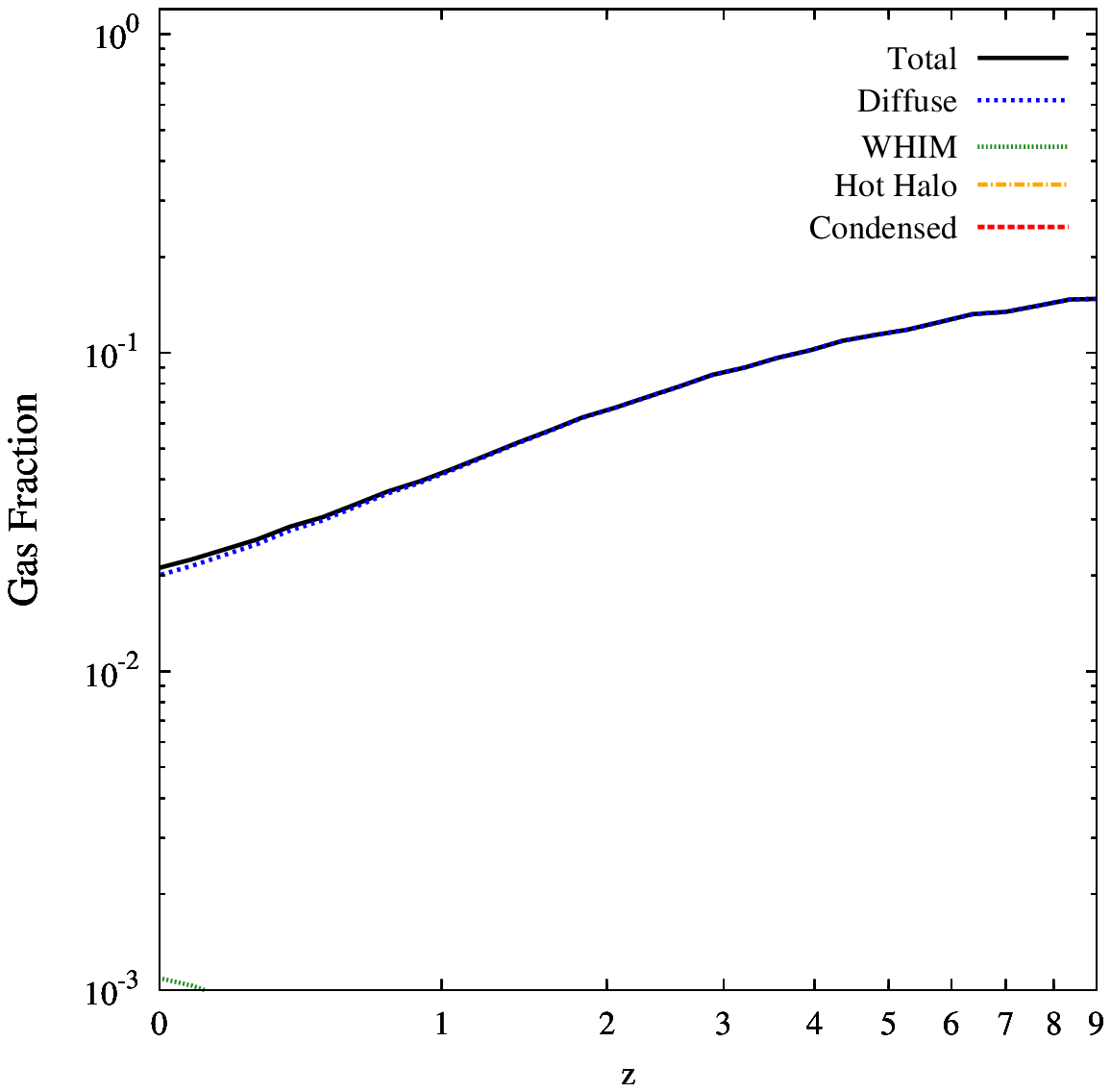}
		\caption{Voids}
		\label{fig:gas_phase_d}
		\end{center}
	\end{subfigure}%
	\quad
	\begin{subfigure}{.33 \textwidth}
		\begin{center}
		\includegraphics[width=2.8in, height=2.7in]{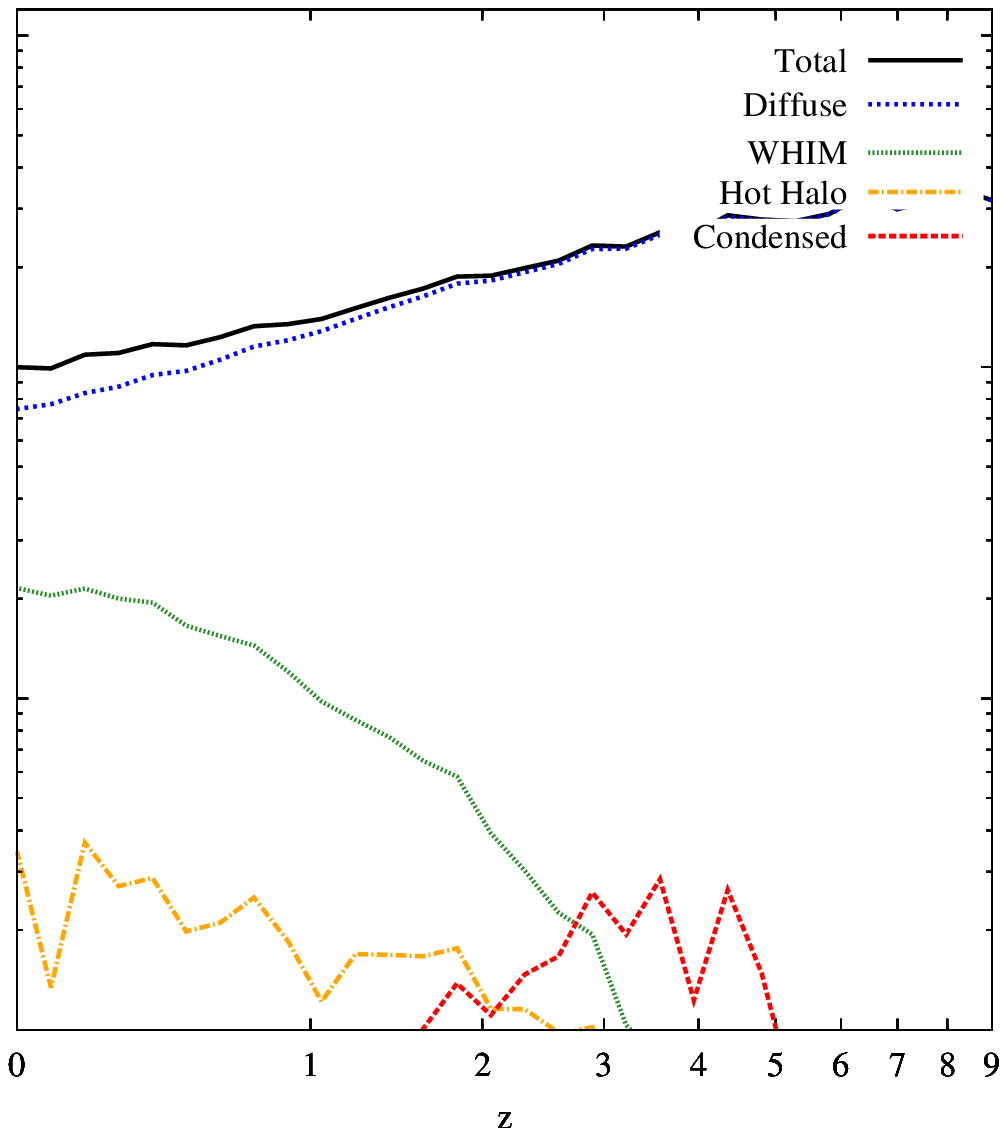}
		\caption{Unassigned}
		\label{fig:gas_phase_e}
		\end{center}
	\end{subfigure}
	\end{array}$
	%\end{center}
\caption{Gas Fraction vs. Redshift using criterion 3.  This illustrates the gas mass fraction in the diffuse, WHIM, hot halo and condensed gas phases as a function of structure.}
\label{fig:gas_phase}
\end{figure*}

\begin{figure*}
	\psfrag{-2}[cc][cc][0.8]{-2}
	\psfrag{2}[cc][cc][0.8]{2}
	\psfrag{3}[cc][cc][0.8]{3}
	\psfrag{4}[cc][cc][0.8]{4}
	\psfrag{5}[cc][cc][0.8]{5}
	\psfrag{6}[cc][cc][0.8]{6}
	\psfrag{7}[cc][cc][0.8]{7}
	\psfrag{8}[cc][cc][0.8]{8}
	\psfrag{overdensity}[tc][tc][0.9]{log$_{\text{10}}$($\rho_{\text{gas}} / \bar{\rho}$)}
	\psfrag{Log(T)}[tc][tc][0.90]{log$_{\text{10}}$(T)}

	$\begin{array}{cc}
	\begin{subfigure}{.33 \textwidth}
		\begin{center}
		\includegraphics[width=2.7in]{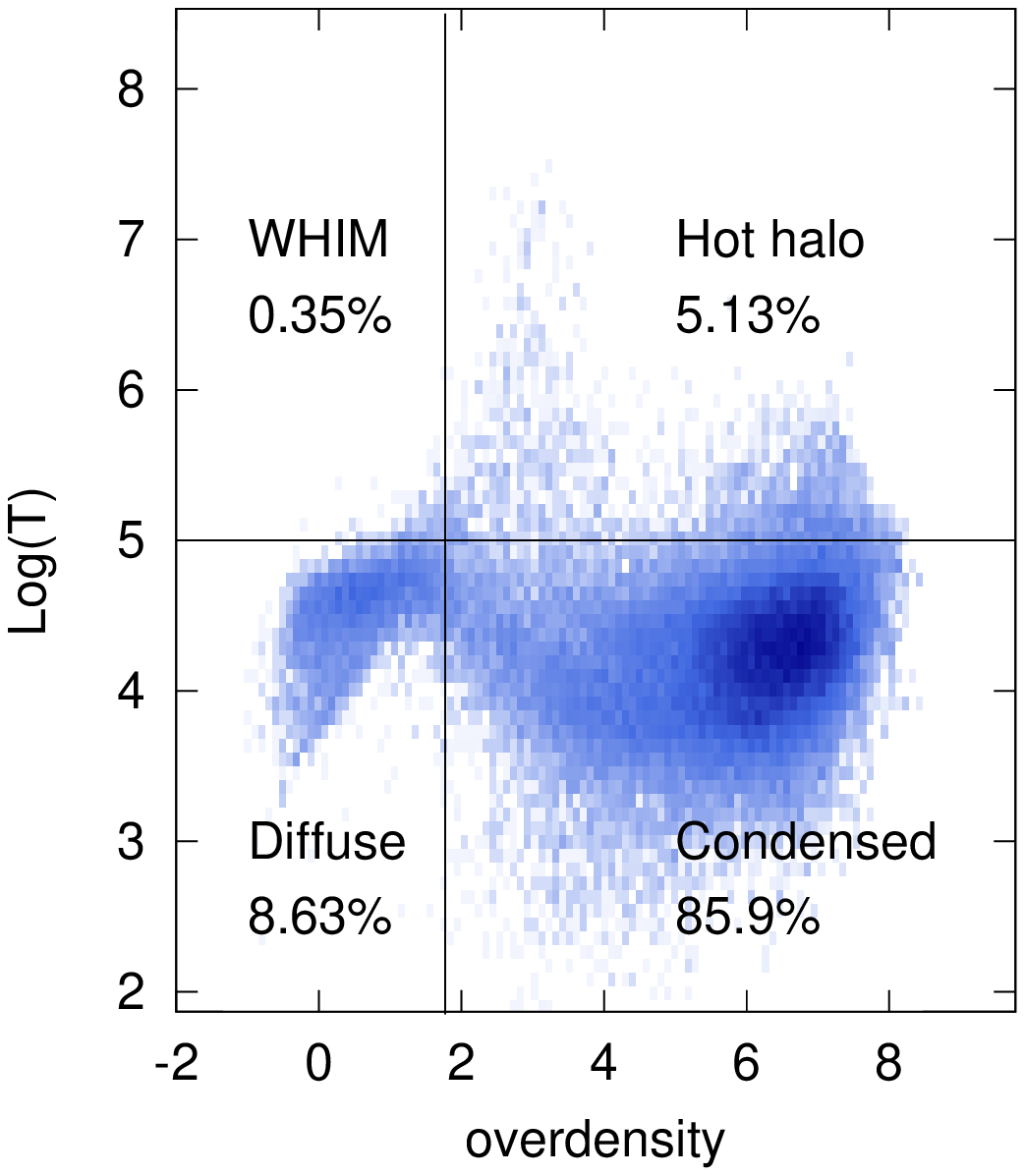}
		\caption{Total}
		\label{fig:sfr_z3_a}
		\end{center}
	\end{subfigure}%
	\begin{subfigure}{.33 \textwidth}
		\begin{center}
		\includegraphics[width=2.7in]{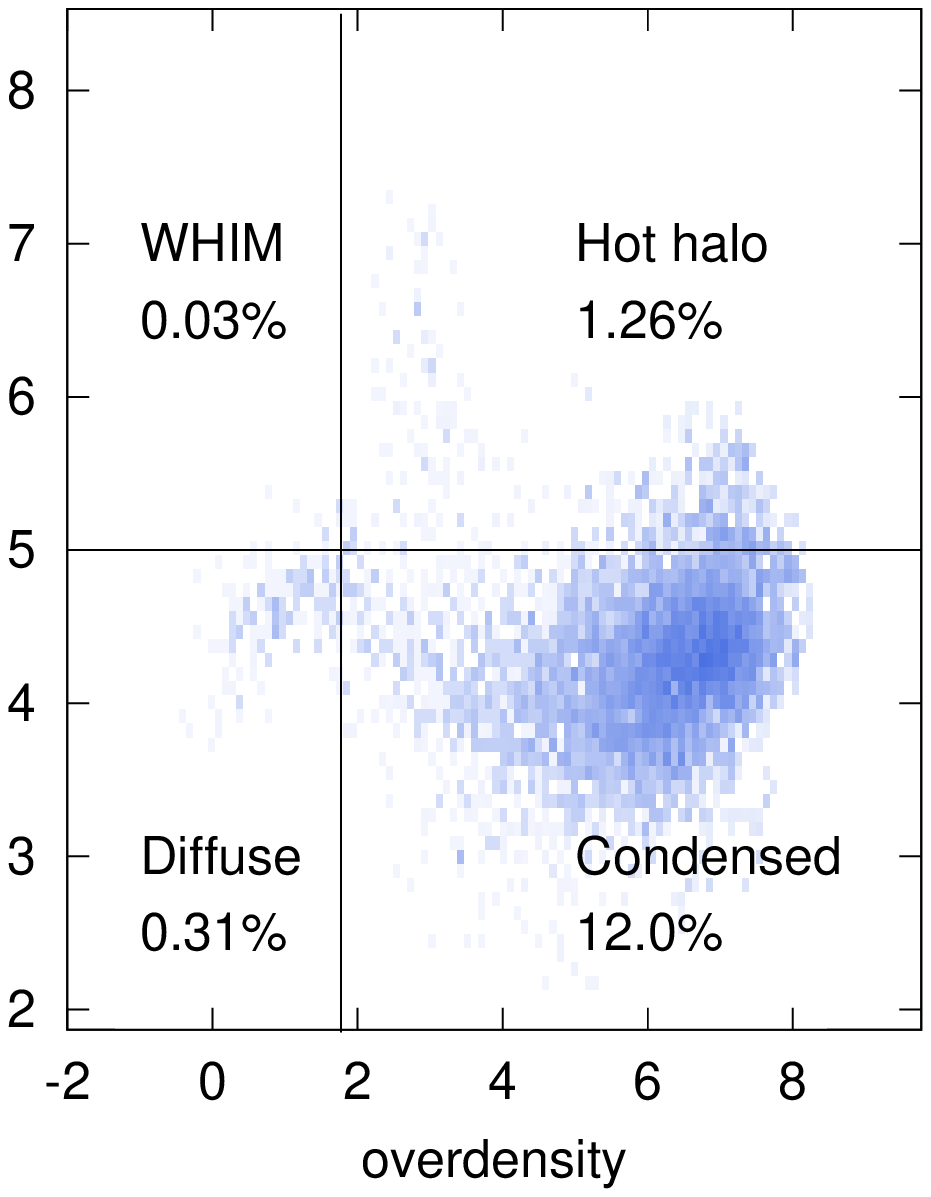}
		\caption{Clusters}
		\label{fig:sfr_z3_b}
		\end{center}
	\end{subfigure}
	\quad
	\begin{subfigure}{.33 \textwidth}
		\begin{center}
		\includegraphics[width=2.8in, height=2.7in]{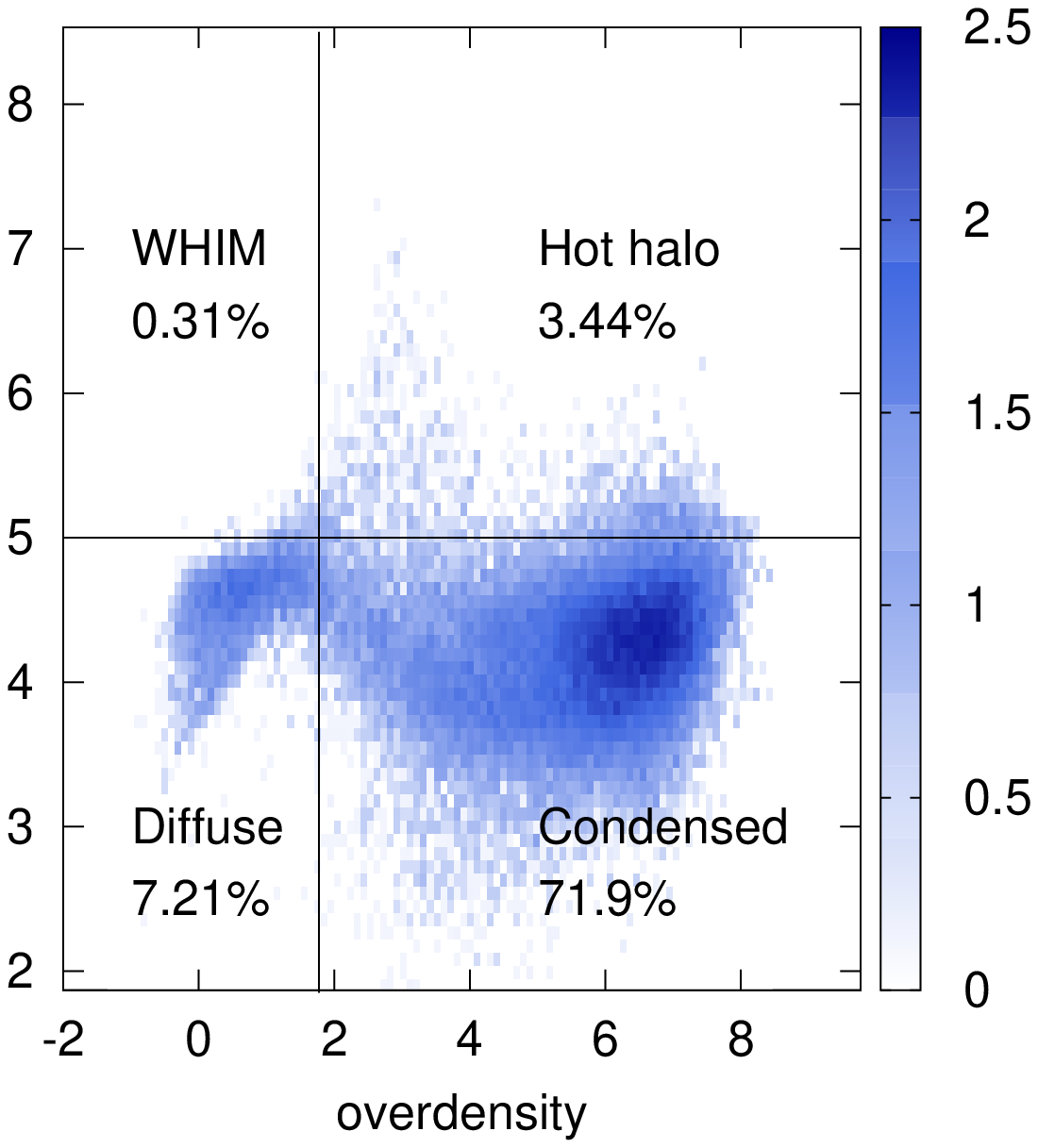}
		\caption{Filaments}
		\label{fig:sfr_z3_c}
		\end{center}
	\end{subfigure}
	\\
	\begin{subfigure}{.33 \textwidth}
		\begin{center}
		\includegraphics[width=2.7in]{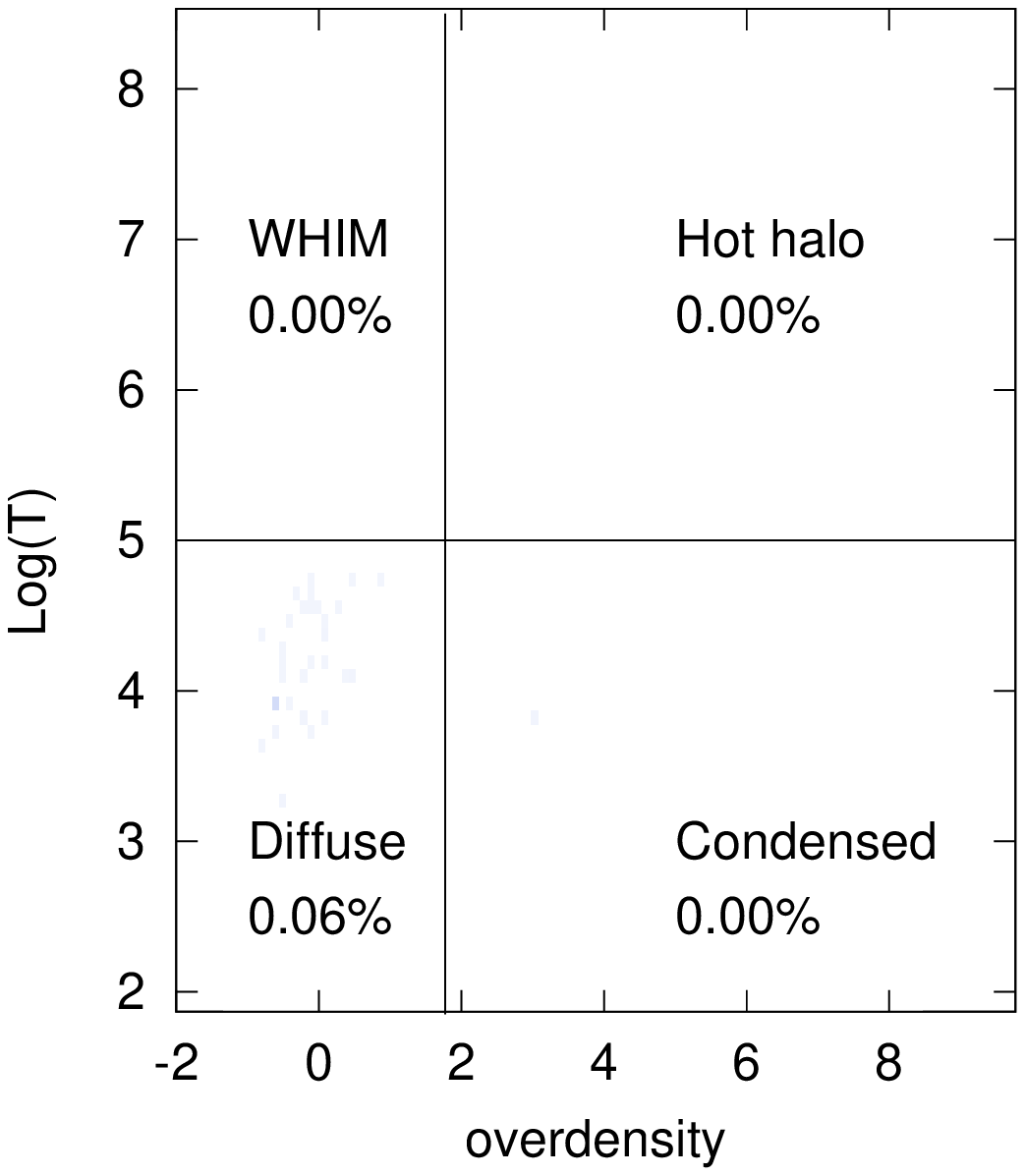}
		\caption{Voids}
		\label{fig:sfr_z3_d}
		\end{center}
	\end{subfigure}%
	\quad
	\begin{subfigure}{.33 \textwidth}
		\begin{center}
		\includegraphics[width=2.8in, height=2.7in]{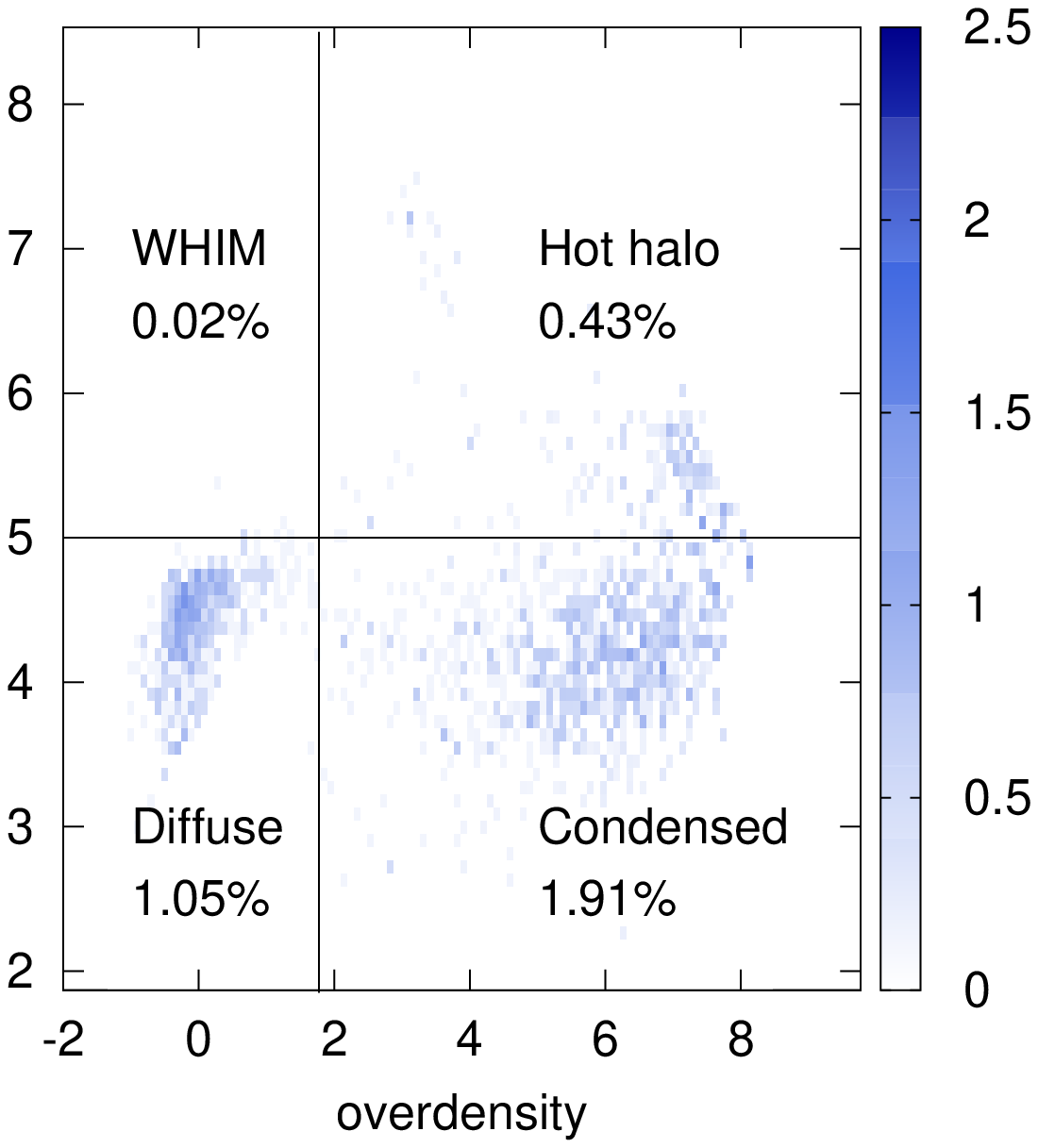}
		\caption{Unassigned}
		\label{fig:sfr_z3_e}
		\end{center}
	\end{subfigure}
	\end{array}$
	%\end{center}
\caption{Star Formation Rate, Redshift z = 3.2 using criterion 3.  The color bar represents the log$_{\text{10}}$ of the star formation rate in M$_{\odot}$ yr$^{-1}$ at a particular density and temperature. The fraction of the total stars formed is listed for each phase and structure.  13.6\% of stars are forming in the clusters, 82.9\% in filaments, 0.06\% in voids and 3.4\% in material unassigned to any structure.}
\label{fig:sfr_z3}
\end{figure*}

\begin{figure*}
	\psfrag{-2}[cc][cc][0.8]{-2}
	\psfrag{2}[cc][cc][0.8]{2}
	\psfrag{3}[cc][cc][0.8]{3}
	\psfrag{4}[cc][cc][0.8]{4}
	\psfrag{5}[cc][cc][0.8]{5}
	\psfrag{6}[cc][cc][0.8]{6}
	\psfrag{7}[cc][cc][0.8]{7}
	\psfrag{8}[cc][cc][0.8]{8}
	\psfrag{overdensity}[tc][tc][0.9]{log$_{\text{10}}$($\rho_{\text{gas}} / \bar{\rho}$)}
	\psfrag{Log(T)}[tc][tc][0.90]{log$_{\text{10}}$(T)}

	$\begin{array}{cc}
	\begin{subfigure}{.33 \textwidth}
		\begin{center}
		\includegraphics[width=2.7in]{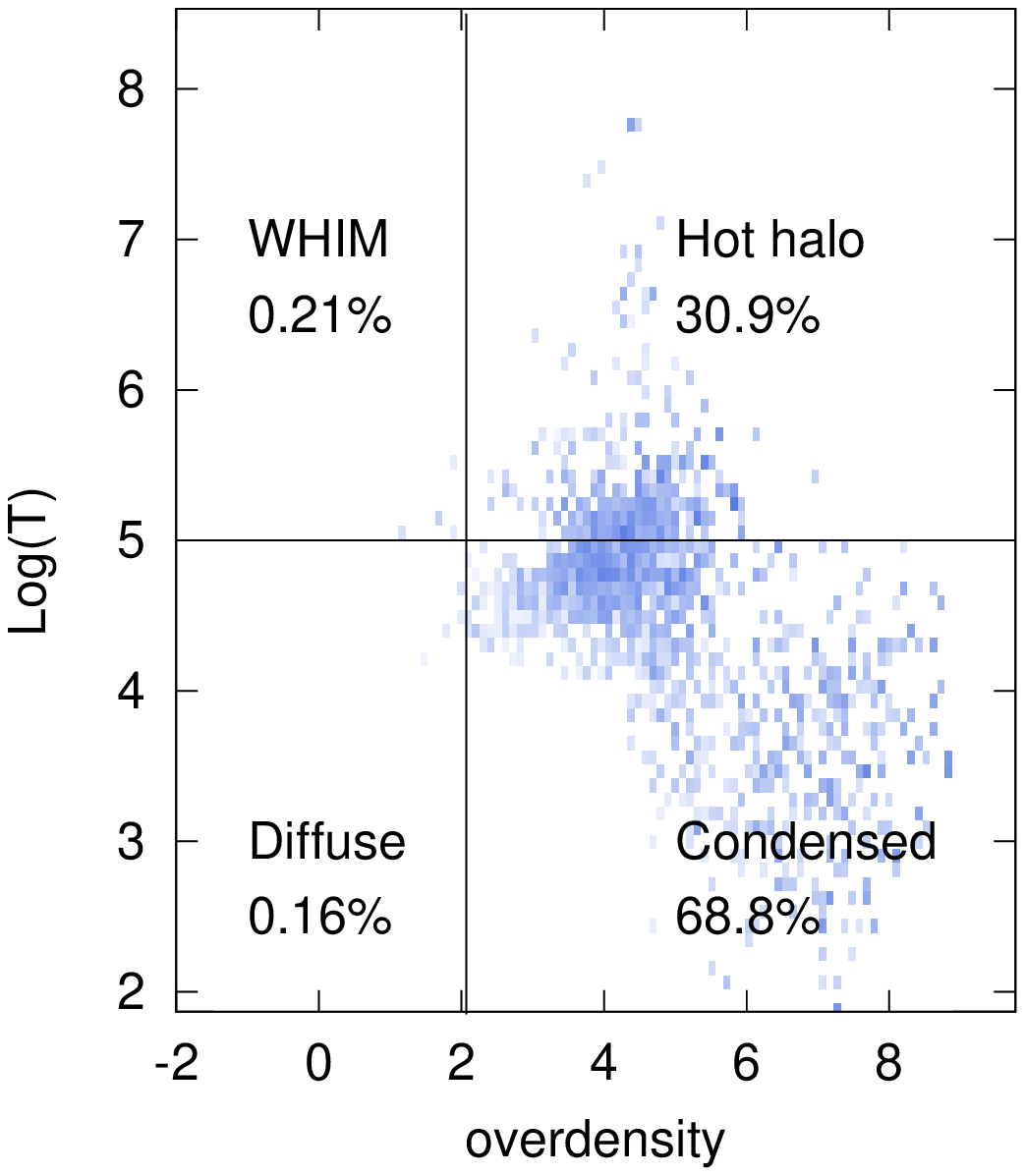}
		\caption{Total}
		\label{fig:sfr_z0_a}
		\end{center}
	\end{subfigure}%
	\begin{subfigure}{.33 \textwidth}
		\begin{center}
		\includegraphics[width=2.7in]{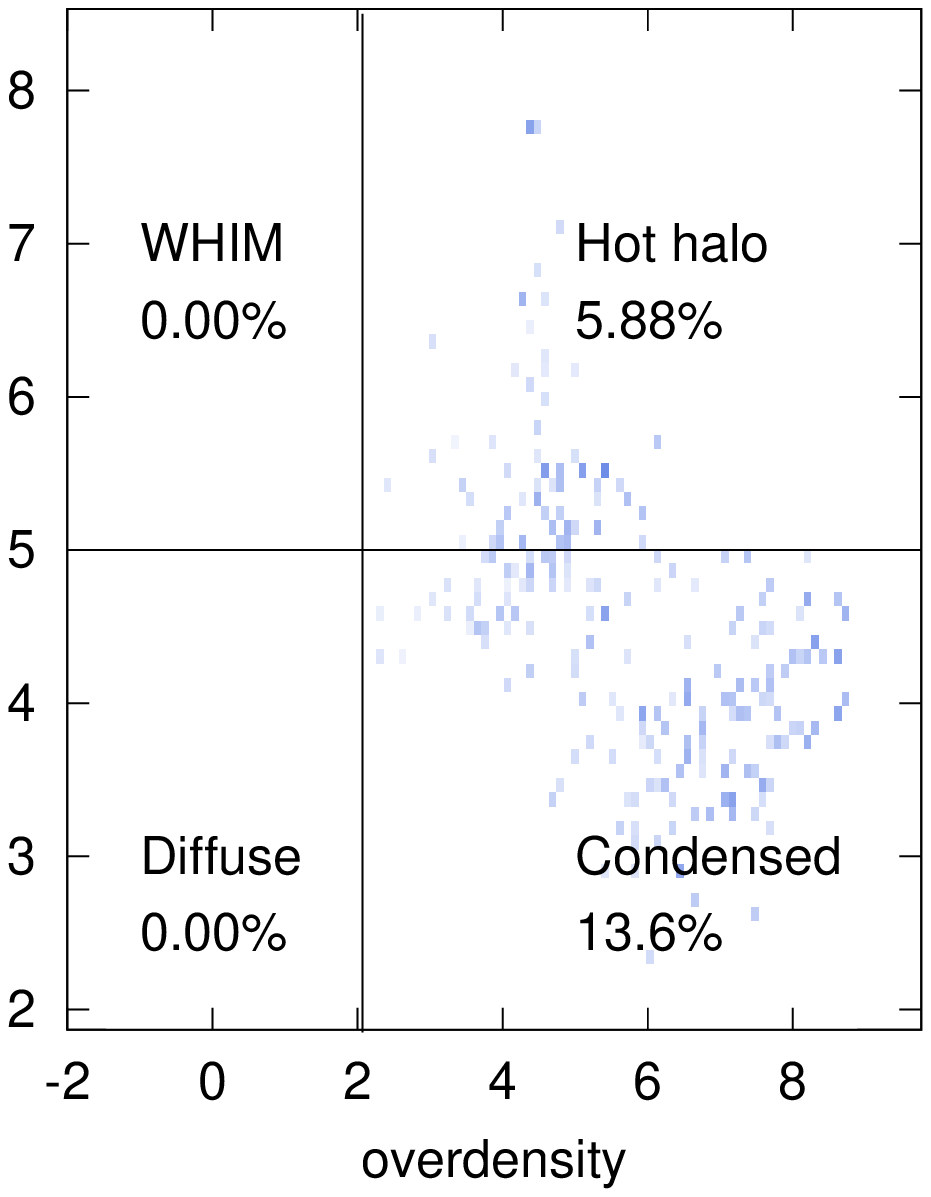}
		\caption{Clusters}
		\label{fig:sfr_z0_b}
		\end{center}
	\end{subfigure}
	\quad
	\begin{subfigure}{.33 \textwidth}
		\begin{center}
		\includegraphics[width=2.8in, height=2.7in]{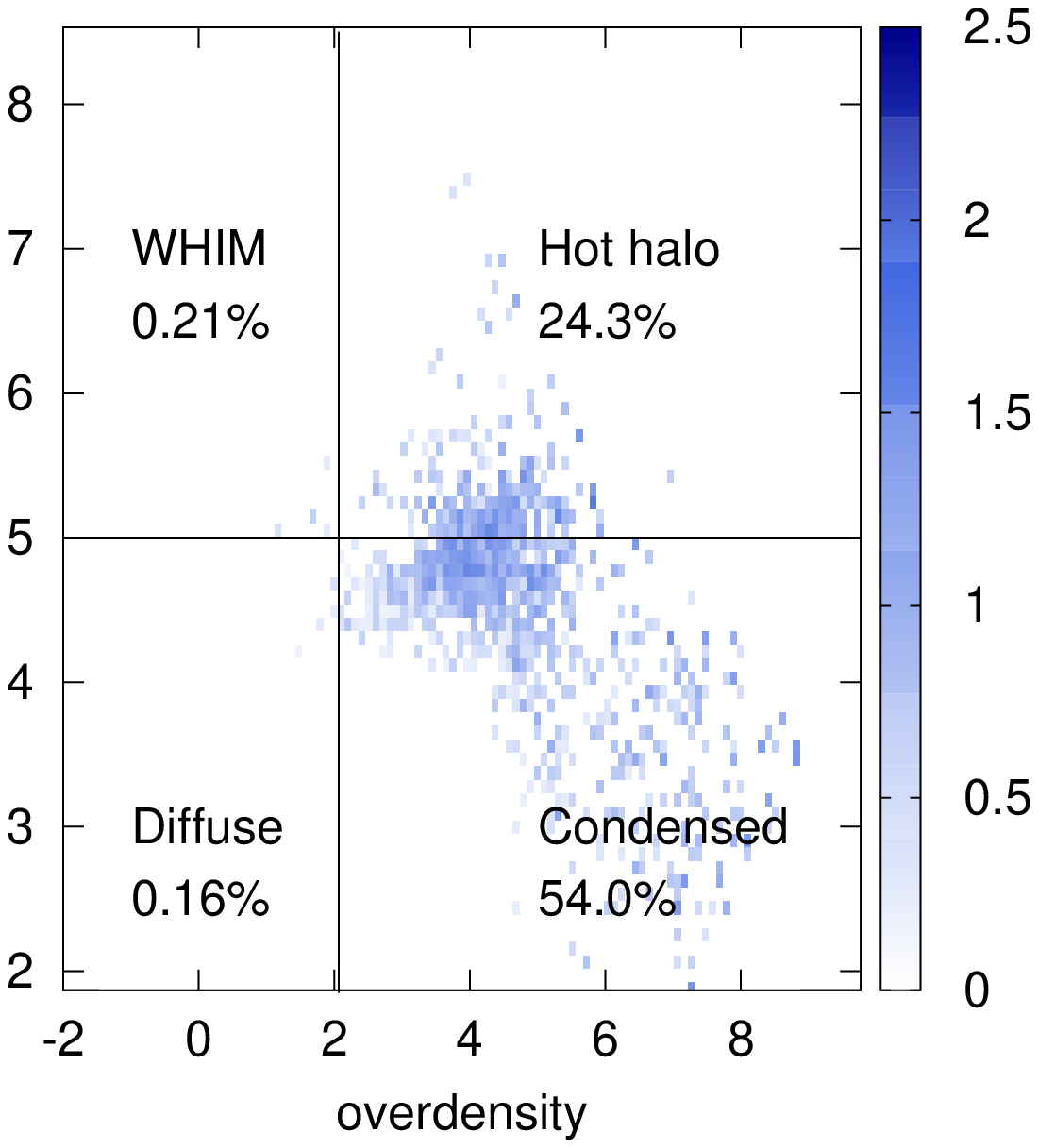}
		\caption{Filaments}
		\label{fig:sfr_z0_c}
		\end{center}
	\end{subfigure}
	\\
	\begin{subfigure}{.33 \textwidth}
		\begin{center}
		\includegraphics[width=2.7in]{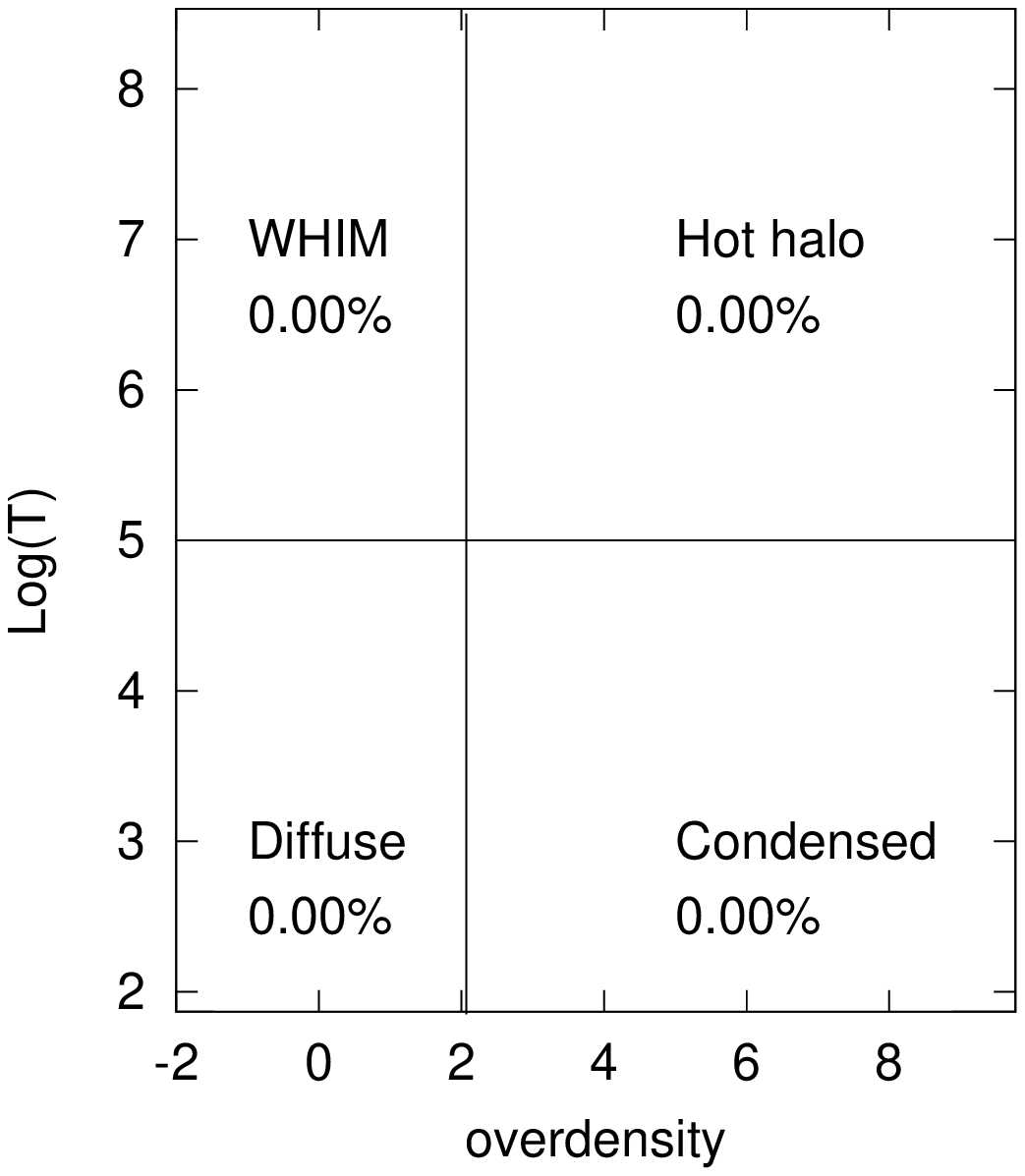}
		\caption{Voids}
		\label{fig:sfr_z0_d}
		\end{center}
	\end{subfigure}%
	\quad
	\begin{subfigure}{.33 \textwidth}
		\begin{center}
		\includegraphics[width=2.8in, height=2.7in]{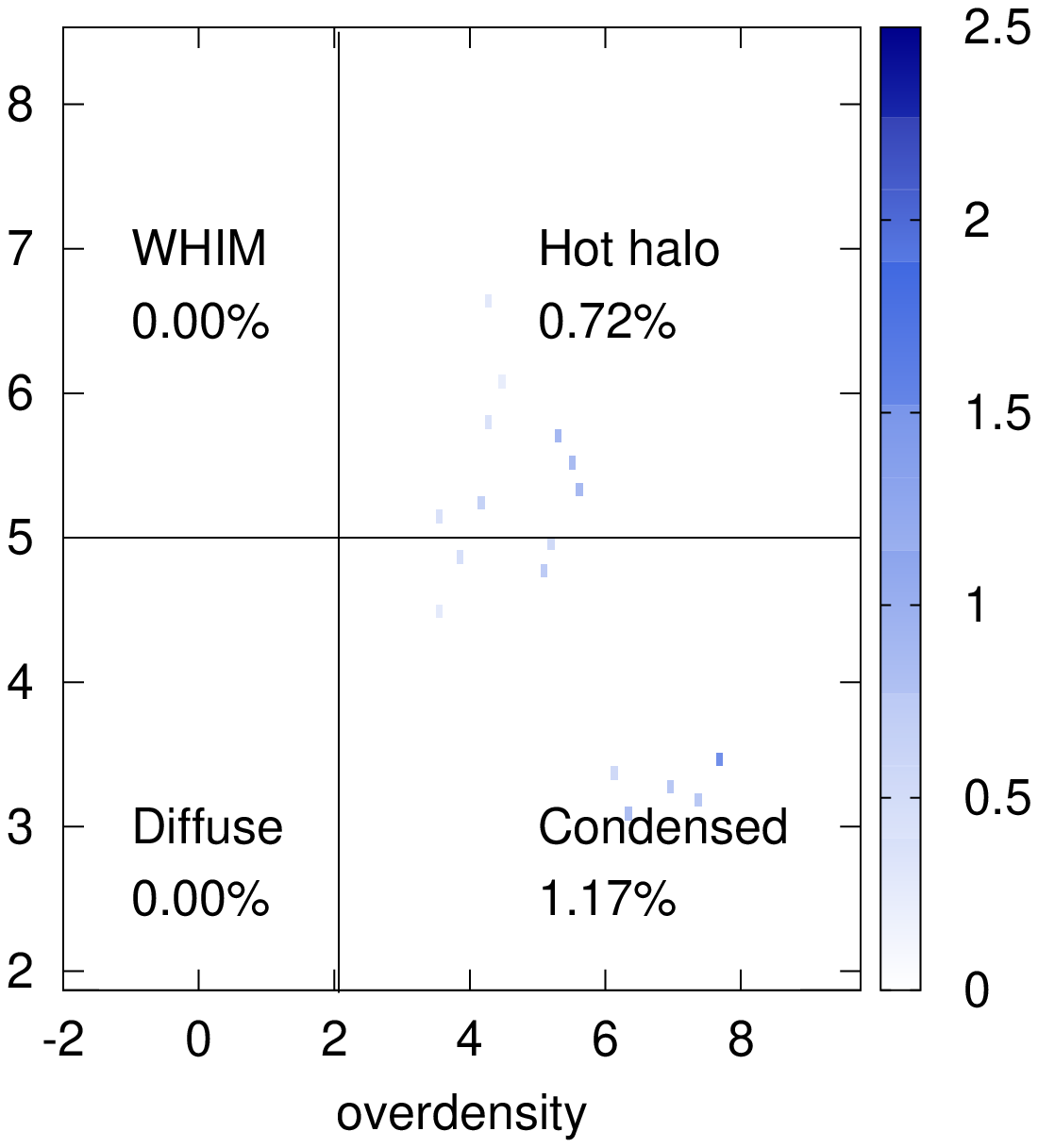}
		\caption{Unassigned}
		\label{fig:sfr_z0_e}
		\end{center}
	\end{subfigure}
	\end{array}$
	%\end{center}
\caption{The star formation rate at redshift z = 0, using criterion 3.  The color bar represents the log$_{\text{10}}$ of the star formation rate in M$_{\odot}$ yr$^{-1}$ at a particular density and temperature.  The fraction of the total star mass formed is listed for each phase and structure.  19.8\% of stars are forming in the clusters, 78.3\% in filaments, 0.00\% in voids and 1.89\% in in material unassigned to any structure.}
\label{fig:sfr_z0}
\end{figure*}

\begin{figure*}
	\psfrag{SFR}[c][c][0.7]{SFR (M$_{\odot}$ yr$^{-1}$)}
	\psfrag{z}[c][c][0.70]{z}	
	$\begin{array}{cc}
	\begin{subfigure}{.33 \textwidth}
		\begin{center}
		\includegraphics[width=2.7in]{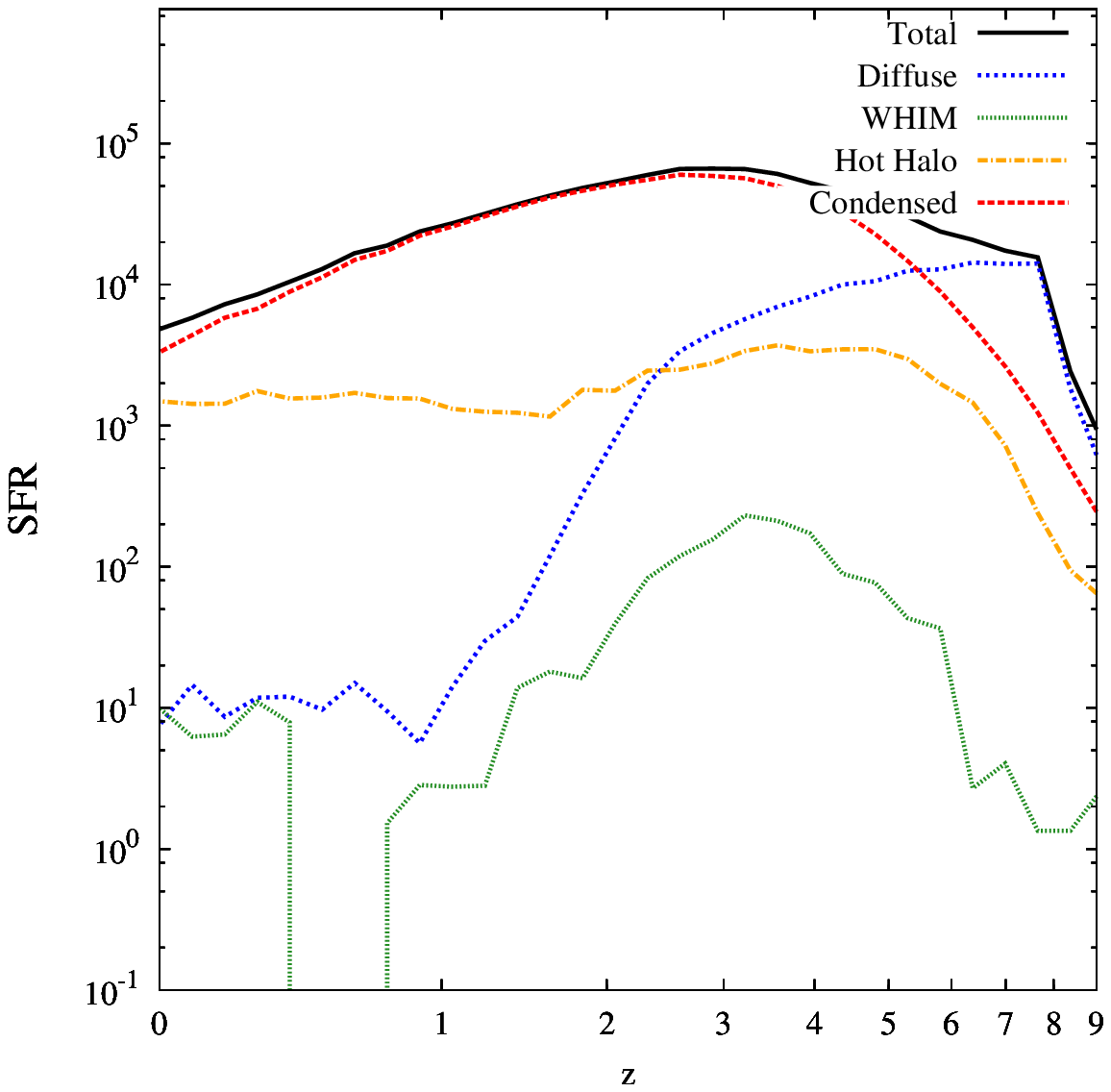}
		\caption{Total}
		\label{fig:sfr_phase_a}
		\end{center}
	\end{subfigure}%
	\quad
	\begin{subfigure}{.33 \textwidth}
		\begin{center}
		\includegraphics[width=2.7in]{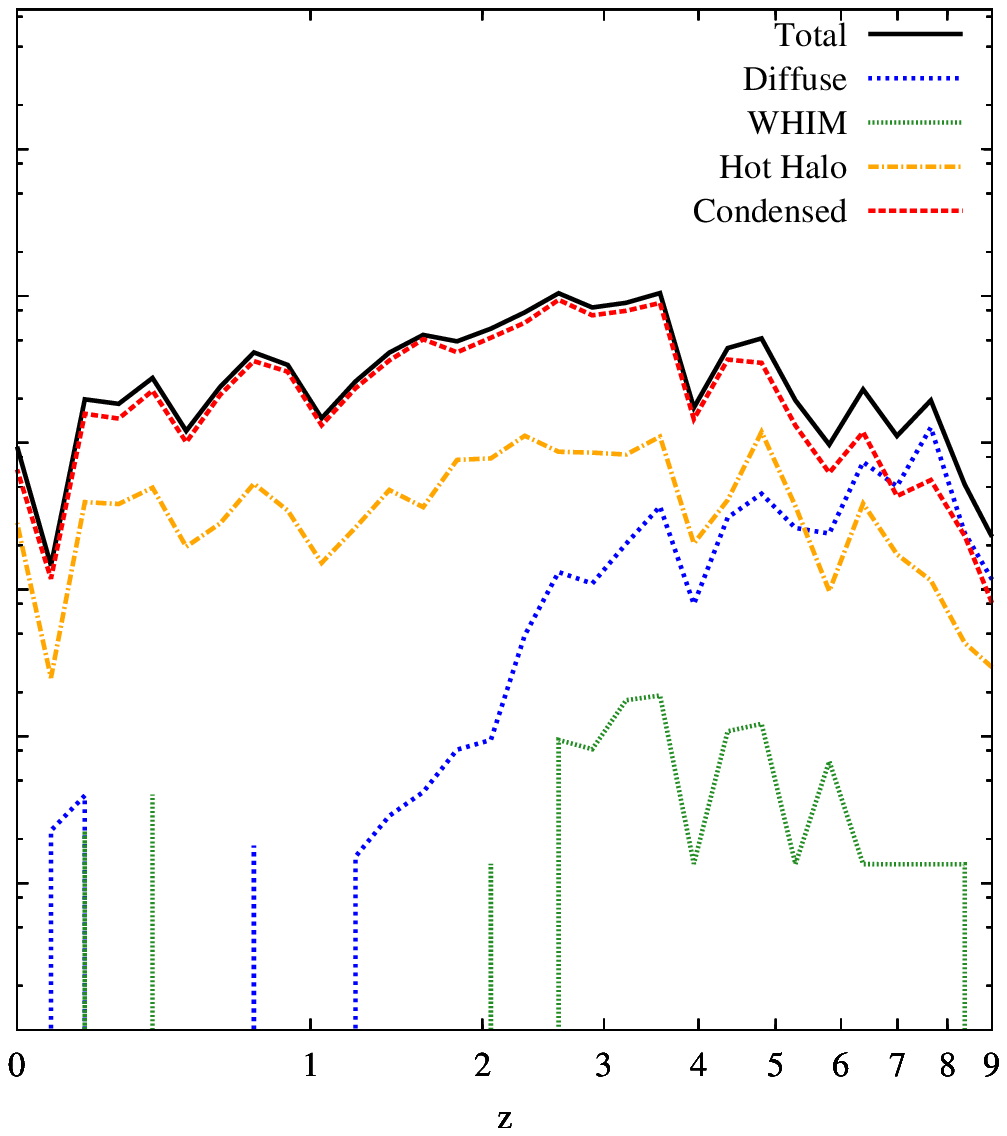}
		\caption{Clusters}
		\label{fig:sfr_phase_b}
		\end{center}
	\end{subfigure}
	\quad
	\begin{subfigure}{.33 \textwidth}
		\begin{center}
		\includegraphics[width=2.8in, height=2.7in]{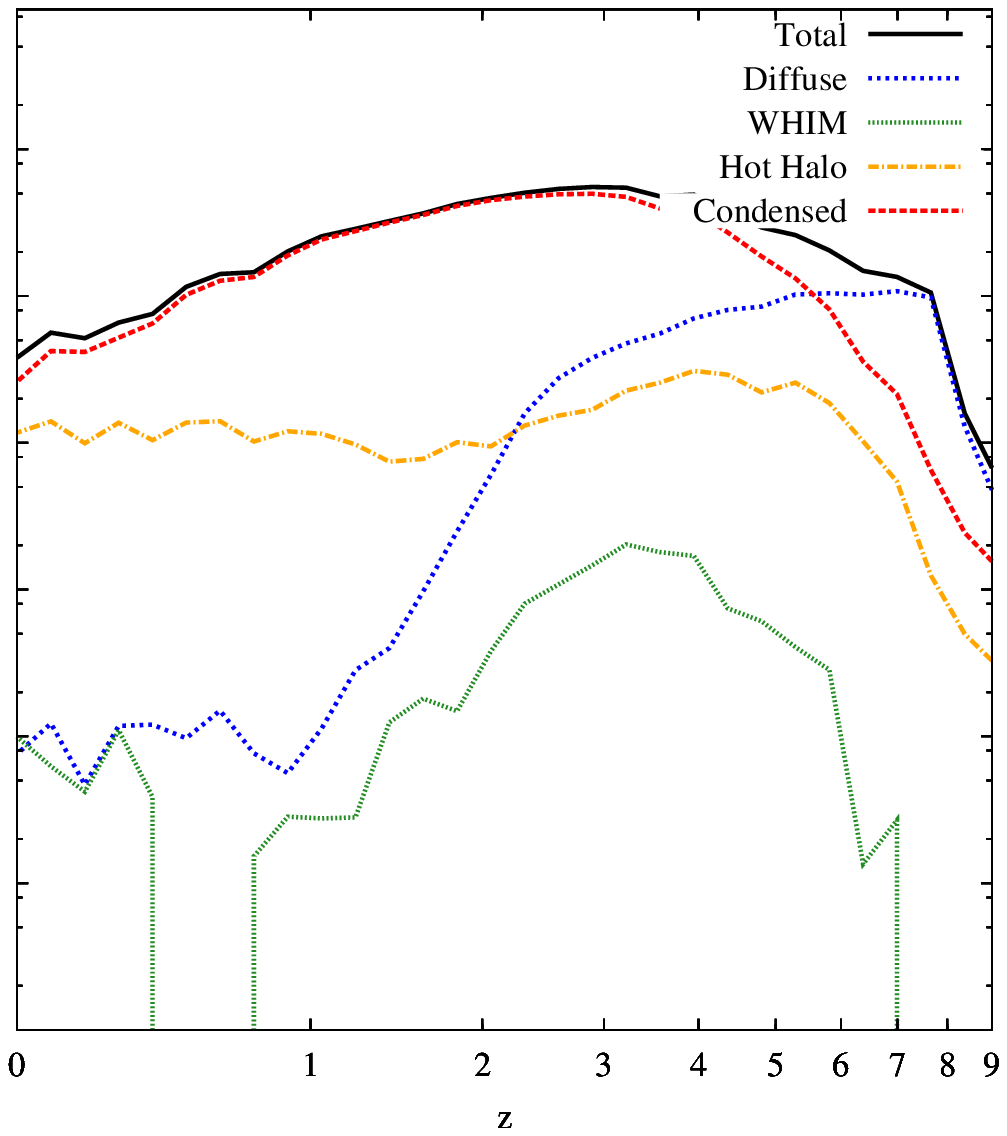}
		\caption{Filaments}
		\label{fig:sfr_phase_c}
		\end{center}
	\end{subfigure}
	\\
	\begin{subfigure}{.33 \textwidth}
		\begin{center}
		\includegraphics[width=2.7in]{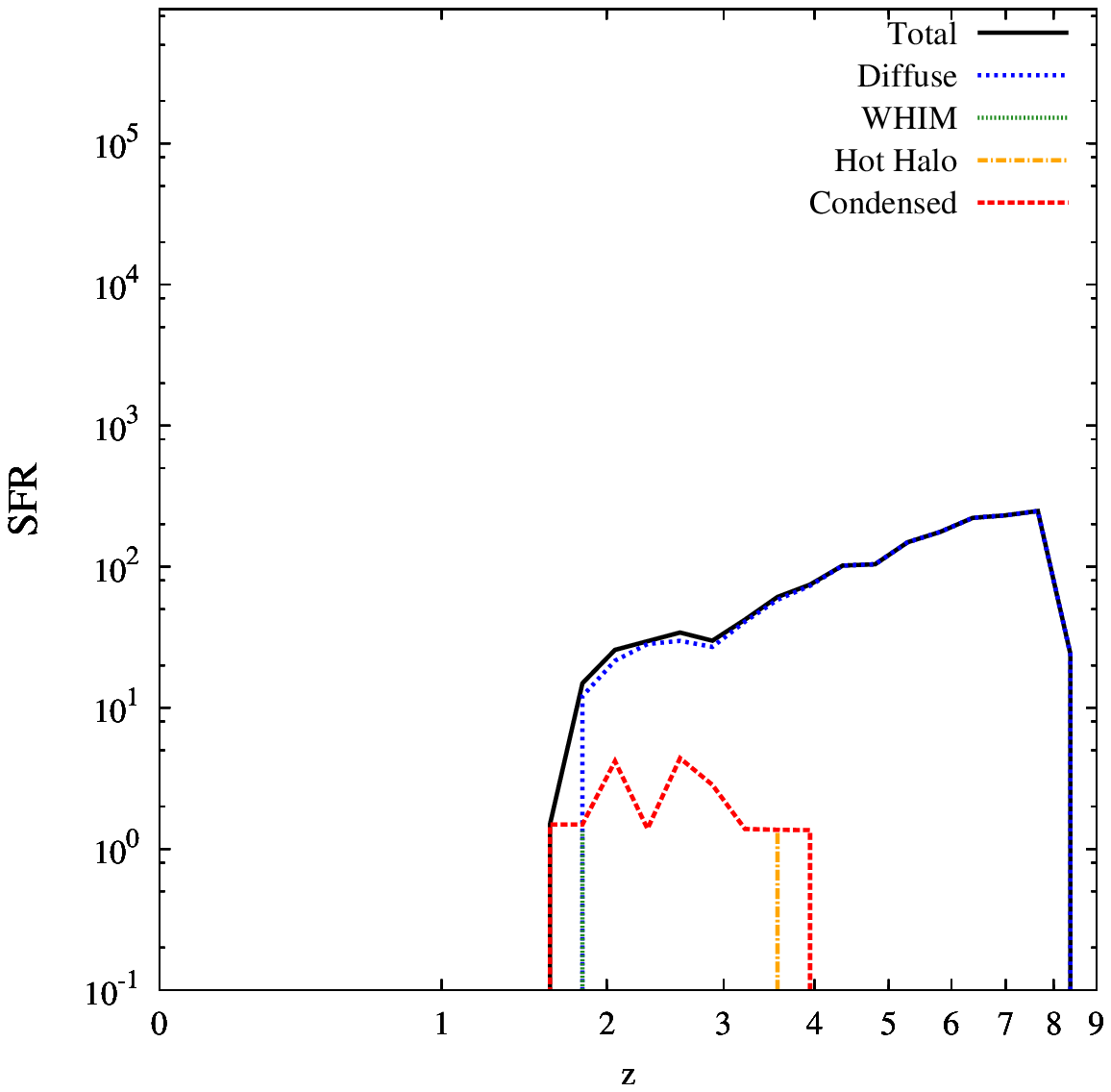}
		\caption{Voids}
		\label{fig:sfr_phase_d}
		\end{center}
	\end{subfigure}%
	\quad
	\begin{subfigure}{.33 \textwidth}
		\begin{center}
		\includegraphics[width=2.8in, height=2.7in]{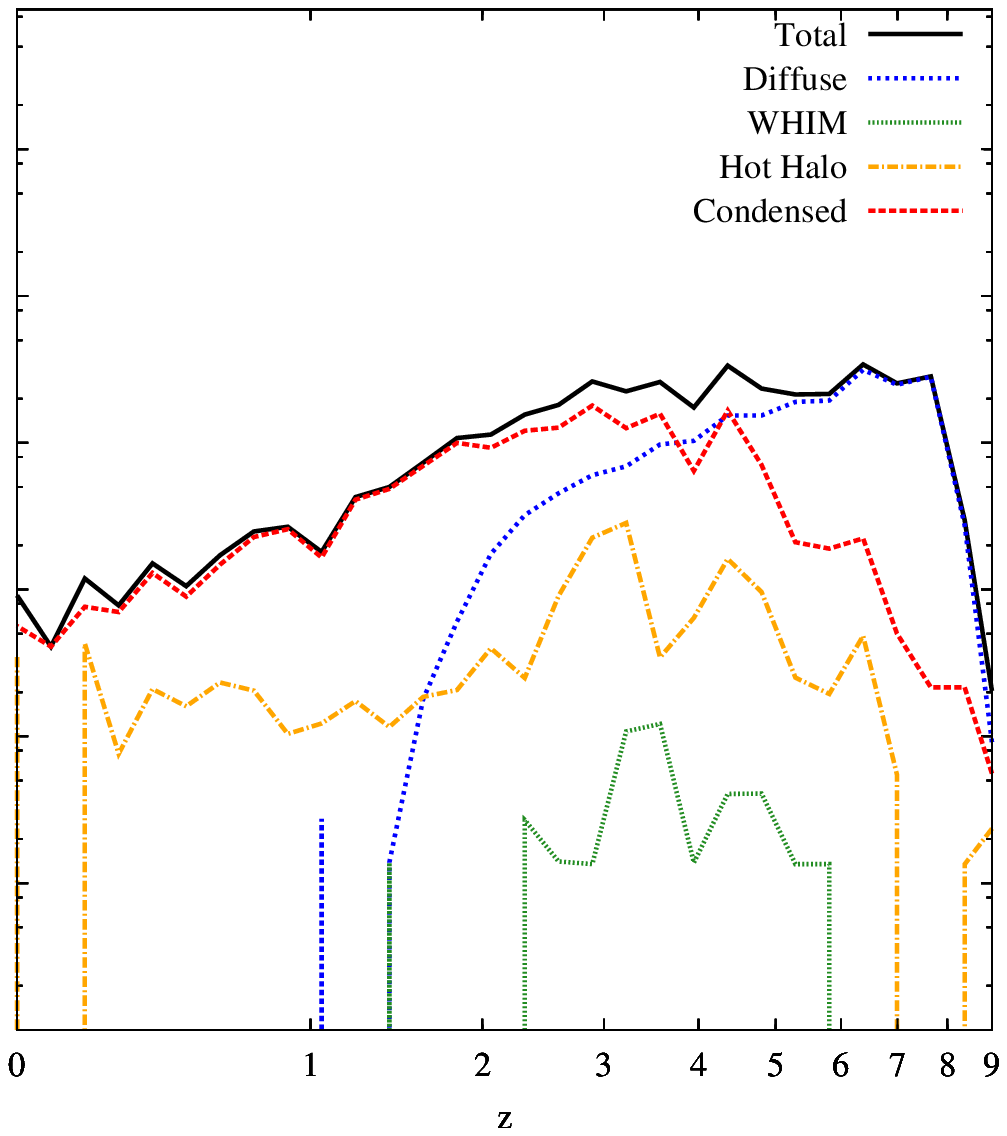}
		\caption{Unassigned}
		\label{fig:sfr_phase_e}
		\end{center}
	\end{subfigure}
	\end{array}$
	%\end{center}
\caption{SFR vs. Redshift using criterion 3.  Illustration of the star formation rate in the diffuse, WHIM, hot halo and condensed gas phases as a function of structure.}
\label{fig:sfr_phase}
\end{figure*}

\begin{description}[leftmargin=*]
\item {\bf Clusters :} The star formation rate peaks in the condensed phase at a redshift z = 2.6 - 3.6.  The star formation rate in the hot halo phase peaks at a similar redshift.  
At redshift z = 3, there is negligible star formation in the WHIM and diffuse gas phases (see Figs.\ref{fig:sfr_z3_b},~\ref{fig:sfr_z0_b} and~\ref{fig:sfr_phase_b}).  The fraction of the total stars forming in the diffuse, WHIM, hot halo and condensed cluster phases are 0.31\%, 0.03\%, 1.26\% and 12.0\%  respectively (see Fig.~\ref{fig:sfr_z3_b}).   
In this environment, at a redshift z = 0, 5.86\% and 13.9\% of the total stars form in the hot halo and condensed cluster gas phases respectively (see Fig.~\ref{fig:sfr_z0_b}).
\smallskip

\item {\bf Filaments :} Just as in the clusters, the star formation peaks at a redshift z $\sim$ 3 in the condensed phase (see Fig.~\ref{fig:sfr_phase_c}).  The diffuse phase dominates star formation at redshifts greater than z = 6.  The star formation rate in the hot halo phase and WHIM increase to a maximum at a redshift z = 4 and a redshift z = 3 respectively.  
From Fig.~\ref{fig:sfr_z3_c}, the strongest star formation is occurring in the regime where 10$^{4}$ $\leq$ T $\leq$ 10$^{5}$ and 6 $\leq$ $\rho_{\text{gas}} / \bar{\rho}$ $\leq$ 8.  This is above the cooling branch, due to the thermal steady state between UV-heating and radiative cooling, (e.g. Fig.~\ref{fig:gas_phase_z3_c} where T $\approx$ 10$^{4}$ and 2 $\leq$ $\rho_{\text{gas}} / \bar{\rho}$ $\leq$ 6).  This suggests that either the nearby gas is heated above this branch in regions surrounding star forming regions or the gas must be above this steady state branch to actively cool and form stars or that possibly, both these conditions hold. 

At a redshift z = 3.2, the fraction of the total stars forming in the diffuse, WHIM, hot halo and condensed cluster phases are 7.18\%, 0.31\%, 3.48\% and 71.9\% respectively (see Fig.~\ref{fig:sfr_z3_c}).  This demonstrates weakly trimodal star formation structure distribution, dominated by the condensed phase.  

At a redshift z = 0, the fraction of total stars forming in the diffuse, WHIM, hot halo and condensed filament phases (see Fig.~\ref{fig:sfr_z0_c}) are 0.16\%,  0.16\%, 24.2\% and 53.8\% respectively.  There is clearly a shift away from the trimodal distribution in the diffuse, hot halo and condensed phases to a bimodal star formation structure in the hot halo and condensed phases.  
\smallskip

\item {\bf Voids :} The star formation rate peaks in the diffuse phase at a redshift z = 7.7 (see Fig.~\ref{fig:sfr_phase_d}).  At a redshift z = 3.2, only 0.06\% of stars are being formed in voids, entirely in gas in the diffuse phase (see Fig.~\ref{fig:sfr_z3_d}).  After a redshift z = 1.6, there is no star formation in voids because they are being vacated of mass (see Figs.~\ref{fig:sfr_z0_d} and~\ref{fig:sfr_phase_d}).
\smallskip

\item {\bf Unassigned :} The star formation peaks at a redshift z = 6.4 in the diffuse phase (see Fig.~\ref{fig:sfr_phase_e}).  At a redshift z = 3.2 the fraction of total stars formed in the unassigned diffuse, WHIM, hot halo and condensed unassigned phases are 1.05\%, 0.02\%, 0.42\% and 1.91\% respectively (see Fig.~\ref{fig:sfr_z3_e}).  The trimodal distribution is similar to the filaments' star formation phase distribution.  An interesting difference between the unassigned and the  clusters / filaments, is that near (3 $\leq$ z $\leq$ 4) the maximum unassigned star formation rate of the condensed and diffuse phases are almost equivalent (see Fig.~\ref{fig:sfr_phase_e}).  
\end{description}

In Figs.~\ref{fig:sfr_z3} and~\ref{fig:sfr_z0}, the star formation is very clearly dominated by filaments at densities greater than 5 times the mean gas density and temperatures below 10$^5$K.  It is primarily occurring in the condensed gas phase.  The next most important regimes for star formation are the condensed cluster and the filament hot halo phases.

\begin{figure*}
	\psfrag{Gas Fraction}[tc][tc][0.7]{Gas fraction converted to stars (yr$^{-1}$)}
	\psfrag{z}[tc][tc][0.70]{z}	
	$\begin{array}{cc}
	\begin{subfigure}{.33 \textwidth}
		\begin{center}
		\includegraphics[width=2.7in]{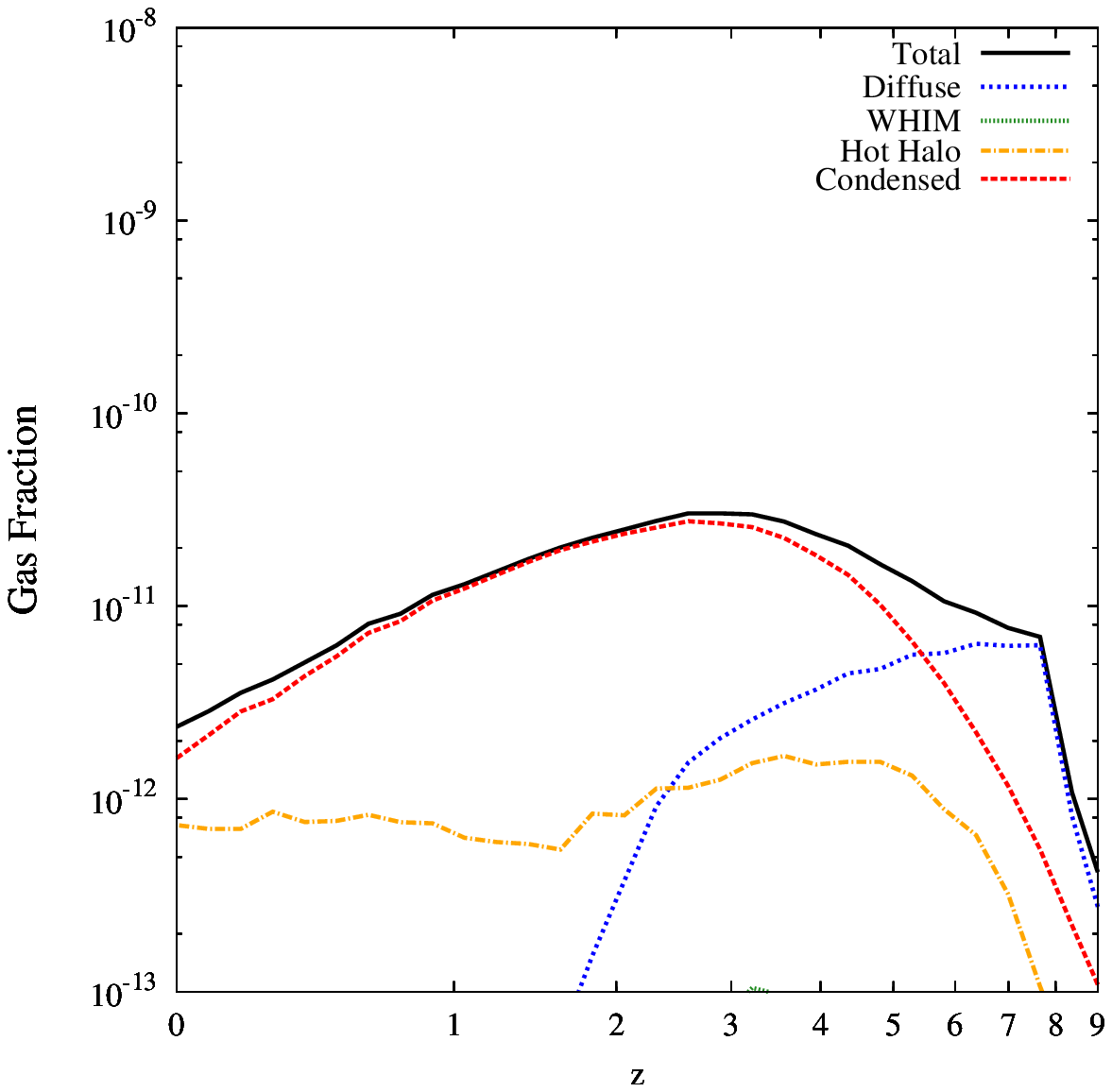}
		\caption{Total}
		\label{fig:combined_phase_a}
		\end{center}
	\end{subfigure}%
	\quad
	\begin{subfigure}{.33 \textwidth}
		\begin{center}
		\includegraphics[width=2.7in]{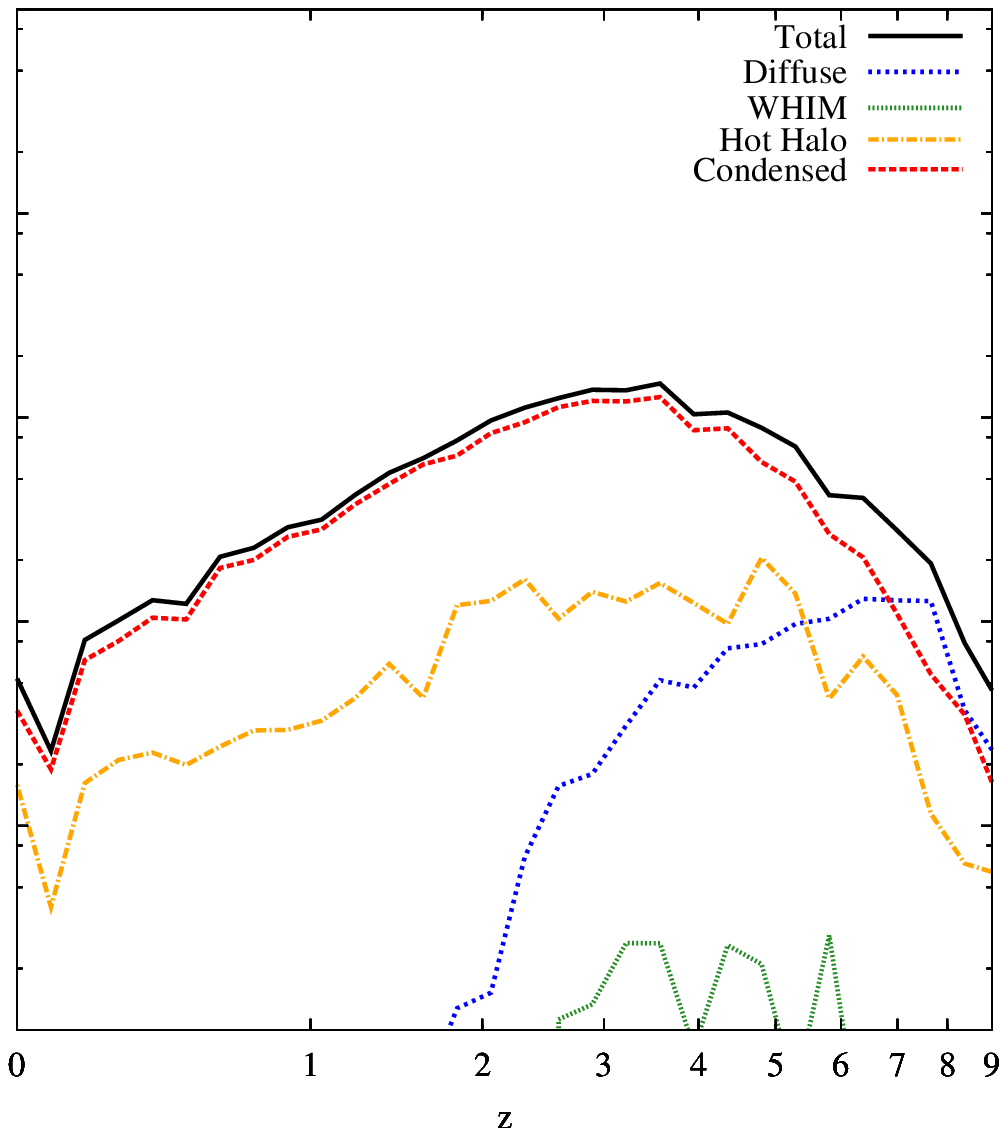}
		\caption{Clusters}
		\label{fig:combined_phase_b}
		\end{center}
	\end{subfigure}
	\quad
	\begin{subfigure}{.33 \textwidth}
		\begin{center}
		\includegraphics[width=2.8in, height=2.7in]{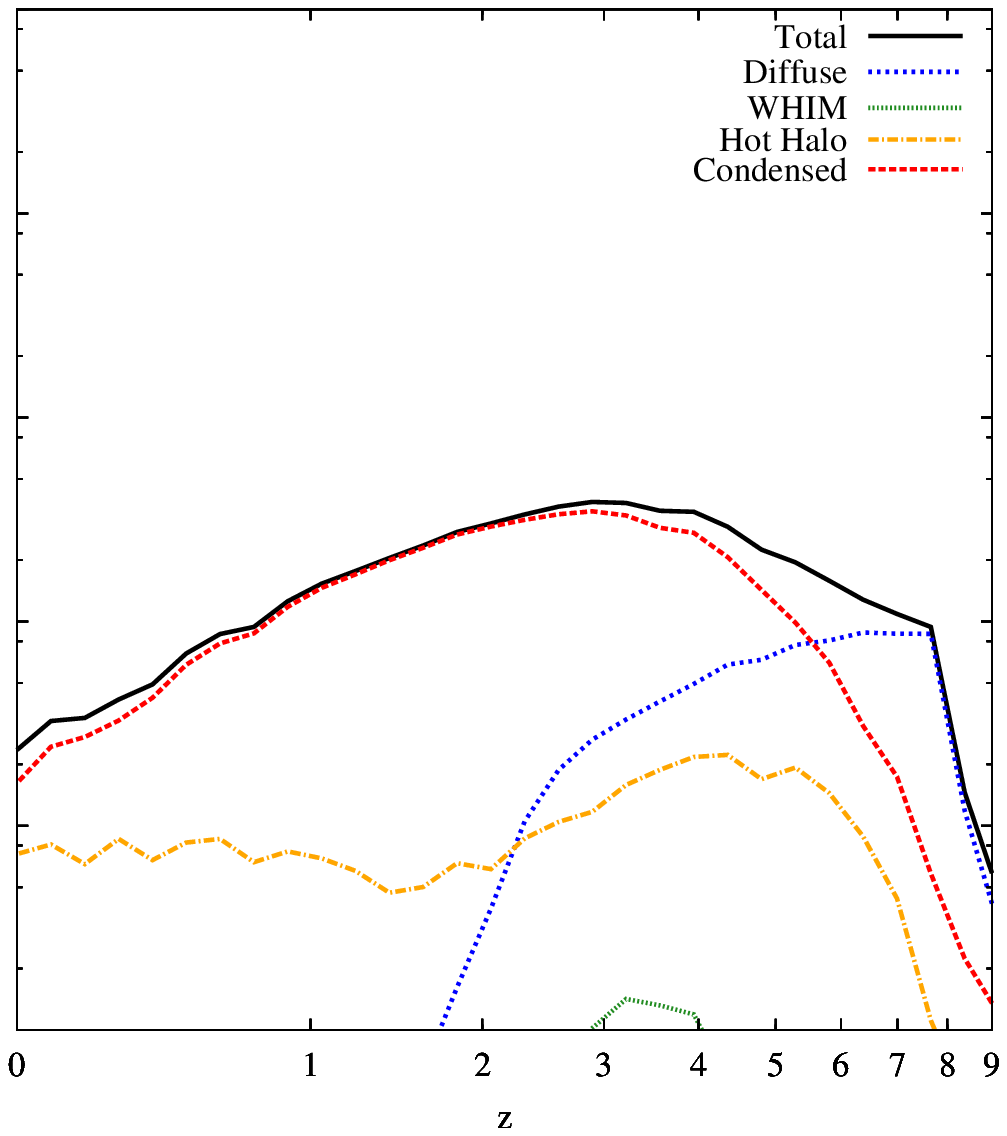}
		\caption{Filaments}
		\label{fig:combined_phase_c}
		\end{center}
	\end{subfigure}
	\\
	\begin{subfigure}{.33 \textwidth}
		\begin{center}
		\includegraphics[width=2.7in]{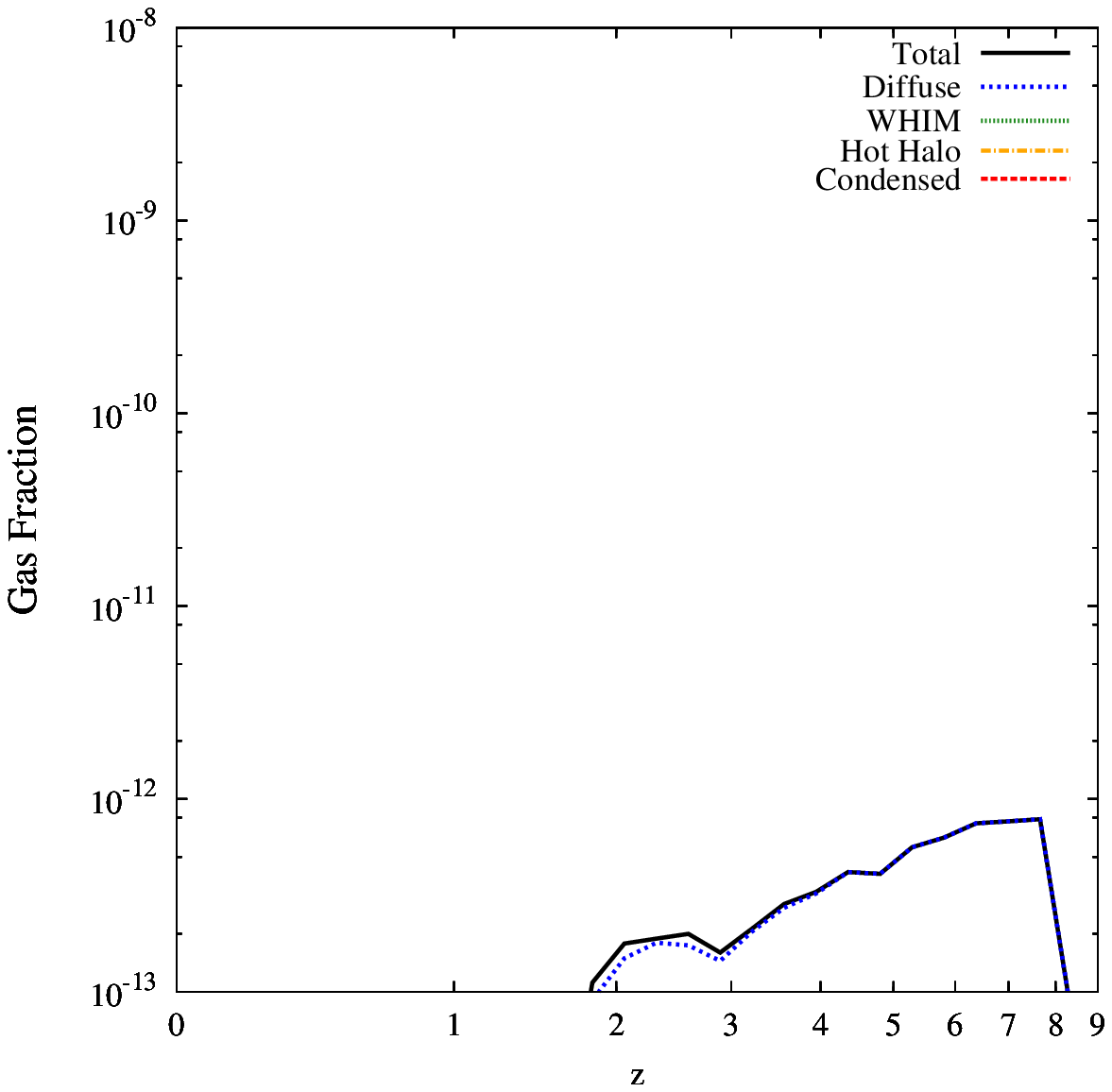}
		\caption{Voids}
		\label{fig:combined_phase_d}
		\end{center}
	\end{subfigure}%
	\quad
	\begin{subfigure}{.33 \textwidth}
		\begin{center}
		\includegraphics[width=2.8in, height=2.7in]{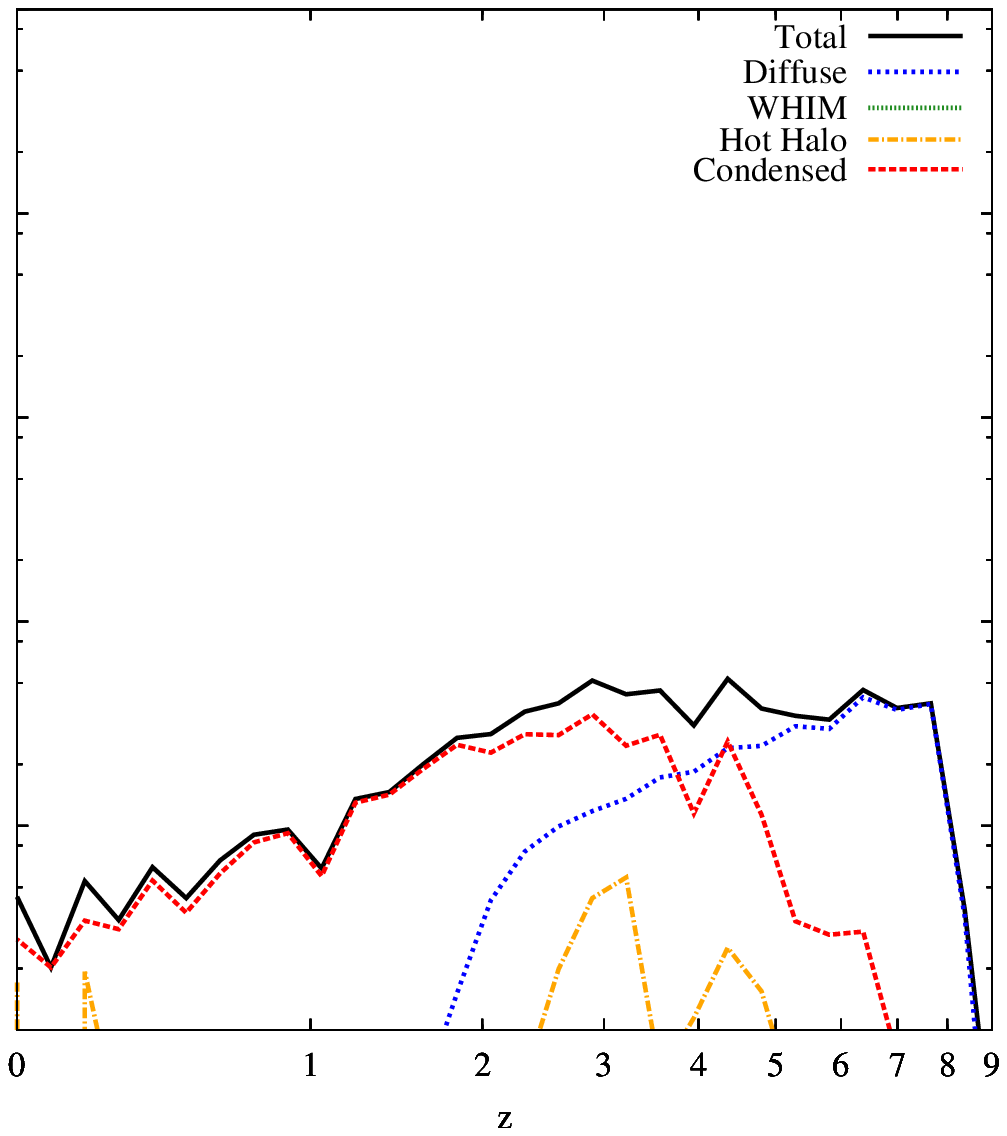}
		\caption{Unassigned}
		\label{fig:combined_phase_e}
		\end{center}
	\end{subfigure}
	\end{array}$
	%\end{center}
\caption{The fraction of structure gas converted by different gas phases using criterion 3.  This illustrates the fraction of the total gas in a structure type that is converted into stars per year in the diffuse, WHIM, hot halo and condensed gas phases as a function of structure.}
\label{fig:combined_phase}
\end{figure*}

We investigate the fraction of total gas belonging to a particular structure type that is converted into stars per year in Fig.~\ref{fig:combined_phase}.  Fig.~\ref{fig:combined_phase_b} shows that the clusters convert gas in the condensed phase into stars about 4 times more efficiently than the filaments do at a redshift z = 3.6 (see Fig.~\ref{fig:combined_phase_c}).  This is due to the increased gas densities in clusters (and metallicities) enabling the gas to rapidly cool and form stars.  In clusters, filaments and unassigned regions the condensed phase is where the most efficient conversion of gas into stars  occurs.  In clusters at redshifts z $<$ 5, the hot halo gas becomes more efficient than the diffuse gas at converting gas into stars.  While for filaments it takes until a redshift z = 2 for the hot halo to become more efficient than the diffuse phase.  In Fig.~\ref{fig:combined_phase}, it appears that the diffuse phase dominates the efficiency of converting gas into stars at very high redshifts.  However as structures form in the universe, the condensed and the hot halo gas begin to dominate.

\begin{figure*}
	\psfrag{meas}[tc][tc][0.7]{Structure Measure}
	\psfrag{Gas Fraction}[tc][tc][0.7]{Gas fraction converted to stars (yr$^{-1}$)}

%	\psfrag{Log(T)}[tc][tc][0.90]{Log$_{\text{10}}$(T)}	
	$\begin{array}{cc}
%	\begin{subfigure}{.33 \textwidth}
%		\begin{center}
%		\includegraphics[width=2.7in]{./Figures/salA/phase_vs_thresh/totl_combined_phase_z3.2.eps}
%		\caption{Total}
%		\label{fig:combined_phase_a}
%		\end{center}
%	\end{subfigure}%
%	\quad
	\begin{subfigure}{.5 \textwidth}
		\begin{center}
		\includegraphics[width=3.1in]{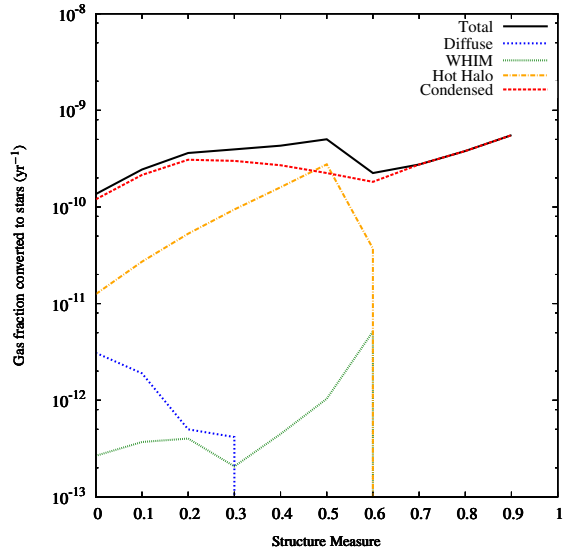}
		\caption{Clusters}
		\label{fig:phase_vs_thresh_z3.2_a}
		\end{center}
	\end{subfigure}
	\quad
	\begin{subfigure}{.5 \textwidth}
		\begin{center}
		\includegraphics[width=3.1in]{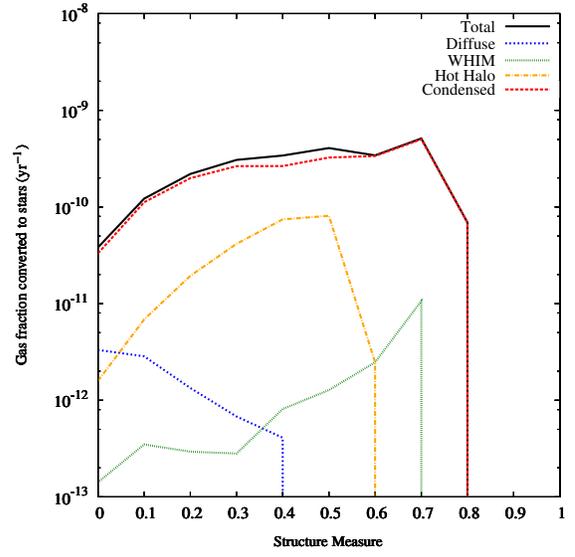}
		\caption{Filaments}
		\label{fig:phase_vs_thresh_z3.2_b}
		\end{center}
	\end{subfigure}
	\\
	\begin{subfigure}{.5 \textwidth}
		\begin{center}
		\includegraphics[width=3.1in]{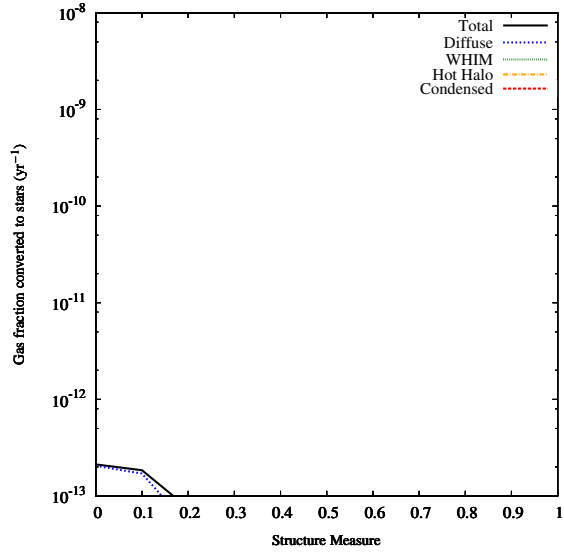}
		\caption{Voids}
		\label{fig:phase_vs_thresh_z3.2_c}
		\end{center}
	\end{subfigure}%
	\quad
	\begin{subfigure}{.5\textwidth}
		\begin{center}
		\includegraphics[width=3.1in]{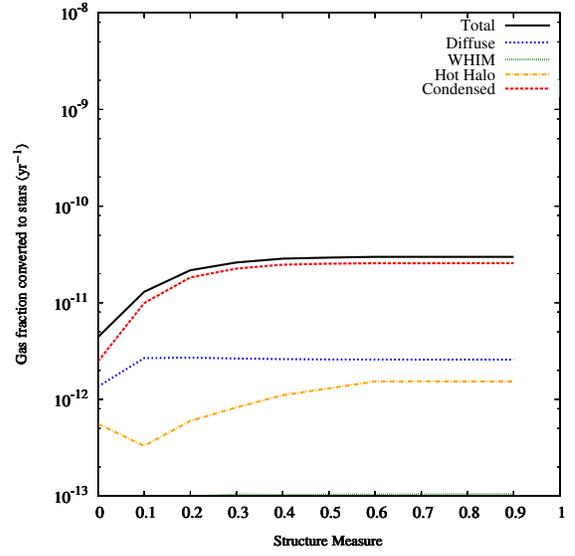}
		\caption{Unassigned}
		\label{fig:phase_vs_thresh_z3.2_d}
		\end{center}
	\end{subfigure}
	\end{array}$
	%\end{center}
\caption{Fraction of structure gas converted vs. structure measure at z = 3.2 without the density criterion.  The fraction of the total gas in a structure type converted into stars per year in the diffuse, WHIM, hot halo and condensed gas phases as a function of structure measure is illustrated.}
\label{fig:phase_vs_thresh_z3.2}
\end{figure*}

In Fig.~\ref{fig:phase_vs_thresh_z3.2} we investigate the efficiency at which gas is converted into stars per structure type and as a function of the structure measure threshold at z = 3.2.  By changing the threshold value, various levels of contrast can be probed within the structures.  The higher the structure measure, the higher the contrast.  Fig.~\ref{fig:phase_vs_thresh_z3.2_a} and~\ref{fig:phase_vs_thresh_z3.2_b} indicate that the star formation efficiency in cluster and filament hot halos peaks between the centers and edges of clusters / filaments.  

\begin{figure*}
	\psfrag{meas}[tc][tc][0.7]{Structure Measure}
%	\psfrag{Gas Fraction}[tc][tc][0.7]{Gas fraction converted to stars (yr$^{-1}$)}

%	\psfrag{Log(T)}[tc][tc][0.90]{Log$_{\text{10}}$(T)}	
	$\begin{array}{cc}
%	\begin{subfigure}{.33 \textwidth}
%		\begin{center}
%		\includegraphics[width=2.7in]{./Figures/salA/phase_vs_thresh/totl_combined_phase_z3.2.eps}
%		\caption{Total}
%		\label{fig:combined_phase_a}
%		\end{center}
%	\end{subfigure}%
%	\quad
	\begin{subfigure}{.5 \textwidth}
		\begin{center}
		\includegraphics[width=3.1in]{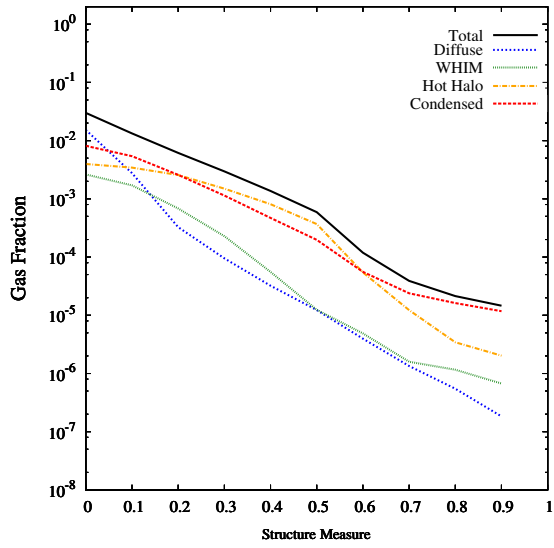}
		\caption{Clusters}
		\label{fig:gas_frac_vs_thresh_z3.2_a}
		\end{center}
	\end{subfigure}
	\quad
	\begin{subfigure}{.5 \textwidth}
		\begin{center}
		\includegraphics[width=3.1in]{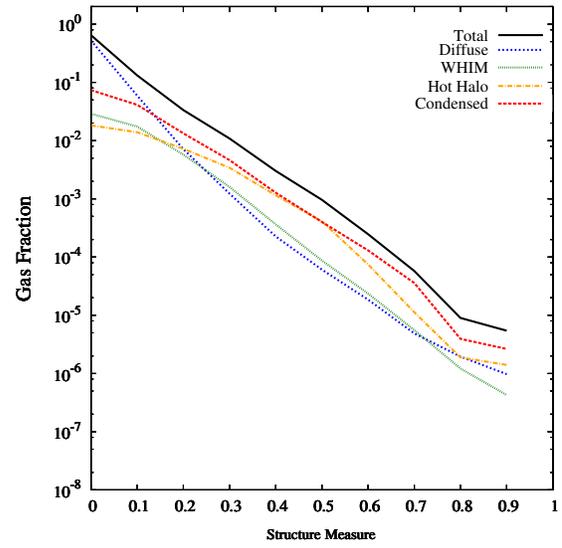}
		\caption{Filaments}
		\label{fig:gas_frac_vs_thresh_z3.2_b}
		\end{center}
	\end{subfigure}
	\\
	\begin{subfigure}{.5 \textwidth}
		\begin{center}
		\includegraphics[width=3.1in]{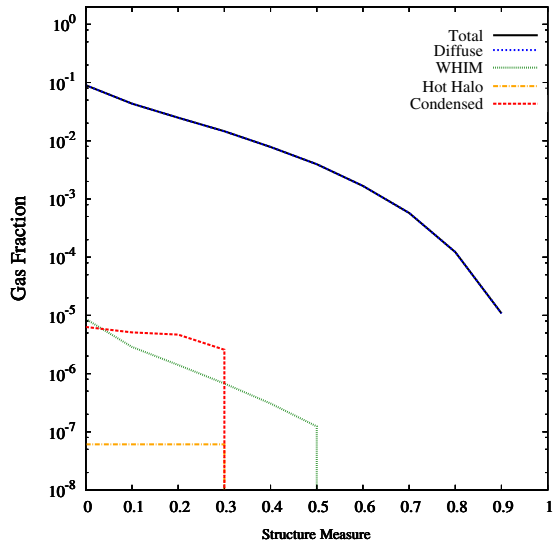}
		\caption{Voids}
		\label{fig:gas_frac_vs_thresh_z3.2_c}
		\end{center}
	\end{subfigure}%
	\quad
	\begin{subfigure}{.5\textwidth}
		\begin{center}
		\includegraphics[width=3.1in]{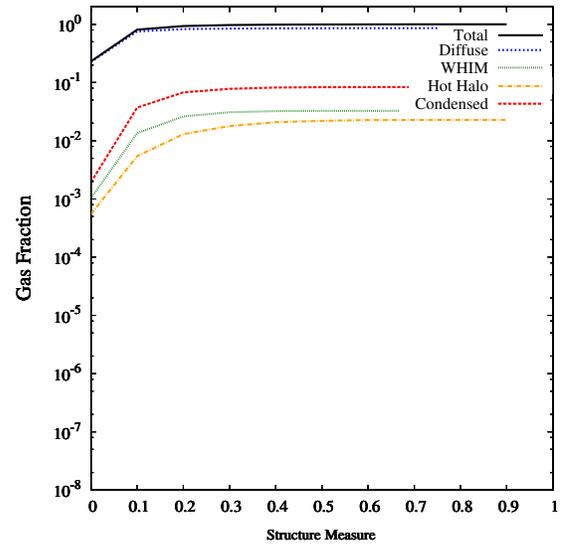}
		\caption{Unassigned}
		\label{fig:gas_frac_vs_thresh_z3.2_d}
		\end{center}
	\end{subfigure}
	\end{array}$
	%\end{center}
\caption{The gas fraction in a particular type of structure vs. the structure measure threshold at z = 3.2, without the density criterion.  The contributions to the diffuse, WHIM, hot halo and condensed gas phases are also                        shown.}
\label{fig:gas_frac_vs_thresh_z3.2}
\end{figure*}

Comparing  Fig.~\ref{fig:phase_vs_thresh_z3.2_a} with Fig.~\ref{fig:phase_vs_thresh_z3.2_b} it is evident that as the structure measure increases for filaments, which corresponds to probing closer to the spine of the filament, the star formation efficiency (e.g. 0.4 - 0.7) starts to match the efficiency in clusters. 
However Fig.~\ref{fig:gas_frac_vs_thresh_z3.2} reminds us that in that regime, the amount of gas in filaments is an order of magnitude higher than that found in clusters. So filaments truly emerge as star formation powerhouses.  In both clusters and filaments, the star formation efficiency in the WHIM rises steeply as the central regions are approached. Towards the outskirts of these structures, the diffuse gas takes over from the WHIM.

%__________________________________________________________________________
%__________________________________________________________________________
%__________________________________________________________________________
%__________________________________________________________________________
%__________________________________________________________________________
%__________________________________________________________________________
%__________________________________________________________________________
%__________________________________________________________________________
%__________________________________________________________________________
%__________________________________________________________________________
%__________________________________________________________________________
%__________________________________________________________________________
%__________________________________________________________________________
%__________________________________________________________________________
%__________________________________________________________________________
%__________________________________________________________________________
%__________________________________________________________________________
%__________________________________________________________________________
%__________________________________________________________________________
%__________________________________________________________________________
%__________________________________________________________________________
%__________________________________________________________________________
%__________________________________________________________________________
%__________________________________________________________________________
%__________________________________________________________________________
%__________________________________________________________________________
%__________________________________________________________________________
%__________________________________________________________________________
%__________________________________________________________________________
%__________________________________________________________________________
%__________________________________________________________________________
%__________________________________________________________________________
%__________________________________________________________________________

%===========================================================================
%================================ SUMMARY ==============================%
%===========================================================================

\section{Conclusions}
\label{sec:conclusions}
\balance
We have introduced a new approach to studying the environment in which galaxies form and evolve.
By coupling tailored simulations with a structure finding algorithm that self-consistently tracks gas in  clusters, filaments and voids, we have begun to examine the redshift evolution of the properties of the different environments of galaxies and their circumgalactic medium. We have have defined and studied poor clusters, filaments and voids focusing on 1) the temperature and density evolution 2) the phase population of the gas in these structures, and 3) the star formation rates and efficiencies in these structures. 

 We find that during the bulk of the cosmic evolution most of the stars and gas inhabit filamentary structures.  At a redshift z = 0 (see Fig.~\ref{fig:gas_phase_z0}), 79.1\% of the gas mass is found in filaments. The gas phase mass fractions are 43.1\%, 30.0\%, 24.7\% and 2.2\% for the diffuse, WHIM, hot halo and condensed phases, respectively. Although most of the WHIM is found in filaments, we caution against equating the filamentary environment with the WHIM since the filamentary gas is in fact multiphase, consisting of almost equal parts hot halo, WHIM and diffuse gas. The condensed gas in these filaments dominates the star forming regions in the universe through all epochs. 
 
 Our definition of structure allows us to probe the inner and outer regions of clusters and filaments. At high redshifts, the filaments are have low structure measure values, corresponding to a relatively low contrast with the background.  At redshift z = 0, the filamentary material has transitioned into higher contrast regions as we would expect from the growth of cosmic structure.

At redshift z = 0, 75.4\% of the star mass is found in the filamentary neighborhood.  The unassigned material does house significant star formation at higher redshifts when we use a structure threshold = 0.1. However, since this star formation is reassigned to filamentary regions when we relax this criterion, we suspect that this star formation is occurring in low contrast regions, or the outskirts, of the filaments.  The star mass in clusters and voids, is correspondingly less affected by the change in structure threshold, leading us to conclude that the unassigned material is indeed in the low contrast regions in filaments. There is some indication of star formation occurring in voids, but at low redshifts the gas vacates the voids and flow into filaments and clusters.  This leads to a peak of the total stellar mass in voids at a redshift z = 4, which then decreases at lower redshifts.

Our ability to map the temperature-density evolution of the filaments, clusters, and voids allows us to study the different phases of gas in these structures.
We have used this to track the star formation in different phases of the gas. When star formation peaks at a redshift z = 3.2, 8.6\% of the stars form in the diffuse phase, 0.36\% form in the WHIM phase, 5.2\% form in the hot halo phase, and 86\% form in the condensed phase (see Fig.~\ref{fig:sfr_z3_a}).  At that redshift 65\% of the gas is in filaments and it is broken down into 53\%, 2.9\%, 1.8\% and 7.4\% of the total gas is in the diffuse, WHIM, hot halo and condensed phases.  Over half of all star formation  therefore occurs in 7.4\% of the gas.  At a redshift z = 0, 0.16\% of all the stars form in the diffuse phase, 0.16\% form in the WHIM phase, 30.7\% form in the hot halo phase and 68.9\% form in the condensed phase (see Fig.~\ref{fig:sfr_z0_a}).  Figs.~\ref{fig:sfr_z3_a} and~\ref{fig:sfr_z0_a} demonstrate a transition from a trimodal gas phase (diffuse, hot halo and condensed) distribution for star formation at high redshift to a bimodal gas phase star formation distribution in the present era (hot halo and condensed). 
 The transition from the hot halo and diffuse phase containing similar amounts of star formation at z = 3 to the hot halo having over 150 times the star formation rate at z = 0 cannot be completely attributed the decrease in the diffuse phase. At z = 0, the diffuse phase still has more gas than the hot halo phase (see Fig.~\ref{fig:gas_phase_a}).  This indicates that hot halos are simply more efficient at converting gas into stars (see Fig.~\ref{fig:combined_phase_a}).

We also notice in Fig.~\ref{fig:combined_phase} that the poor clusters and groups are overall the most efficient at converting gas into stars.  The higher densities (and higher metallicities) enable gas to cool and form stars.  The condensed phase plays the most significant role in converting structure gas into stars followed by the hot halo at low redshift and the diffuse phase at high redshifts.  However, in the inner regions of filaments that correspond to a high filament measure, star formation efficiency is comparable to that in poor clusters.

We have shown that filaments play a significant role in the history of star formation.  The majority of star formation occurs within cold, condensed gas in filaments at intermediate redshifts (z $\sim$ 3).  We also show that much of the star formation above a redshift z = 3 occurs in low contrast regions of filaments, but as the density contrast increases at lower redshift the star formation switches to high contrast regions, or the inner parts of filaments.  Since filaments bridge the void regions to the cluster regions, it suggests that the majority of star formation occurs in galaxies at intermediate redshifts in filamentary regions prior to the accretion onto clusters.

%========================== ACKNOWLEDGEMENTS ==========================%
\section{Acknowledgements}
This research was supported in part by the Notre Dame
Center for Research Computing. Work also supported in part by DOE grant
DE-FG02-95-ER40934.

%============================ BIBLIOGRAPHY ============================%
%see ~/Library/texmf/bibtex/bib/references.bib
%\bibliography{/Users/ali/Library/texmf/bibtex/bib/references}
\bibliography{references}

\begin{thebibliography}{148}
\expandafter\ifx\csname natexlab\endcsname\relax\def\natexlab#1{#1}\fi

\bibitem[{{Abell}(1965)}]{abell65}
{Abell} G.~O., 1965, \araa, 3, 1

\bibitem[{{Akamatsu} {et~al}\mbox{.}(2011){Akamatsu}, {Hoshino}, {Ishisaki},
  {Ohashi}, {Sato}, {Takei}, \& {Ota}}]{akamatsu11}
{Akamatsu} H., {Hoshino} A., {Ishisaki} Y., {Ohashi} T., {Sato} K., {Takei} Y.,
  {Ota} N., 2011, \pasj, 63, 1019

\bibitem[{{Albrecht} {et~al}\mbox{.}(2009){Albrecht}, {Amendola}, {Bernstein},
  {Clowe}, {Eisenstein}, {Guzzo}, {Hirata}, {Huterer}, {Kirshner}, {Kolb}, \&
  {Nichol}}]{albrecht09}
{Albrecht} A. {et~al.}, 2009, ArXiv e-prints

\bibitem[{{Arag{\'o}n-Calvo} {et~al}\mbox{.}(2007){Arag{\'o}n-Calvo}, {van de
  Weygaert}, {Jones}, \& {van der Hulst}}]{aragon07b}
{Arag{\'o}n-Calvo} M.~A., {van de Weygaert} R., {Jones} B.~J.~T., {van der
  Hulst} J.~M., 2007, \apjl, 655, L5

\bibitem[{{Aragon-Salamanca}, {Baugh} \& {Kauffmann}(1998){Aragon-Salamanca},
  {Baugh}, \& {Kauffmann}}]{aragon98}
{Aragon-Salamanca} A., {Baugh} C.~M., {Kauffmann} G., 1998, \mnras, 297, 427

\bibitem[{{Bacon}, {Refregier} \& {Ellis}(2000){Bacon}, {Refregier}, \&
  {Ellis}}]{bacon00}
{Bacon} D.~J., {Refregier} A.~R., {Ellis} R.~S., 2000, \mnras, 318, 625

\bibitem[{{Baldry} \& {Glazebrook}(2003)}]{baldry03}
{Baldry} I.~K., {Glazebrook} K., 2003, \apj, 593, 258

\bibitem[{{Balogh} {et~al}\mbox{.}(1997){Balogh}, {Morris}, {Yee}, {Carlberg},
  \& {Ellingson}}]{balogh97}
{Balogh} M.~L., {Morris} S.~L., {Yee} H.~K.~C., {Carlberg} R.~G., {Ellingson}
  E., 1997, \apjl, 488, L75

\bibitem[{{Balogh}, {Navarro} \& {Morris}(2000){Balogh}, {Navarro}, \&
  {Morris}}]{balogh00a}
{Balogh} M.~L., {Navarro} J.~F., {Morris} S.~L., 2000, \apj, 540, 113

\bibitem[{{Balogh} {et~al}\mbox{.}(2001){Balogh}, {Pearce}, {Bower}, \&
  {Kay}}]{balogh01}
{Balogh} M.~L., {Pearce} F.~R., {Bower} R.~G., {Kay} S.~T., 2001, \mnras, 326,
  1228

\bibitem[{{Bertschinger} \& {Gelb}(1991)}]{bertschinger91}
{Bertschinger} E., {Gelb} J.~M., 1991, Computers in Physics, 5, 164

\bibitem[{{Blitz}(1993)}]{blitz93}
{Blitz} L., 1993, in Protostars and Planets III, {Levy} E.~H., {Lunine} J.~I.,
  eds., pp. 125--161

\bibitem[{{Bond}, {Kofman} \& {Pogosyan}(1996){Bond}, {Kofman}, \&
  {Pogosyan}}]{bond96}
{Bond} J.~R., {Kofman} L., {Pogosyan} D., 1996, \nat, 380, 603

\bibitem[{{Booth} \& {Schaye}(2009)}]{booth09}
{Booth} C.~M., {Schaye} J., 2009, \mnras, 398, 53

\bibitem[{{Butcher} \& {Oemler}(1978)}]{butcher78}
{Butcher} H., {Oemler}, Jr. A., 1978, \apj, 219, 18

\bibitem[{{Ceccarelli} {et~al}\mbox{.}(2006){Ceccarelli}, {Padilla}, {Valotto},
  \& {Lambas}}]{ceccarelli06}
{Ceccarelli} L., {Padilla} N.~D., {Valotto} C., {Lambas} D.~G., 2006, \mnras,
  373, 1440

\bibitem[{{Cen} \& {Fang}(2006)}]{cen06b}
{Cen} R., {Fang} T., 2006, \apj, 650, 573

\bibitem[{{Cen} \& {Ostriker}(1999)}]{cen99}
{Cen} R., {Ostriker} J.~P., 1999, \apj, 514, 1

\bibitem[{{Cen} \& {Ostriker}(2006)}]{cen06a}
{Cen} R., {Ostriker} J.~P., 2006, \apj, 650, 560

\bibitem[{{Chabrier}(2003)}]{chabrier03a}
{Chabrier} G., 2003, \pasp, 115, 763

\bibitem[{{Chevalier}(1974)}]{chevalier74}
{Chevalier} R.~A., 1974, \apj, 188, 501

\bibitem[{{Christensen} {et~al}\mbox{.}(2010){Christensen}, {Quinn}, {Stinson},
  {Bellovary}, \& {Wadsley}}]{christensen10}
{Christensen} C.~R., {Quinn} T., {Stinson} G., {Bellovary} J., {Wadsley} J.,
  2010, \apj, 717, 121

\bibitem[{{Colless} {et~al}\mbox{.}(2001){Colless}, {Dalton}, {Maddox},
  {Sutherland}, {Norberg}, {Cole}, {Bland-Hawthorn}, {Bridges}, {Cannon},
  {Collins}, {Couch}, {Cross}, {Deeley}, {De Propris}, {Driver}, {Efstathiou},
  {Ellis}, {Frenk}, {Glazebrook}, {Jackson}, {Lahav}, {Lewis}, {Lumsden},
  {Madgwick}, {Peacock}, {Peterson}, {Price}, {Seaborne}, \&
  {Taylor}}]{colless01}
{Colless} M. {et~al.}, 2001, \mnras, 328, 1039

\bibitem[{{Dahlen} {et~al}\mbox{.}(2004){Dahlen}, {Strolger}, {Riess},
  {Mobasher}, {Chary}, {Conselice}, {Ferguson}, {Fruchter}, {Giavalisco},
  {Livio}, {Madau}, {Panagia}, \& {Tonry}}]{dahlen04}
{Dahlen} T. {et~al.}, 2004, \apj, 613, 189

\bibitem[{{Dalla Vecchia} \& {Schaye}(2008)}]{dalla_vecchia08}
{Dalla Vecchia} C., {Schaye} J., 2008, \mnras, 387, 1431

\bibitem[{{Danforth} \& {Shull}(2008)}]{danforth08}
{Danforth} C.~W., {Shull} J.~M., 2008, \apj, 679, 194

\bibitem[{{Dav{\'e}} {et~al}\mbox{.}(2001){Dav{\'e}}, {Cen}, {Ostriker},
  {Bryan}, {Hernquist}, {Katz}, {Weinberg}, {Norman}, \& {O'Shea}}]{dave01}
{Dav{\'e}} R. {et~al.}, 2001, \apj, 552, 473

\bibitem[{{Dav{\'e}} {et~al}\mbox{.}(2010){Dav{\'e}}, {Oppenheimer}, {Katz},
  {Kollmeier}, \& {Weinberg}}]{dave10}
{Dav{\'e}} R., {Oppenheimer} B.~D., {Katz} N., {Kollmeier} J.~A., {Weinberg}
  D.~H., 2010, \mnras, 408, 2051

\bibitem[{{David}, {Forman} \& {Jones}(1990){David}, {Forman}, \&
  {Jones}}]{david90}
{David} L.~P., {Forman} W., {Jones} C., 1990, \apj, 359, 29

\bibitem[{{Davis} {et~al}\mbox{.}(1985){Davis}, {Efstathiou}, {Frenk}, \&
  {White}}]{davis85}
{Davis} M., {Efstathiou} G., {Frenk} C.~S., {White} S.~D.~M., 1985, \apj, 292,
  371

\bibitem[{{de Lapparent}, {Geller} \& {Huchra}(1986){de Lapparent}, {Geller},
  \& {Huchra}}]{delapparent86}
{de Lapparent} V., {Geller} M.~J., {Huchra} J.~P., 1986, \apjl, 302, L1

\bibitem[{{Descoteaux}, {Collins} \& {Siddiqi}(2004){Descoteaux}, {Collins}, \&
  {Siddiqi}}]{descoteaux04}
{Descoteaux} M., {Collins} L., {Siddiqi} K., 2004, in Lecture Notes in Computer
  Science, Vol. 3117, Computer Vision and Mathematical Methods in Medical and
  Biomedical Image Analysis, pp. 169--180

\bibitem[{{Dietrich} {et~al}\mbox{.}(2005){Dietrich}, {Schneider}, {Clowe},
  {Romano-D{\'{\i}}az}, \& {Kerp}}]{dietrich05}
{Dietrich} J.~P., {Schneider} P., {Clowe} D., {Romano-D{\'{\i}}az} E., {Kerp}
  J., 2005, \aap, 440, 453

\bibitem[{{Dressler}(1980)}]{dressler80}
{Dressler} A., 1980, \apj, 236, 351

\bibitem[{{Dressler}, {Thompson} \& {Shectman}(1985){Dressler}, {Thompson}, \&
  {Shectman}}]{dressler85}
{Dressler} A., {Thompson} I.~B., {Shectman} S.~A., 1985, \apj, 288, 481

\bibitem[{{Einasto} {et~al}\mbox{.}(2011){Einasto}, {Liivam{\"a}gi}, {Tempel},
  {Saar}, {Tago}, {Einasto}, {Enkvist}, {Einasto}, {Mart{\'{\i}}nez},
  {Hein{\"a}m{\"a}ki}, \& {Nurmi}}]{einasto11}
{Einasto} M. {et~al.}, 2011, \apj, 736, 51

\bibitem[{{Eldridge} \& {Tout}(2004)}]{eldridge04}
{Eldridge} J.~J., {Tout} C.~A., 2004, \mnras, 353, 87

\bibitem[{{Ellis}, {Maartens} \& {MacCallum}(2012){Ellis}, {Maartens}, \&
  {MacCallum}}]{ellis12}
{Ellis} G., {Maartens} R., {MacCallum} M., 2012, Relativistic Cosmology.
  Cambridge University Press, New York, NY

\bibitem[{{Fang} {et~al}\mbox{.}(2010){Fang}, {Buote}, {Humphrey}, {Canizares},
  {Zappacosta}, {Maiolino}, {Tagliaferri}, \& {Gastaldello}}]{fang10}
{Fang} T., {Buote} D.~A., {Humphrey} P.~J., {Canizares} C.~R., {Zappacosta} L.,
  {Maiolino} R., {Tagliaferri} G., {Gastaldello} F., 2010, \apj, 714, 1715

\bibitem[{{Farouki} \& {Shapiro}(1981)}]{farouki81}
{Farouki} R., {Shapiro} S.~L., 1981, \apj, 243, 32

\bibitem[{{Ferland} {et~al}\mbox{.}(1998){Ferland}, {Korista}, {Verner},
  {Ferguson}, {Kingdon}, \& {Verner}}]{ferland98}
{Ferland} G.~J., {Korista} K.~T., {Verner} D.~A., {Ferguson} J.~W., {Kingdon}
  J.~B., {Verner} E.~M., 1998, \pasp, 110, 761

\bibitem[{{Frangi} {et~al}\mbox{.}(1998){Frangi}, {Niessen}, {Vincken}, \&
  {Viergever}}]{frangi98}
{Frangi} A.~F., {Niessen} W.~J., {Vincken} K.~L., {Viergever} M.~A., 1998, in
  Lecture Notes in Computer Science, Vol. 1496, Medical Image Computing and
  Computer-Assisted Intervention - MICCAI'98, pp. 130--137

\bibitem[{{Fukugita}(2004)}]{fukugita04a}
{Fukugita} M., 2004, in IAU Symposium, Vol. 220, Dark Matter in Galaxies,
  {Ryder} S., {Pisano} D., {Walker} M., {Freeman} K., eds., p. 227

\bibitem[{{Fukugita}, {Hogan} \& {Peebles}(1998){Fukugita}, {Hogan}, \&
  {Peebles}}]{fukugita98}
{Fukugita} M., {Hogan} C.~J., {Peebles} P.~J.~E., 1998, \apj, 503, 518

\bibitem[{{Fukugita} \& {Peebles}(2004)}]{fukugita04b}
{Fukugita} M., {Peebles} P.~J.~E., 2004, \apj, 616, 643

\bibitem[{{Fukui} {et~al}\mbox{.}(1999){Fukui}, {Mizuno}, {Yamaguchi},
  {Mizuno}, {Onishi}, {Ogawa}, {Yonekura}, {Kawamura}, {Tachihara}, {Xiao},
  {Yamaguchi}, {Hara}, {Hayakawa}, {Kato}, {Abe}, {Saito}, {Mano}, {Matsunaga},
  {Mine}, {Moriguchi}, {Aoyama}, {Asayama}, {Yoshikawa}, \& {Rubio}}]{fukui99}
{Fukui} Y. {et~al.}, 1999, \pasj, 51, 745

\bibitem[{{Galassi} {et~al}\mbox{.}(2011){Galassi}, {Davies}, {Theiler},
  {Gough}, {Gerard}, {Alken}, {Booth}, \& {Rossi}}]{galassi11}
{Galassi} M., {Davies} J., {Theiler} J., {Gough} B., {Gerard} J., {Alken} P.,
  {Booth} M., {Rossi} F., 2011, GNU Scientific Library Reference Manual.
  Network Theory Ltd.

\bibitem[{{Geller} \& {Huchra}(1989)}]{geller89}
{Geller} M.~J., {Huchra} J.~P., 1989, Science, 246, 897

\bibitem[{{Gerritsen}(1997)}]{gerritsen97}
{Gerritsen} J.~P.~E., 1997, PhD thesis, , Groningen University, the
  Netherlands, (1997)

\bibitem[{{Gott} {et~al}\mbox{.}(2005){Gott}, {Juri{\'c}}, {Schlegel}, {Hoyle},
  {Vogeley}, {Tegmark}, {Bahcall}, \& {Brinkmann}}]{gott05}
{Gott}, III J.~R., {Juri{\'c}} M., {Schlegel} D., {Hoyle} F., {Vogeley} M.,
  {Tegmark} M., {Bahcall} N., {Brinkmann} J., 2005, \apj, 624, 463

\bibitem[{{Graur} {et~al}\mbox{.}(2014){Graur}, {Rodney}, {Maoz}, {Riess},
  {Jha}, {Postman}, {Dahlen}, {Holoien}, {McCully}, {Patel}, {Strolger},
  {Ben{\'{\i}}tez}, {Coe}, {Jouvel}, {Medezinski}, {Molino}, {Nonino},
  {Bradley}, {Koekemoer}, {Balestra}, {Cenko}, {Clubb}, {Dickinson},
  {Filippenko}, {Frederiksen}, {Garnavich}, {Hjorth}, {Jones}, {Leibundgut},
  {Matheson}, {Mobasher}, {Rosati}, {Silverman}, {U}, {Jedruszczuk}, {Li},
  {Lin}, {Mirmelstein}, {Neustadt}, {Ovadia}, \& {Rogers}}]{graur14}
{Graur} O. {et~al.}, 2014, \apj, 783, 28

\bibitem[{{Greggio}(2005)}]{greggio05}
{Greggio} L., 2005, \aap, 441, 1055

\bibitem[{{Greggio} \& {Renzini}(1983)}]{greggio83}
{Greggio} L., {Renzini} A., 1983, \aap, 118, 217

\bibitem[{{Grogin} \& {Geller}(1999)}]{grogin99}
{Grogin} N.~A., {Geller} M.~J., 1999, \aj, 118, 2561

\bibitem[{{Grogin} \& {Geller}(2000)}]{grogin00}
{Grogin} N.~A., {Geller} M.~J., 2000, \aj, 119, 32

\bibitem[{{Gunn} \& {Gott}(1972)}]{gunn72}
{Gunn} J.~E., {Gott}, III J.~R., 1972, \apj, 176, 1

\bibitem[{{Gupta} {et~al}\mbox{.}(2012){Gupta}, {Mathur}, {Krongold},
  {Nicastro}, \& {Galeazzi}}]{gupta12}
{Gupta} A., {Mathur} S., {Krongold} Y., {Nicastro} F., {Galeazzi} M., 2012,
  \apjl, 756, L8

\bibitem[{{Gursky} {et~al}\mbox{.}(1971){Gursky}, {Kellogg}, {Murray}, {Leong},
  {Tananbaum}, \& {Giacconi}}]{gursky71}
{Gursky} H., {Kellogg} E., {Murray} S., {Leong} C., {Tananbaum} H., {Giacconi}
  R., 1971, \apjl, 167, L81

\bibitem[{{Haardt} \& {Madau}(2001)}]{haardt01}
{Haardt} F., {Madau} P., 2001, in Clusters of Galaxies and the High Redshift
  Universe Observed in X-rays, {Neumann} D.~M., {Tran} J.~T.~V., eds.

\bibitem[{{Hahn} {et~al}\mbox{.}(2007){Hahn}, {Carollo}, {Porciani}, \&
  {Dekel}}]{hahn07}
{Hahn} O., {Carollo} C.~M., {Porciani} C., {Dekel} A., 2007, \mnras, 381, 41

\bibitem[{{Heymans} {et~al}\mbox{.}(2008){Heymans}, {Gray}, {Peng}, {van
  Waerbeke}, {Bell}, {Wolf}, {Bacon}, {Balogh}, {Barazza}, {Barden},
  {B{\"o}hm}, {Caldwell}, {H{\"a}u{\ss}ler}, {Jahnke}, {Jogee}, {van Kampen},
  {Lane}, {McIntosh}, {Meisenheimer}, {Mellier}, {S{\'a}nchez}, {Taylor},
  {Wisotzki}, \& {Zheng}}]{heymans08}
{Heymans} C. {et~al.}, 2008, \mnras, 385, 1431

\bibitem[{{Hopkins} \& {Beacom}(2006)}]{hopkins06}
{Hopkins} A.~M., {Beacom} J.~F., 2006, \apj, 651, 142

\bibitem[{{Hoyle} \& {Vogeley}(2004)}]{hoyle04}
{Hoyle} F., {Vogeley} M.~S., 2004, \apj, 607, 751

\bibitem[{{Ivezic} {et~al}\mbox{.}(2008){Ivezic}, {Tyson}, {Abel}, {Acosta},
  {Allsman}, {AlSayyad}, {Anderson}, {Andrew}, {Angel}, {Angeli}, {Ansari},
  {Antilogus}, {Arndt}, {Astier}, {Aubourg}, {Axelrod}, {Bard}, {Barr},
  {Barrau}, {Bartlett}, {Bauman}, {Beaumont}, {Becker}, {Becla}, {Beldica},
  {Bellavia}, {Blanc}, {Blandford}, {Bloom}, {Bogart}, {Borne}, {Bosch},
  {Boutigny}, {Brandt}, {Brown}, {Bullock}, {Burchat}, {Burke}, {Cagnoli},
  {Calabrese}, {Chandrasekharan}, {Chesley}, {Cheu}, {Chiang}, {Claver},
  {Connolly}, {Cook}, {Cooray}, {Covey}, {Cribbs}, {Cui}, {Cutri}, {Daubard},
  {Daues}, {Delgado}, {Digel}, {Doherty}, {Dubois}, {Dubois-Felsmann},
  {Durech}, {Eracleous}, {Ferguson}, {Frank}, {Freemon}, {Gangler}, {Gawiser},
  {Geary}, {Gee}, {Geha}, {Gibson}, {Gilmore}, {Glanzman}, {Goodenow},
  {Gressler}, {Gris}, {Guyonnet}, {Hascall}, {Haupt}, {Hernandez}, {Hogan},
  {Huang}, {Huffer}, {Innes}, {Jacoby}, {Jain}, {Jee}, {Jernigan},
  {Jevremovic}, {Johns}, {Jones}, {Juramy-Gilles}, {Juric}, {Kahn}, {Kalirai},
  {Kallivayalil}, {Kalmbach}, {Kantor}, {Kasliwal}, {Kessler}, {Kirkby},
  {Knox}, {Kotov}, {Krabbendam}, {Krughoff}, {Kubanek}, {Kuczewski},
  {Kulkarni}, {Lambert}, {Le Guillou}, {Levine}, {Liang}, {Lim}, {Lintott},
  {Lupton}, {Mahabal}, {Marshall}, {Marshall}, {May}, {McKercher}, {Migliore},
  {Miller}, {Mills}, {Monet}, {Moniez}, {Neill}, {Nief}, {Nomerotski},
  {Nordby}, {O'Connor}, {Oliver}, {Olivier}, {Olsen}, {Ortiz}, {Owen}, {Pain},
  {Peterson}, {Petry}, {Pierfederici}, {Pietrowicz}, {Pike}, {Pinto}, {Plante},
  {Plate}, {Price}, {Prouza}, {Radeka}, {Rajagopal}, {Rasmussen}, {Regnault},
  {Ridgway}, {Ritz}, {Rosing}, {Roucelle}, {Rumore}, {Russo}, {Saha},
  {Sassolas}, {Schalk}, {Schindler}, {Schneider}, {Schumacher}, {Sebag},
  {Sembroski}, {Seppala}, {Shipsey}, {Silvestri}, {Smith}, {Smith}, {Strauss},
  {Stubbs}, {Sweeney}, {Szalay}, {Takacs}, {Thaler}, {Van Berg}, {Vanden Berk},
  {Vetter}, {Virieux}, {Xin}, {Walkowicz}, {Walter}, {Wang}, {Warner},
  {Willman}, {Wittman}, {Wolff}, {Wood-Vasey}, {Yoachim}, {Zhan}, \& {for the
  LSST Collaboration}}]{ivezic08}
{Ivezic} Z. {et~al.}, 2008, ArXiv e-prints

\bibitem[{{J{\~o}eveer}, {Einasto} \& {Tago}(1978){J{\~o}eveer}, {Einasto}, \&
  {Tago}}]{joeveer78}
{J{\~o}eveer} M., {Einasto} J., {Tago} E., 1978, \mnras, 185, 357

\bibitem[{{Jauzac} {et~al}\mbox{.}(2012){Jauzac}, {Jullo}, {Kneib}, {Ebeling},
  {Leauthaud}, {Ma}, {Limousin}, {Massey}, \& {Richard}}]{jauzac12}
{Jauzac} M. {et~al.}, 2012, \mnras, 426, 3369

\bibitem[{{Jones}, {van de Weygaert} \& {Arag{\'o}n-Calvo}(2010){Jones}, {van
  de Weygaert}, \& {Arag{\'o}n-Calvo}}]{jones10}
{Jones} B.~J.~T., {van de Weygaert} R., {Arag{\'o}n-Calvo} M.~A., 2010, \mnras,
  408, 897

\bibitem[{{Kaiser}, {Wilson} \& {Luppino}(2000){Kaiser}, {Wilson}, \&
  {Luppino}}]{kaiser00}
{Kaiser} N., {Wilson} G., {Luppino} G.~A., 2000, ArXiv Astrophysics e-prints

\bibitem[{{Katz}(1992)}]{katz92}
{Katz} N., 1992, \apj, 391, 502

\bibitem[{{Katz}, {Weinberg} \& {Hernquist}(1996){Katz}, {Weinberg}, \&
  {Hernquist}}]{katz96}
{Katz} N., {Weinberg} D.~H., {Hernquist} L., 1996, \apjs, 105, 19

\bibitem[{{Kauffmann}, {White} \& {Guiderdoni}(1993){Kauffmann}, {White}, \&
  {Guiderdoni}}]{kauffmann93}
{Kauffmann} G., {White} S.~D.~M., {Guiderdoni} B., 1993, \mnras, 264, 201

\bibitem[{{Kennicutt}(1998)}]{kennicutt98a}
{Kennicutt}, Jr. R.~C., 1998, \apj, 498, 541

\bibitem[{{Kirshner} {et~al}\mbox{.}(1981){Kirshner}, {Oemler}, {Schechter}, \&
  {Shectman}}]{kirshner81}
{Kirshner} R.~P., {Oemler}, Jr. A., {Schechter} P.~L., {Shectman} S.~A., 1981,
  \apjl, 248, L57

\bibitem[{{Kitayama} \& {Suto}(1996)}]{kitayama96}
{Kitayama} T., {Suto} Y., 1996, \apj, 469, 480

\bibitem[{{Klypin} \& {Shandarin}(1983)}]{klypin83}
{Klypin} A.~A., {Shandarin} S.~F., 1983, \mnras, 204, 891

\bibitem[{{Kobayashi}(2004)}]{kobayashi04}
{Kobayashi} C., 2004, \mnras, 347, 740

\bibitem[{{Kobayashi} {et~al}\mbox{.}(1998){Kobayashi}, {Tsujimoto}, {Nomoto},
  {Hachisu}, \& {Kato}}]{kobayashi98}
{Kobayashi} C., {Tsujimoto} T., {Nomoto} K., {Hachisu} I., {Kato} M., 1998,
  \apjl, 503, L155

\bibitem[{{Kobayashi} {et~al}\mbox{.}(2006){Kobayashi}, {Umeda}, {Nomoto},
  {Tominaga}, \& {Ohkubo}}]{kobayashi06}
{Kobayashi} C., {Umeda} H., {Nomoto} K., {Tominaga} N., {Ohkubo} T., 2006,
  \apj, 653, 1145

\bibitem[{{Komatsu} {et~al}\mbox{.}(2011){Komatsu}, {Smith}, {Dunkley},
  {Bennett}, {Gold}, {Hinshaw}, {Jarosik}, {Larson}, {Nolta}, {Page},
  {Spergel}, {Halpern}, {Hill}, {Kogut}, {Limon}, {Meyer}, {Odegard}, {Tucker},
  {Weiland}, {Wollack}, \& {Wright}}]{komatsu11}
{Komatsu} E. {et~al.}, 2011, \apjs, 192, 18

\bibitem[{{Kreckel} {et~al}\mbox{.}(2011){Kreckel}, {Platen},
  {Arag{\'o}n-Calvo}, {van Gorkom}, {van de Weygaert}, {van der Hulst},
  {Kova{\v c}}, {Yip}, \& {Peebles}}]{kreckel11}
{Kreckel} K. {et~al.}, 2011, \aj, 141, 4

\bibitem[{{Larson}(1974)}]{larson74}
{Larson} R.~B., 1974, \mnras, 169, 229

\bibitem[{{Leisawitz}, {Bash} \& {Thaddeus}(1989){Leisawitz}, {Bash}, \&
  {Thaddeus}}]{leisawitz89}
{Leisawitz} D., {Bash} F.~N., {Thaddeus} P., 1989, \apjs, 70, 731

\bibitem[{{Leitherer}, {Robert} \& {Drissen}(1992){Leitherer}, {Robert}, \&
  {Drissen}}]{leitherer92}
{Leitherer} C., {Robert} C., {Drissen} L., 1992, \apj, 401, 596

\bibitem[{{Mannucci}, {Della Valle} \& {Panagia}(2006){Mannucci}, {Della
  Valle}, \& {Panagia}}]{mannucci06}
{Mannucci} F., {Della Valle} M., {Panagia} N., 2006, \mnras, 370, 773

\bibitem[{{Maoz}, {Mannucci} \& {Brandt}(2012){Maoz}, {Mannucci}, \&
  {Brandt}}]{maoz12a}
{Maoz} D., {Mannucci} F., {Brandt} T.~D., 2012, \mnras, 426, 3282

\bibitem[{{Marigo}(2001)}]{marigo01}
{Marigo} P., 2001, \aap, 370, 194

\bibitem[{{Massey} {et~al}\mbox{.}(2007){Massey}, {Heymans}, {Berg{\'e}},
  {Bernstein}, {Bridle}, {Clowe}, {Dahle}, {Ellis}, {Erben}, {Hetterscheidt},
  {High}, {Hirata}, {Hoekstra}, {Hudelot}, {Jarvis}, {Johnston}, {Kuijken},
  {Margoniner}, {Mandelbaum}, {Mellier}, {Nakajima}, {Paulin-Henriksson},
  {Peeples}, {Roat}, {Refregier}, {Rhodes}, {Schrabback}, {Schirmer}, {Seljak},
  {Semboloni}, \& {van Waerbeke}}]{massey07}
{Massey} R. {et~al.}, 2007, \mnras, 376, 13

\bibitem[{{Mathews} {et~al}\mbox{.}(2014){Mathews}, {Snedden}, {Phillips},
  {Suh}, {Coughlin}, {Bhattacharya}, {Zhao}, \& {Lan}}]{mathews14}
{Mathews} G.~J., {Snedden} A., {Phillips} L.~A., {Suh} I.-S., {Coughlin} J.,
  {Bhattacharya} A., {Zhao} X., {Lan} N.~Q., 2014, Modern Physics Letters A,
  29, 30012

\bibitem[{{Matteucci} {et~al}\mbox{.}(2006){Matteucci}, {Panagia}, {Pipino},
  {Mannucci}, {Recchi}, \& {Della Valle}}]{matteucci06}
{Matteucci} F., {Panagia} N., {Pipino} A., {Mannucci} F., {Recchi} S., {Della
  Valle} M., 2006, \mnras, 372, 265

\bibitem[{{McKee} \& {Ostriker}(1977)}]{mckee77}
{McKee} C.~F., {Ostriker} J.~P., 1977, \apj, 218, 148

\bibitem[{{Mo}, {van den Bosch} \& {White}(2010){Mo}, {van den Bosch}, \&
  {White}}]{mo10}
{Mo} H., {van den Bosch} F., {White} S., 2010, Galaxy Formation and Evolution.
  Cambridge University Press, New York, NY

\bibitem[{{Moore}, {Lake} \& {Katz}(1998){Moore}, {Lake}, \& {Katz}}]{moore98}
{Moore} B., {Lake} G., {Katz} N., 1998, \apj, 495, 139

\bibitem[{{Mosconi} {et~al}\mbox{.}(2001){Mosconi}, {Tissera}, {Lambas}, \&
  {Cora}}]{mosconi01}
{Mosconi} M.~B., {Tissera} P.~B., {Lambas} D.~G., {Cora} S.~A., 2001, \mnras,
  325, 34

\bibitem[{{Narayanan} {et~al}\mbox{.}(2011){Narayanan}, {Savage}, {Wakker},
  {Danforth}, {Yao}, {Keeney}, {Shull}, {Sembach}, {Froning}, \&
  {Green}}]{narayanan11}
{Narayanan} A. {et~al.}, 2011, \apj, 730, 15

\bibitem[{{Navarro}, {Abadi} \& {Steinmetz}(2004){Navarro}, {Abadi}, \&
  {Steinmetz}}]{navarro04}
{Navarro} J.~F., {Abadi} M.~G., {Steinmetz} M., 2004, \apjl, 613, L41

\bibitem[{{Navarro} \& {White}(1993)}]{navarro93}
{Navarro} J.~F., {White} S.~D.~M., 1993, \mnras, 265, 271

\bibitem[{{Nicastro}(2003)}]{nicastro03}
{Nicastro} F., 2003, ArXiv Astrophysics e-prints

\bibitem[{{Nomoto} {et~al}\mbox{.}(1997{\natexlab{a}}){Nomoto}, {Hashimoto},
  {Tsujimoto}, {Thielemann}, {Kishimoto}, {Kubo}, \& {Nakasato}}]{nomoto97b}
{Nomoto} K., {Hashimoto} M., {Tsujimoto} T., {Thielemann} F.-K., {Kishimoto}
  N., {Kubo} Y., {Nakasato} N., 1997{\natexlab{a}}, Nuclear Physics A, 616, 79

\bibitem[{{Nomoto} {et~al}\mbox{.}(1997{\natexlab{b}}){Nomoto}, {Iwamoto},
  {Nakasato}, {Thielemann}, {Brachwitz}, {Tsujimoto}, {Kubo}, \&
  {Kishimoto}}]{nomoto97a}
{Nomoto} K., {Iwamoto} K., {Nakasato} N., {Thielemann} F.-K., {Brachwitz} F.,
  {Tsujimoto} T., {Kubo} Y., {Kishimoto} N., 1997{\natexlab{b}}, Nuclear
  Physics A, 621, 467

\bibitem[{{Oemler}(1974)}]{oemler74}
{Oemler}, Jr. A., 1974, \apj, 194, 1

\bibitem[{{Okamoto} {et~al}\mbox{.}(2005){Okamoto}, {Eke}, {Frenk}, \&
  {Jenkins}}]{okamoto05}
{Okamoto} T., {Eke} V.~R., {Frenk} C.~S., {Jenkins} A., 2005, \mnras, 363, 1299

\bibitem[{{Oppenheimer} \& {Dav{\'e}}(2006)}]{oppenheimer06}
{Oppenheimer} B.~D., {Dav{\'e}} R., 2006, \mnras, 373, 1265

\bibitem[{{Oppenheimer} {et~al}\mbox{.}(2012){Oppenheimer}, {Dav{\'e}}, {Katz},
  {Kollmeier}, \& {Weinberg}}]{oppenheimer12}
{Oppenheimer} B.~D., {Dav{\'e}} R., {Katz} N., {Kollmeier} J.~A., {Weinberg}
  D.~H., 2012, \mnras, 420, 829

\bibitem[{{Oppenheimer} {et~al}\mbox{.}(2010){Oppenheimer}, {Dav{\'e}},
  {Kere{\v s}}, {Fardal}, {Katz}, {Kollmeier}, \& {Weinberg}}]{oppenheimer10}
{Oppenheimer} B.~D., {Dav{\'e}} R., {Kere{\v s}} D., {Fardal} M., {Katz} N.,
  {Kollmeier} J.~A., {Weinberg} D.~H., 2010, \mnras, 406, 2325

\bibitem[{{Paz}, {Stasyszyn} \& {Padilla}(2008){Paz}, {Stasyszyn}, \&
  {Padilla}}]{paz08}
{Paz} D.~J., {Stasyszyn} F., {Padilla} N.~D., 2008, \mnras, 389, 1127

\bibitem[{{Peebles}(2001{\natexlab{a}})}]{peebles01b}
{Peebles} P.~J.~E., 2001{\natexlab{a}}, International Journal of Modern Physics
  A, 16, 4223

\bibitem[{{Peebles}(2001{\natexlab{b}})}]{peebles01a}
{Peebles} P.~J.~E., 2001{\natexlab{b}}, \apj, 557, 495

\bibitem[{{Piontek} \& {Steinmetz}(2011)}]{piontek11}
{Piontek} F., {Steinmetz} M., 2011, \mnras, 410, 2625

\bibitem[{{Podsiadlowski} {et~al}\mbox{.}(2008){Podsiadlowski}, {Mazzali},
  {Lesaffre}, {Han}, \& {F{\"o}rster}}]{podsiadlowski08}
{Podsiadlowski} P., {Mazzali} P., {Lesaffre} P., {Han} Z., {F{\"o}rster} F.,
  2008, \nar, 52, 381

\bibitem[{{Portinari}, {Chiosi} \& {Bressan}(1998){Portinari}, {Chiosi}, \&
  {Bressan}}]{portinari98}
{Portinari} L., {Chiosi} C., {Bressan} A., 1998, \aap, 334, 505

\bibitem[{{Richter} {et~al}\mbox{.}(2004){Richter}, {Savage}, {Tripp}, \&
  {Sembach}}]{richter04}
{Richter} P., {Savage} B.~D., {Tripp} T.~M., {Sembach} K.~R., 2004, \apjs, 153,
  165

\bibitem[{{Rojas} {et~al}\mbox{.}(2004){Rojas}, {Vogeley}, {Hoyle}, \&
  {Brinkmann}}]{rojas04}
{Rojas} R.~R., {Vogeley} M.~S., {Hoyle} F., {Brinkmann} J., 2004, \apj, 617, 50

\bibitem[{{Rosati}, {Borgani} \& {Norman}(2002){Rosati}, {Borgani}, \&
  {Norman}}]{rosati02}
{Rosati} P., {Borgani} S., {Norman} C., 2002, \araa, 40, 539

\bibitem[{{Rubin} {et~al}\mbox{.}(2014){Rubin}, {Hennawi}, {Prochaska},
  {Simcoe}, {Myers}, \& {Wingyee Lau}}]{rubin14b}
{Rubin} K.~H.~R., {Hennawi} J.~F., {Prochaska} J.~X., {Simcoe} R.~A., {Myers}
  A., {Wingyee Lau} M., 2014, ArXiv e-prints

\bibitem[{{Salpeter}(1955)}]{salpeter55}
{Salpeter} E.~E., 1955, \apj, 121, 161

\bibitem[{{Sarazin}(1986)}]{sarazin86}
{Sarazin} C.~L., 1986, Reviews of Modern Physics, 58, 1

\bibitem[{{Savage}, {Tripp} \& {Lu}(1998){Savage}, {Tripp}, \& {Lu}}]{savage98}
{Savage} B.~D., {Tripp} T.~M., {Lu} L., 1998, \aj, 115, 436

\bibitem[{{Schaye} \& {Dalla Vecchia}(2008)}]{schaye08}
{Schaye} J., {Dalla Vecchia} C., 2008, \mnras, 383, 1210

\bibitem[{{Schaye} {et~al}\mbox{.}(2010){Schaye}, {Dalla Vecchia}, {Booth},
  {Wiersma}, {Theuns}, {Haas}, {Bertone}, {Duffy}, {McCarthy}, \& {van de
  Voort}}]{schaye10}
{Schaye} J. {et~al.}, 2010, \mnras, 402, 1536

\bibitem[{{Schmidt}(1959)}]{schmidt59}
{Schmidt} M., 1959, \apj, 129, 243

\bibitem[{{Scoccimarro} {et~al}\mbox{.}(2012){Scoccimarro}, {Hui}, {Manera}, \&
  {Chan}}]{scoccimarro12}
{Scoccimarro} R., {Hui} L., {Manera} M., {Chan} K.~C., 2012, \prd, 85, 083002

\bibitem[{{Sembach} {et~al}\mbox{.}(2004){Sembach}, {Tripp}, {Savage}, \&
  {Richter}}]{sembach04}
{Sembach} K.~R., {Tripp} T.~M., {Savage} B.~D., {Richter} P., 2004, \apjs, 155,
  351

\bibitem[{{Shen}, {Wadsley} \& {Stinson}(2010){Shen}, {Wadsley}, \&
  {Stinson}}]{shen10}
{Shen} S., {Wadsley} J., {Stinson} G., 2010, \mnras, 407, 1581

\bibitem[{{Shull}, {Smith} \& {Danforth}(2012){Shull}, {Smith}, \&
  {Danforth}}]{shull12}
{Shull} J.~M., {Smith} B.~D., {Danforth} C.~W., 2012, \apj, 759, 23

\bibitem[{{Snedden} {et~al}\mbox{.}(2014){Snedden}, {Arielle Phillips},
  {Mathews}, {Coughlin}, {Suh}, \& {Bhattacharya}}]{snedden14}
{Snedden} A., {Arielle Phillips} L., {Mathews} G.~J., {Coughlin} J., {Suh}
  I.-S., {Bhattacharya} A., 2014, ArXiv e-prints

\bibitem[{{Snedden} \& {Phillips}(2012)}]{snedden12}
{Snedden} A., {Phillips} L., 2012, in American Astronomical Society Meeting
  Abstracts, Vol. 219, American Astronomical Society Meeting Abstracts \#219,
  p. \#336.03

\bibitem[{{Sommer-Larsen}, {G{\"o}tz} \& {Portinari}(2003){Sommer-Larsen},
  {G{\"o}tz}, \& {Portinari}}]{sommer03}
{Sommer-Larsen} J., {G{\"o}tz} M., {Portinari} L., 2003, \apj, 596, 47

\bibitem[{{Springel}(2005)}]{springel05c}
{Springel} V., 2005, \mnras, 364, 1105

\bibitem[{{Springel} \& {Hernquist}(2002)}]{springel02}
{Springel} V., {Hernquist} L., 2002, \mnras, 333, 649

\bibitem[{{Springel} \& {Hernquist}(2003{\natexlab{a}})}]{springel03a}
{Springel} V., {Hernquist} L., 2003{\natexlab{a}}, \mnras, 339, 289

\bibitem[{{Springel} \& {Hernquist}(2003{\natexlab{b}})}]{springel03b}
{Springel} V., {Hernquist} L., 2003{\natexlab{b}}, \mnras, 339, 312

\bibitem[{{Springel}, {Yoshida} \& {White}(2001){Springel}, {Yoshida}, \&
  {White}}]{springel01}
{Springel} V., {Yoshida} N., {White} S.~D.~M., 2001, \na, 6, 79

\bibitem[{{Stinson} {et~al}\mbox{.}(2006){Stinson}, {Seth}, {Katz}, {Wadsley},
  {Governato}, \& {Quinn}}]{stinson06}
{Stinson} G., {Seth} A., {Katz} N., {Wadsley} J., {Governato} F., {Quinn} T.,
  2006, \mnras, 373, 1074

\bibitem[{{Thacker} \& {Couchman}(2000)}]{thacker00}
{Thacker} R.~J., {Couchman} H.~M.~P., 2000, \apj, 545, 728

\bibitem[{{Thacker} \& {Couchman}(2001)}]{thacker01}
{Thacker} R.~J., {Couchman} H.~M.~P., 2001, \apjl, 555, L17

\bibitem[{{Tittley} \& {Henriksen}(2001)}]{tittley01}
{Tittley} E.~R., {Henriksen} M., 2001, \apj, 563, 673

\bibitem[{{Tripp} \& {Savage}(2000)}]{tripp00}
{Tripp} T.~M., {Savage} B.~D., 2000, \apj, 542, 42

\bibitem[{{Trujillo}, {Carretero} \& {Patiri}(2006){Trujillo}, {Carretero}, \&
  {Patiri}}]{trujillo06}
{Trujillo} I., {Carretero} C., {Patiri} S.~G., 2006, \apjl, 640, L111

\bibitem[{{van Gorkom}(2004)}]{vangorkom04}
{van Gorkom} J.~H., 2004, Clusters of Galaxies: Probes of Cosmological
  Structure and Galaxy Evolution, 305

\bibitem[{{Van Waerbeke} {et~al}\mbox{.}(2000){Van Waerbeke}, {Mellier},
  {Erben}, {Cuillandre}, {Bernardeau}, {Maoli}, {Bertin}, {McCracken}, {Le
  F{\`e}vre}, {Fort}, {Dantel-Fort}, {Jain}, \& {Schneider}}]{vanwaerbeke00}
{Van Waerbeke} L. {et~al.}, 2000, \aap, 358, 30

\bibitem[{{Vogeley} {et~al}\mbox{.}(1994){Vogeley}, {Geller}, {Park}, \&
  {Huchra}}]{vogeley94}
{Vogeley} M.~S., {Geller} M.~J., {Park} C., {Huchra} J.~P., 1994, \aj, 108, 745

\bibitem[{{Vogelsberger} {et~al}\mbox{.}(2013){Vogelsberger}, {Genel},
  {Sijacki}, {Torrey}, {Springel}, \& {Hernquist}}]{vogelsberger13}
{Vogelsberger} M., {Genel} S., {Sijacki} D., {Torrey} P., {Springel} V.,
  {Hernquist} L., 2013, \mnras, 436, 3031

\bibitem[{{White} \& {Rees}(1978)}]{white78}
{White} S.~D.~M., {Rees} M.~J., 1978, \mnras, 183, 341

\bibitem[{{Wiersma} {et~al}\mbox{.}(2009){Wiersma}, {Schaye}, {Theuns}, {Dalla
  Vecchia}, \& {Tornatore}}]{wiersma09b}
{Wiersma} R.~P.~C., {Schaye} J., {Theuns} T., {Dalla Vecchia} C., {Tornatore}
  L., 2009, \mnras, 399, 574

\bibitem[{{Williams}, {Blitz} \& {McKee}(2000){Williams}, {Blitz}, \&
  {McKee}}]{williams00}
{Williams} J.~P., {Blitz} L., {McKee} C.~F., 2000, Protostars and Planets IV,
  97

\bibitem[{{Wittman} {et~al}\mbox{.}(2000){Wittman}, {Tyson}, {Kirkman},
  {Dell'Antonio}, \& {Bernstein}}]{wittman00}
{Wittman} D.~M., {Tyson} J.~A., {Kirkman} D., {Dell'Antonio} I., {Bernstein}
  G., 2000, \nat, 405, 143

\bibitem[{{Xu}(1995)}]{xu95}
{Xu} G., 1995, \apjs, 98, 355

\bibitem[{{Zhang} {et~al}\mbox{.}(2009){Zhang}, {Yang}, {Faltenbacher},
  {Springel}, {Lin}, \& {Wang}}]{zhang09}
{Zhang} Y., {Yang} X., {Faltenbacher} A., {Springel} V., {Lin} W., {Wang} H.,
  2009, \apj, 706, 747

\end{thebibliography}

%\label{lastpage}

\end{document}